\documentstyle[12pt]{article}
\def\lromn#1{\uppercase\expandafter{\romannumeral#1}}

\makeatletter

\renewcommand{\theequation}%
{\arabic{section}.\arabic{equation}}
\@addtoreset{equation}{section}
\renewcommand{\appendix}{\par
  \setcounter{section}{0}
  \setcounter{subsection}{0}
  \renewcommand{\thesection}{Appendix~\Alph{section}}
  \renewcommand{\theequation}{\Alph{section}.\arabic{equation}}}

\makeatother

\begin{document}

\begin{flushright}
TU/98/555\\
RCNS-98-18\\
\end{flushright}

\begin{center}
\begin{large}

\bf{
Quantum Kinetic Equation and \\
Cosmic Pair Annihilation
}

\end{large}

\vspace{24pt}

\begin{large}
Sh. Matsumoto and M. Yoshimura

Department of Physics, Tohoku University\\
Sendai 980-8578 Japan\\
\end{large}

\vspace{48pt}

{\bf ABSTRACT}
\end{center}

Pair annihilation of heavy stable particle that occurs in the
early universe is investigated, and
quantum kinetic equation for the momentum distribution of the annihilating
particle is derived, using the influence functional method.
A bosonic field theory model is used to  describe the
pair annihilation in the presence of 
decay product particles making up a thermal environment.
A crossing symmetric Hartree approximation that determines
self-consistently the equilibrium distribution is developed
for an otherwise intractable theory.
The time evolution equation and its Markovian approximation is derived,
to give a generalized Boltzmann equation including off-shell effects.
The narrow width approximation to an energy integral in this
equation gives the usual Boltzmann equation 
in a thermal bath of light particles.
The off-shell effect is a correction to the Boltzmann equation at high
temperatures, but is dominant at low temperatures.
The effect changes the equilibrium distribution from the
familiar $1/(\,e^{\omega _{k}/T} - 1\,)$ to a modified one given by
a Gibbs formula.
Integrated over momenta, the particle number density becomes roughly
of order (coupling)$ \times \sqrt{T/M}\cdot T^{3}$ at low temperatures
for the S-wave annihilation.
The relic mass density in the present universe
is insensitive to the coupling strength in a large range of
the mass and the coupling parameters,
and scales with the WIMP mass as 
\( \:
\approx 6 \times 10^{4}\,eV\,cm^{-3}\,(M/GeV)^{4/3} \,.
\: \)
The bound from the closure density gives an
upper WIMP mass bound roughly of order 1 $GeV$ in the present model.

\newpage

\section{Introduction}

\vspace{0.5cm} 
\hspace*{0.5cm} 
The Boltzmann equation for the particle distribution function
is described in terms of 
S-matrix elements defined on the mass shell,
such as the cross section and the decay rate. 
This integro-differential equation is intuitively
appealing since relation to the classical theory is evident.
In the quantum mechanical context its foundation and its possible 
generalization have extensively been discussed in the past.
We may only mention two approaches; the real-time thermal Green's function
method \cite{real-time th green} 
and the closed time path method \cite{closed time path}.
Both use as the fundamental quantity the Green's function that contains
information off the mass shell.
It is generally believed that some sort of coarse graining or reduction 
of detailed quantum mechanical information is needed to derive
a quantum version of the Boltzmann equation.
There is however no unique generalization of the Boltzmann equation 
known to us.
Perhaps correction to the Bolzmann equation differs depending on
a physical situation one has in mind.

We investigate in the present work
the non-equilibrium process of pair annihilation
of some heavy particles in the presence of
a thermal bath of lighter particles.
Time evolution of the occupation number for the heavy particle
in a particular momentum mode is derived and analyzed from a new
perspective.

The problem is of great interest in cosmology, and
occurs for pair annihilation of stable particles such
as the anti-nucleon and the positron. 
For instance, the electron-positron annihilation has a great impact on
nucleosynthesis that takes place immediately after this
annihilation. The annihilation rate has been calculated using
a thermally averaged Boltzmann equation, but
estimate of the off-shell effect to the annihilation process
has never been worked out (see, however, 
\cite{finite temp effect on nucleosynthesis} for a limited calculation
of finite temperature effects).
This list of interesting candidates for the cosmic pair annihilation
should be expanded to include WIMP (weakly interacting
massive particle) or LSP (lightest supersymmetric particle)
\cite{lsp} that has been hypothesized, with a good motivation, as a 
solution to the dark matter problem in cosmology.

We would like to obtain a suitable and useful generalization of the
Boltzmann equation in these circumstances.
The physical situation we have in mind
is characterized by the presence of annihilating
heavy particles ($e^{\pm }$ in
the case of electron-positron annihilation) interacting among
themselves ($e^{\pm }$ scattering and 
pair annihilation into two photons) and with lighter
particles (scattering with photons).
Although we work out a particular toy model for this,
our method is general enough for extension to other cases, and
moreover our result is simple enough for  practical use.

We believe that for derivation of the generalized Boltzmann equation
it is important to use a general and flexible
framework allowing for a consistent approximation scheme.
It is often practically difficult to estimate correction to the Boltzmann
equation in the Green's function or the operator method.
Our experience in a similar, but a much simpler problem of the unstable
particle decay in thermal medium \cite{jmy-98-1} $-$ \cite{jmy-98-2}
suggests that the influence functional
method invented by Feynman and Vernon \cite{feynman-vernon} 
may be useful to the present problem.
The quantum mechanical problem of the excited state decay
\cite{caldeira-leggett 83},
analogous to, but simpler than, the unstable particle decay in field theory,
is completely solvable at the operator level \cite{jmy-98-1}, 
but at the same time
the influence functional method gives an identical result 
\cite{jmy-97} to the operator approach.
In the present work we develop a new Hartree type of approximation
of the annihilation-scattering problem within the influence functional
approach. We also utilize the operator identity to facilitate
our analysis.

In order to quantitatively discuss the role of scattering in the annihilation
process, it is important to simultaneously deal with the two processes,
the annihilation and the scattering.
It is thus best to employ a fully relativistic field theory that
respects the crossing symmetry.
For simplicity we take throughout this paper
an interaction Lagrangian density of
the form, $\varphi ^{2}\chi ^{2}$, where $\varphi $ is a heavy bosonic
field and $\chi $ a lighter bosonic field.
We assume that the mass of $\varphi \gg $  the mass of $\chi $.
The annihilation channel $\varphi \varphi \rightarrow \chi \chi $ is related
to the scattering channel $\varphi \chi \rightarrow \varphi \chi $ by
a crossing symmetry. The inverse processes to these and the other
1 to 3 processes, $\varphi \leftrightarrow \varphi \chi \chi $ and
$\chi \leftrightarrow \varphi \varphi \chi $, that may also occur
in thermal medium of $\chi $ particles are treated here symmetrically.
It is important to recall that a finite time behavior of the quantum
system in thermal medium allows the process such as
\( \:
\varphi \leftrightarrow \varphi \chi \chi \,, 
\: \)
even if it is kinematically forbidden for the on-shell S-matrix element.

We first derive an integral equation that self-consistently determines
the equilibrium distribution function $f(\vec{k})$ 
within the Hartree approximation.
Derivation of this equation is based on the assumption that
time variation of the distribution function proceeds
more slowly than indivisual microscopic reactions occur.
Under this assumption the result \cite{jmy-98-1}, \cite{jmy-97} of 
the completely solvable model may be used, and the large time 
($\gg $ relaxation time ) limit can be taken.
The resultant equilibrium distribution function deviates from
the ideal gas form, $1/(e^{\beta \omega _{k}} - 1)$, but
may be understood by the Gibbs formula $e^{-\,\beta H_{{\rm tot}}}$
with $H_{{\rm tot}}$ the total Hamiltonian including interaction
between the $\varphi $ system and the $\chi $ environment.

We then derive a time evolution equation for the distribution function.
This equation contains the initial memory term,
hence is non-Markovian, as any exact treatment of the quantum mechanical
behavior would demand.
We next devise a useful Markovian approximation to the Hartree
model and examine this approximation in detail.

The on-shell Boltzmann equation arises as a result of
the narrow resonance approximation for a Breit-Wigner type of energy
integral in our Hartree-Markovian model.
The Boltzmann approximation is excellent in many practical cases, but
it fails when the main contributing part to the energy integral 
includes the region off the resonance pole 
at $\omega \approx $ the $\varphi $
energy. Thus, the Boltzmann equation is
modified significantly for the environment temperature $\ll $ the
$\varphi $ mass $M$. 
We present a complete kinetic equation for the momentum
distribution function, which may be used at low temeratures.

Integration over the momentum of the distribution function 
gives a rate equation for the number density which is of
prime interest in the annihilation-scattering problem.
One may naively expect that 
the scattering process conserves the particle number, thus
scattering terms cancel in the Boltzmann equation.
This is explicitly confirmed for the on-shell part of our
quantum kinetic equation.
But it is not clear whether a similar complete cancellation occurs for
the off-shell quantity. This is because this conservation law
is based on the commutability of the particle number with
the scattering part of the effective Hamiltonian, along with
the unitary evolution in the quantum system.
Dissipation due to the thermal medium and its associated fluctuation however
causes a non-unitary evolution for a part of the entire system.
It is thus an interesting open question whether the scattering-related
contribution remains for the off-shell part.
In any event we find that the scattering and its inverse process 
do contribute, but with a very small rate.
A large number density of order,
\( \:
O[10^{-3}]\,\lambda \,\sqrt{\frac{T}{M}}\,T^{3}
\,, 
\: \)
is derived for the off-shell contribution of the inverse annihilation
process, where
$\lambda $ is the relevant $\varphi ^{2}\chi ^{2}$ coupling.
At low temperatures this becomes much larger than the usual one,
\( \:
\approx (MT/2\pi )^{3/2}\,e^{-M/T} \,, 
\: \)
namely the thermal number density of zero chemical potential.

In cosmology the thermal environment gradually changes according to
the adiabatic law; the temperature $\propto $ the cosmic scale factor
$1/a(t)$. Along with the obvious change of the number density 
$\propto  1/a^{3}(t)$,
the thermally averaged rate in the generalized Boltzmann equation
decreases as the temperature decreases.
Thus, the decoupling or the freeze-out
of annihilation takes place roughly at the temperature
when the thermal rate is equal to the Hubble rate 
\cite{heavy particle decoupling}.
What is left after the freeze-out is then the relic abundance of heavy
stable particles.
It is important to accurately estimate the relic abundance
of WIMP or LSP for the dark matter problem. 

The off-shell effect considered in the present work gives a 
relic abundance of order
\( \:
Y = n_{\varphi }/n_{\gamma } \approx O[0.1]\,
(\frac{M}{m_{{\rm pl}}})^{1/3}
\,,
\: \)
with $m_{{\rm pl}}$ the Planck mass.
Unlike the estimate based on the on-shell Boltzmann equation,
this abundance does not suffer from the suppression factor 
\( \:
e^{-M/T_{f}}
\: \)
where $T_{f}$ is the freeze-out temperature, usually much less than $M$.
Moreover, the relic fraction $Y$ is insensitive to the $\varphi ^{2}\chi ^{2}$
coupling, if the coupling is not too small, that is if
\( \:
\lambda > 9.3 \times 10^{-5}\,(M/GeV)^{0.32}
\: \)
for $10^{-3}\,GeV < M < 1\,TeV$.
Translated to the mass density of the present universe, this gives
an allowed WIMP mass range for the dark matter; $M < 1\,GeV$.
It is of considerable interest to work out the allowed parameter
region for LSP in realistic supersymmetric theories.

The rest of this paper is organized as follows. 
In Section 2 the influence functional method is briefly
explained, taking our model of $\varphi ^{2}\chi ^{2}$ interaction.
A non-perturbative approximation of the Hartree type is then introduced
in connection to the influence functional.
The kernel function in the reduced Hartree model is determined
in terms of the spectral function that itself contains the correlator
to be determined.
In Section 3 we explain how the concept of the slow variation
of the particle distribution helps to derive 
a self-consistent equation that determines
the equilibrium distribution function. 
This is achieved by using the result of
the exactly solvable model of the excited state decay.
In particular, the explicit form of the coincident time limit
of two-body correlators is required for its derivation, and 
we calculate the correlator 
by using a new generating functional method 
within the influence functional formalism.

In Section 4
the time evolution equation is derived for the occupation
number. This equation is non-Markovian and
the initial memory term is identified.
A simple Markovian approximation becomes possible, 
again under the assumption that the time scale of the $\varphi $ number
density variation is larger than that of the microscopic reaction time.
The Markovian equation thus derived
makes its relation to the usual Boltzmann equation transpant.
We then point out that this Markovian equation contains
the off-shell effect which becomes dominant at low temperatures.
Cancellation of the scattering process
in the on-shell contribution, when one integrates over momenta,
becomes important to get a large off-shell contribution to
the number density at low temperatures.
In Section 5 the cosmological estimate of the relic abundance 
of heavy stable particles is given in the present model.
The off-shell effect prolongs the freeze-out epoch, but 
due to a power law decrease ($\propto T^{3.5}$) of the number density
with temperature, 
the net effect gives a larger freeze-out density.
Finally a simple estimate of the present mass density is given;
\( \:
\rho _{0} \approx O[6\times 10^{4}]\,(M/GeV)^{4/3}\,eV\,cm^{-3} \,
\: \)
in a certain range of the parameter space $(M\,, \lambda )$.

In four Appendices we explain technical points relegated in
the main text; (A) generating functional method in the influence
functional framework, (B) renormalization of the off-shell distribution
function, (C) some details for the unstable
particle decay treated in the Hartree approximation, and
(D) technical details for computing various integrals 
needed to obtain the off-shell distribution function.

A short summary of our result stressing the estimate of relic
abundance has been given in a previous note \cite{my 98-1}.
In the present longer paper we give all details stated there and
present further new results not given there.

\vspace{0.5cm} 
\section{Hartree approximation}

\vspace{0.5cm} 
\hspace*{0.5cm} 
We consider a relativistic field theory model of the Lagrangian density
given by
\begin{eqnarray}
&&
{\cal L} = {\cal L}_{\varphi } + {\cal L}_{\chi } +
{\cal L}_{{\rm int}}  \,, 
\\ &&
{\cal L}_{\varphi } + {\cal L}_{\chi } = \frac{1}{2}\,
(\partial _{\mu }\varphi )^{2} - \frac{1}{2}\, M^{2}\,\varphi ^{2}
+ \frac{1}{2}\, (\partial _{\mu }\chi )^{2} - 
\frac{1}{2}\, m^{2}\,\chi ^{2} \,,
\\ &&
{\cal L}_{{\rm int}} = -\,\frac{\lambda }{2}\,\varphi ^{2}\,\chi ^{2}
\,.
\end{eqnarray}
It is later explained how to renormalize and introduce counter terms
when it becomes necessary.
The heavy particle $\varphi $ can pair-annihilate into a $\chi $ pair;
$\varphi \varphi \rightarrow \chi \chi $ with the dimensionless
interaction strength $\lambda $.
This coupling $\lambda $ must be less than unity for our method
to be useful.
We assume that the lighter particle $\chi $ makes up a thermal 
environment of temperature $T = 1/\beta $ in our unit of
the Boltzmann constant
\( \:
k_{B} = 1
\: \).
It is thus assumed that although not written explicitly, there
is interaction among $\chi $ particles themselves or with some
other light particles to maintain a thermal equilibrium.

In the problem of our interest
one focusses on a particular
dynamical degree of freedom (the $\varphi $ field in our case)
and integrates out the environment part (the $\chi $ field)
altogether. 
In the influence functional method \cite{feynman-vernon} 
this integration is carried out for the squared amplitude,
namely for the probability function directly,
by using the path integral technique.
This way one has to deal with the conjugate field variable $\varphi '$
along with $\varphi $, since the complex conjugated quantity is
multiplied in the probability.

Define first the influence functional ${\cal F}$
by integrating out the $\chi $
field degree of freedom during a fixed time interval between
$t_{i}$ and $t_{f}$;
\begin{eqnarray}
&& 
\int\,{\cal D}\chi  \,\int\,{\cal D}\chi '\,\exp 
\left[\, i \int\,dx\,
\left( \,{\cal L}_{\chi }(x) - {\cal L}_{\chi '}(x)
\right.
\right.
\nonumber \\ && \hspace*{0.5cm} \left. \left.
+\, {\cal L}_{{\rm int}}(\varphi(x) \,, \chi (x))
- {\cal L}_{{\rm int}}(\varphi'(x) \,, \chi' (x))
\,\right) \right] \,.
\end{eqnarray}
We convolute with this the initial and and the final density matrix
of the $\chi  $ system.
For the initial state it is assumed that the entire system
is described by an uncorrelated product of the system and the
environment density matrix,
\begin{equation}
\rho _{i}^{(\varphi )} \times \rho _{i}^{(\chi )} \,, \hspace{0.5cm} 
\rho _{i}^{(\chi )} = 
\rho _{\beta } = e^{-\,\beta H_{0}(\chi )}/{\rm tr}\,
e^{-\,\beta H_{0}(\chi )} \,.
\end{equation}
Here $\rho _{\beta }$ is the density matrix for a thermal
environment, written in the operator notation using 
the environment Hamiltonian $H_{0}(\chi )$. 
To avoid a possible confusion, we sometimes write the operator 
explicitly by the tilded letter such as $\tilde{\chi }$ here.
The density matrix $\rho _{i}^{(\varphi )}$
for the $\varphi $ system is arbitrary, except that it is assumed to
commute with the $\varphi $ Hamiltonian,
\begin{equation}
\left[ \,\rho _{i}^{(\varphi )} \,, H_{0}(\varphi )\,\right] = 0
\,.
\end{equation}
We believe that this restriction on the $\varphi $ state
does not exclude many practical cases of interest.
At the final time $t_{f}$ the $\chi $ integration
is performed taking the condition of non-observation for the environment,
\begin{eqnarray}
\int\,d\chi _{f}\,\int\,d\chi _{f}'\,\delta (\chi _{f} - \chi_{f}')\,
(\cdots ) \,, 
\end{eqnarray}
with the understanding that the environment is totally unspecified
at the time $t_{f}$.

The result of $\chi $ integration is given by a Gaussian integral
and the influence functional to order $\lambda ^{2}$ 
is of the form,
\begin{eqnarray}
&& \hspace*{1cm}
{\cal F}_{4}[\varphi \,, \varphi '] =
\nonumber \\ && \hspace*{-1cm}
\exp [\,-\,\frac{1}{4}\,\int_{x_{0} > y_{0}}\,dx\,dy\,
\left( \,\xi _{2}(x)\,\alpha _{R}(x\,,  y)\,\xi _{2}(y) + i\,
\xi _{2}(x)\,\alpha _{I}(x \,,  y)\,X_{2}(y)\,\right)\,] \,, 
\label{original influence functional} 
\\ &&
\hspace*{1cm} 
X_{2}(x) \equiv  \varphi ^{2}(x) + \varphi '\,^{2}(x) 
\,, \hspace{0.5cm} 
\xi _{2}(x) \equiv \varphi ^{2}(x) - \varphi '\,^{2}(x) \,, 
\\ && 
\alpha (x \,, y) = \alpha _{R}(x \,, y) + i\alpha _{I}(x \,, y)
= \lambda ^{2}\,
{\rm tr}\;\left( T[\widetilde{\chi }^{2}(x)\widetilde{\chi }^{2}(y)\,
\rho _{\beta }]\right) \,.
\label{2-point kernel} 
\end{eqnarray}
Note the presence of the time ordering, $x_{0} > y_{0}$, in the above
formula.
Unless a confusion occurs, we shall simplify the four dimensional
integral such as
\( \:
\int\,d^{4}x\,(\cdots )
\: \)
by writing $\int\,dx\,(\cdots )$.
We note that the kernel function $\alpha (x\,,  y)$ satisfies
the time translation invariance, thus may be written
as $\alpha (x - y)$, 
as can explicitly be shown 
by taking the complete set of $H_{0}(\chi )$ eigenstates;
\begin{eqnarray}
&&
{\rm tr}\;\left( \tilde{A}(x_{0}) \,\tilde{B}(y_{0})\,\rho _{\beta }\right)
= \sum_{n\,, i}\,e^{-\,i(E_{n} - E_{i})(x_{0} - y_{0})}
\,(\rho _{\beta })_{ii}
\,\langle i|\tilde{A}|n \rangle\,\langle n|\tilde{B}|i \rangle
\,.
\end{eqnarray}
An explicit form of the
kernel function $\alpha (x)$ or its Fourier transform is given later.

Higher order terms in $\lambda ^{2}$ are actually present in the 
exponent of the influence functional. These contribute either to
many-body processes we are not interested in or to higher order
terms in our process. Moreover, Feynman and Vernon proved that
the above form of the influence functional satisfies the required 
fundamental property such as the causality and the unitarity.
We may thus safely ignore these higher order contributions
in the weak coupling limit.

Another point is the tadpole contribution and how the kernel $\alpha_{i} $
is modified due to the tadpole, which we discuss later.
(See eq.(\ref{modified 2-point kernel by tadpole}) for the
modification.)

The convolution with the system variable $\varphi $ gives the
reduced density matrix for the system at any given time $t_{f}$;
\begin{eqnarray}
&& \hspace*{-1cm}
\rho ^{(R)}(\varphi _{f} \,, \varphi _{f}') =
\,\int\,d\varphi _{i}\,\int\,d\varphi_{i}'
\int\,{\cal D}\varphi \,\int\,{\cal D}\varphi '
\,e^{iS(\varphi ) - iS(\varphi ')}\,
{\cal F}[\varphi \,, \varphi ']\,\rho _{i}^{(\varphi )}
(\varphi _{i} \,, \varphi _{i}')
 \,, \nonumber \\ &&
\end{eqnarray}
from which one can deduce physical quantities for the $\varphi $ system.
Here $S(\varphi )$ is the action for the $\varphi $ system obtained
from the basic Lagrangian.

It is often useful to introduce a notation for the
correlator in the influence functional method;
for $x_{0} > y_{0}$,
\begin{eqnarray}
&& 
\langle \,\varphi (x) \varphi (y)\, \rangle =
\int\,d\varphi (x)\,\int\,d\varphi'(x)
\,\int\,d\varphi (y)\,\int\,d\varphi '(y)
\nonumber \\ && \hspace*{-1cm} 
\cdot \int\,{\cal D}\varphi \,\int\,{\cal D}\varphi '
\,e^{iS(\varphi ) - iS(\varphi ')}\,
{\cal F}[\varphi \,, \varphi ']\,\varphi (x)\varphi (y)
\,\rho ^{(R)}(\varphi (y) \,, \varphi' (y))
\,. 
\end{eqnarray}
In this formula the functional integration is performed for
$\varphi $ during the time interval, $x_{0} > t > y_{0}$, where
$t_{f} > x_{0} > y_{0} > t_{i}$.
The path integral prior to the time $y_{0}$ gives the reduced density matrix
$\rho ^{(R)}$ at the time $y_{0}$, evolved from $\rho _{i}$ at 
the time $t_{i}$. In general, the correlation function 
\( \:
\langle \varphi (x)\varphi (y) \rangle
\: \)
does not obey the time translation invariance,
reflecting that the initial memory is never completely
erased in the quantum system.
A nice feature of this correlator formula is that the initial memory
effect appears compactly via the reduced density matrix $\rho ^{(R)}$.
One may generalize the concept of the expectation value 
$\langle \cdots  \rangle$
to any multiple of local operators including also the conjugate $\varphi '$.

The model thus specified is difficult to solve due to the appearance of
the quartic term of $\varphi $ in the influence functional 
(\ref{original influence functional}). 
The situation is however simplified when one
considers a mean field approximation. 
In the mean field or the Hartree approximation one replaces
a product of multi-field operators by a product of two operators with
several averaged two-body correlators.
This approximation is good if one can ignore a higher order correlation
than that of two-body. 
The Hartree model is expected to work well in a dilute system.
The dilute system is defined by the low occupation
number for each mode. It should not be confused by a possibly large
value of the number density which is a mode-summed quantity. 
For the bose system we consider here this is a circumstance very far
from the bose condensed state. The dilute approximation seems good
in most cosmological application.

The Hartree approximation we now introduce is a Gaussian truncation to
the influence functional; we replace the original one by properly defining
a new kernel function $\beta (x\,, y)$ in the quadratic form;
\begin{eqnarray}
&& \hspace*{-1cm}
{\cal F}_{2}[\varphi \,, \varphi '] =
\exp [\,-\,\int_{x_{0} > y_{0}}\,dx\,dy\,
\left( \,\xi (x)\,\beta  _{R}(x\,, y)\,\xi (y) + i\,
\xi (x)\,\beta  _{I}(x\,, y)\,X(y)\,\right)\,] \,, 
\nonumber 
\\ && \label{influence functional in r-model} 
\\ &&
\hspace*{1cm} 
X(x) \,, \hspace{0.3cm}
\xi (x) \equiv \varphi (x) \pm \varphi '(x) \,.
\end{eqnarray}
As for the correlator, the new kernel $\beta _{i}$
in the truncated
model does not satisfy the time translation invariance.
The identification of the new kernel $\beta (x\,,  y)$
is made by comparing two point correlators to any arbitrary
order $\lambda ^{2}$;\\
\( \:
\langle X(x)\xi (y) \rangle \,, \langle X(x)X(y) \rangle
\,, \langle \xi (x) \xi (y)\rangle \,.
\: \)
It turns out that the last correlator $\langle \xi (x) \xi (y)\rangle $
vanishes both for the original and the Hartree-approximated model,
which can be regarded as a consistency check of our approach.
Once the Hartree model is determined by the kernel $\beta (x \,, y)$,
one can work out its consequences to all order of $\lambda $.

Let us first introduce two types of propagator for $\varphi $ field;
\begin{eqnarray}
&&
G(x \,, y) = 
i\,\langle \varphi (x)\varphi (y) \rangle_{{\cal F} = {\cal F}_{4}} \,,
\end{eqnarray}
and
\( \:
i\,\langle \varphi (x)\varphi (y) \rangle_{{\cal F} = 1} \,.
\: \)
The first one is the full propagator taking into
account the environment interaction.
The second one, on the other hand, 
is an extension of the free propagator in perturbative field theory
accommodated to a non-trivial $\varphi $ state in our problem.
Without the non-local interaction in the influence functional
${\cal F}$, one may use for 
$i\,\langle \varphi (x)\varphi (y) \rangle_{{\cal F} = 1}$
the complete set of eigenstates of the $\varphi $ Hamiltonian
$H_{0}(\varphi )$, to show that this quantity is translationally
invariant; introducing a new notation for the propagator,
\begin{eqnarray}
&& \hspace*{0.5cm} 
G_{0}(x - y) \equiv 
i\,\langle \varphi (x)\varphi (y) \rangle_{{\cal F} = 1} 
\nonumber 
\\ &&
\hspace*{-1.5cm} 
= \,
i\,\int_{-\infty }^{\infty }\,dk_{0}\,e^{-ik_{0}(x_{0} - y_{0})}\,
\sum_{n \,, i}\,\delta (k_{0} - E_{n} + E_{i})\,\rho ^{(\varphi )}_{i}\,
\langle i|\tilde{\varphi }(0 \,, \vec{x})|n \rangle\,
\langle n|\tilde{\varphi }(0 \,, \vec{y})|i \rangle \,.
\label{free propagator} 
\end{eqnarray}
Note  that $\rho _{i}^{(\varphi )}$ is time independent.
It is always useful to decompose this into independent momentum modes;
for $x_{0} > 0$
\begin{eqnarray}
-\,iG_{0}(x ) = \int\,\frac{d^{3}k}{(2\pi )^{3}2\omega _{k}}\,
\left( \, f_{0}(\vec{k})e^{ik\cdot x} +
(\,1 + f_{0}(\vec{k})\,)e^{- ik\cdot x}\,\right)
\,,
\end{eqnarray}
where $k_{0} = \omega _{k} = \sqrt{k^{2} + M^{2}}$.
One may use the harmonic oscillator basis $|n\rangle _{\vec{k}}$
($n = 0 \,, 1 \,, 2 \,, \cdots $)
in this plane wave decomposition to show
\begin{equation}
f_{0}(\vec{k}) = 
|\langle 1|\tilde{\varphi }(\vec{k})|0 \rangle_{\vec{k}}|^{2}\,
\sum_{n}\,n\,\rho _{n }^{(\vec{k})}  \,.
\end{equation}
In the usual language of the creation and the annihilation operators,
\begin{eqnarray}
&& \hspace*{-1cm}
\langle a_{1}a_{2} \rangle_{{\cal F} = 1} 
= 0  = \langle a^{\dag }_{1}a^{\dag }_{2} \rangle_{{\cal F} = 1} 
\,, \hspace{0.5cm} \langle a^{\dag }(\vec{k}')a(\vec{k}) \rangle
_{{\cal F} = 1} 
= f_{0}(\vec{k})\,\delta (\vec{k} - \vec{k}')
\,.
\end{eqnarray}

We compute correlators in two theories, using the two
different forms of the influence functional, 
${\cal F}_{4}$ and ${\cal F}_{2}$,
and identify two results. One then has an equality;
\begin{eqnarray}
&& 
\int\,dx'\,dy'\,G_{0}(x - x')\beta (x'\,, y')G_{0}(y' - y) 
\nonumber \\ 
&=& -\,i\,\int\,dx'\,dy'\,G_{0}(x - x')\alpha (x' - y')G(x'\,, y') 
G_{0}(y' - y)
\,. \label{hartree identity} 
\end{eqnarray}
This identity leads to a crucial relation of the two kernels;
\begin{eqnarray}
&&
\beta (x\,, y) = -\,i\,\alpha (x - y)\,G(x \,, y)
\,.
\label{relation between 2 kernels} 
\end{eqnarray}
It is important in the self-consistent
Hartree approximation that the full propagator $G$
instead of $G_{0}$ appears in this equation, since the truncated
model should contain the full $\varphi $ propagator to be determined
self-consistently.
Equivalently, the full propagator has a self-consistency
condition in the Hartree approximation; in the matrix form,
\begin{equation}
G = G_{0} - i\,G_{0}\beta G_{0} + \cdots = 
G_{0} - G_{0}(\alpha G)G_{0} + \cdots \,.
\end{equation}

The next task is to derive a useful relation between the spectral
functions that appear in eq.(\ref{relation between 2 kernels}).
The full propagator $G$ has two spacetime arguments, for which
we use the center of mass (CM) variable $(x + y)/2$ and
the relative coordinate $x - y$. We observe that there is no
dependence on the CM space coordinate $(\vec{x} + \vec{y})/2$ due
to the spatial homogeneity, hence
Fourier-transform the relative coordinate to get
\begin{eqnarray}
&&
-\,i\,G(x \,, y) = \int\,\frac{d^{4}k}{(2\pi )^{4}}\,
\tilde{f}(- \,k \,, \frac{x_{0} + y_{0}}{2})\,e^{ik\cdot (x - y)}
\,. 
\end{eqnarray}
The spectral $\tilde{f}$ here may depend on the CM time.

The occupation number $f(\vec{k} \,, t)$ that becomes important
in subsequent discussion is defined from
the coincident time limit of the Fourier transformed full propagator.
We first define the spatial Fourier decomposition of the full
propagator,
\begin{eqnarray}
&&
\tilde{G}(x_{0} \,, y_{0} \,; \vec{k})
\equiv \int\,d^{3}(x - y)\,G(x_{0} \,, \vec{x} \,; y_{0} \,, \vec{y})
\,e^{-i\vec{k}\cdot (\vec{x} - \vec{y})}
\\ && \hspace*{1cm} 
= \,
i\,\langle \tilde{\varphi }(\vec{k} \,, x_{0})\tilde{\varphi }(-\vec{k}\,, 
y_{0}) \rangle \,, \nonumber 
\\ && 
\tilde{\varphi }(\vec{k} \,, x_{0}) \equiv 
\int\,d^{3}x\,\varphi (\vec{x} \,, x_{0})\,e^{-i\vec{k}\cdot \vec{x}}
\,.
\end{eqnarray}
Due to the hermiticity of the field $\varphi $,
\begin{equation}
\tilde{\varphi }^{\dag }(\vec{k}\,, x_{0}) = \tilde{\varphi }
(-\,\vec{k} \,, x_{0}) \,.
\end{equation}
Using the Heisenberg equation for $\varphi $, 
one has for the occupation number
\begin{eqnarray}
&& \hspace*{-1cm}
\overline{\omega}_{k}
\,\left( f(\vec{k} \,, x_{0}) + \frac{1}{2} \right)
\equiv  
\frac{1}{2}\, 
\langle \,\frac{d\tilde{\varphi }^{\dag }(\vec{k}\,, x_{0})}{dx_{0}}
\frac{d\tilde{\varphi }(\vec{k}\,, x_{0})}{dx_{0}}
+ \overline{\omega}_{k}^{2}\,
\tilde{\varphi }^{\dag }(\vec{k}\,, x_{0})\tilde{\varphi }(\vec{k}\,, x_{0})
\, \rangle
\\ && \hspace*{-1cm}
=\, \frac{i}{4}\,\left[ \,(\frac{\partial }{\partial x_{0}}
- \frac{\partial }{\partial y_{0}})^{2}\,\tilde{G}(x_{0}\,, 
y_{0}\,; -\,\vec{k})\,\right]_{x_{0} \rightarrow y_{0}^{+}}
- \frac{\lambda }{2}\,
\langle \tilde{\varphi }^{\dag }
(\vec{k}\,, x_{0})\,\widetilde{\varphi \chi ^{2}}(\vec{k}\,, x_{0})
\rangle
 \,. 
\end{eqnarray}
The first quantity in the right hand side is further related to
$\tilde{f}$ by
\begin{equation}
f(\vec{k} \,, t) + \frac{1}{2} =
\frac{1}{\overline{\omega}_{k}}
\int\,\frac{dk_{0}}{2\pi }\,k_{0}^{2}\,\tilde{f}(k\,, t) 
- \frac{\lambda }{2\overline{\omega}_{k}}\,
\langle \tilde{\varphi }^{\dag }(\vec{k}\,, t)\,
\widetilde{\varphi \chi^{2}}(\vec{k}\,, t)
\rangle
\,.
\end{equation}
Here $\overline{\omega}_{k}$ is what we may call a reference energy
to define the occupation number for the mode $\vec{k}$.
We take for the time being $\overline{\omega}_{k} = \omega _{k}$,
the energy of free particle,
and later modify the definition of the occupation number $f(\vec{k}\,, t)$
slightly by allowing the temperature dependent $\varphi $ mass in
$\overline{\omega }_{k}$.

We define the spectral weight for the two kernels
in terms of the Fourier decomposition of the relative spacetime
coordinate;
for $x_{0} > y_{0}$
\begin{eqnarray}
&&
\alpha (x - y) = \int\,\frac{d^{4}k}{(2\pi )^{3}}\,\frac{2}{1 - e^{-\beta 
k_{0}}}\,r_{\chi }(k)\,e^{-ik\cdot (x - y)} \,, 
\label{fv kernel for 2-body} 
\\ &&
\beta (x \,, y) \equiv \int\,\frac{d^{4}k}{(2\pi )^{3}}\,
r(k \,, \frac{x_{0} + y_{0}}{2})\,e^{-ik\cdot (x - y)} \,.
\end{eqnarray}
A strange looking factor $1/(1 - e^{-\beta k_{0}})$ is inserted 
for the kernel $\alpha (x)$, 
since this factor compensates the non-analytic
property of the real-time thermal Green's function $\alpha (x)$ defined in
eq.(\ref{2-point kernel}), 
relative to the analytic imaginary-time Green's function.
The factor is well understood \cite{fetter-walecka} and is also 
explained in our previous paper \cite{hjmy-97}.
We shall shortly give an explicit formula of $r_{\chi }(k)$.
The spectral weight $r$ for the Hartree model is determined from
eq.(\ref{relation between 2 kernels}); it is of the form of
a convolution integral;
\begin{eqnarray}
&&
r(k \,, t) = \int\,\frac{d^{4}k'}{(2\pi )^{4}}\,
\tilde{f}(-\,k'\,,t)
\frac{2\,r_{\chi }(k + k')}{1 - e^{-\beta (k_{0} + k_{0}')}}
\,.
\label{original Fourier spectral} 
\end{eqnarray}

The four dimensional integral (\ref{original Fourier spectral})
is not very useful in practice. In the following we shall derive
an important relation of the two spectral functions, $r_{\chi }$ and 
$r$, using the occupation number $f$ and another quantity $v$;
\begin{eqnarray}
&&
r(k \,, t) = 
2\,\int\,\frac{d^{3}k'}{(2\pi )^{3}2\omega _{k'}}\,
\left( \,
\frac{r_{\chi }(k_{0} + \omega _{k'} \,, \vec{k} + \vec{k}')}
{1 - e^{-\beta (k_{0} + \omega _{k'})}}\,f(\vec{k'} \,, t)
\right.
\nonumber \\ && \hspace*{-1cm}
\left. + \,
\frac{r_{\chi }(k_{0} - \omega _{k'} \,, \vec{k} - \vec{k}')}
{1 - e^{-\beta (k_{0} - \omega _{k'})}}\,(1 + f(\vec{k'}\,, t)\,) 
- 
\frac{2\,r_{\chi }(k_{0} \,, \vec{k} - \vec{k'})}{1 - e^{-\beta k_{0}}}\,
v(\vec{k'} \,, t) \,\right) 
\,, \label{discontinuity formula} 
\\ &&
v(\vec{k'} \,, t) = \frac{1}{2}\, 
\langle \,\frac{1}{\omega _{k'}}
\frac{d\tilde{\varphi }^{\dag }(\vec{k'} \,, t)}{dt} 
\frac{d\tilde{\varphi }(\vec{k'} \,, t)}{dt}
- 
\omega _{k'}\,\tilde{\varphi }^{\dag }(\vec{k'} \,, t)
\tilde{\varphi }(\vec{k'} \,, t) \, \rangle
\,. \label{v definition} 
\end{eqnarray}

In order to derive this relation, we first 
note that the quantity of the form,
\begin{equation}
\int\,\frac{dk_{0}}{2\pi }\,\tilde{f}(-k_{0} \,, x_{0})\,F(k_{0}) \,,
\end{equation}
appears in eq.(\ref{original Fourier spectral}).
This is related to a coincident limit of the full propagator,
\begin{eqnarray}
&& \hspace*{1cm} 
\int\,\frac{dk_{0}}{2\pi }\,\tilde{f}(-k_{0} \,, x_{0})\,F(k_{0})
\nonumber \\ &&
=\, \lim_{x_{0} \rightarrow y_{0}^{+}}\,F\left( -\,\frac{i}{2}
(\frac{\partial }{\partial x_{0}} - \frac{\partial }{\partial y_{0}}
)\right)
\,\int\,\frac{dk_{0}}{2\pi }\,
\tilde{f}(-\,k_{0} \,, \frac{x_{0} + y_{0}}{2})\,e^{ik_{0}(x_{0} - y_{0})}
\nonumber \\ && \hspace*{0.5cm} 
= \, \lim_{x_{0} \rightarrow y_{0}^{+}}\,
F\left(-\,\frac{i}{2}
(\frac{\partial }{\partial x_{0}} - \frac{\partial }{\partial y_{0}}
)\right)
\,\left( -\,i\,G(x_{0} \,, y_{0})\right) \,.
\end{eqnarray}
For simplicity we omitted the spatial coordinate or its Fourier
component.

We expand in the Taylor series the derivative operation applied to
\\
\( \:
-\,i\,G(x\,, y) = \langle \varphi (x)\varphi (y) \rangle
\: \)
in this equation and 
evaluate derivatives using the Heisenberg equation for $\varphi $.
For instance,
\begin{eqnarray}
&&
\lim_{x_{0} \rightarrow y_{0}^{+}}\,
(\frac{\partial }{\partial x_{0}} - \frac{\partial }{\partial y_{0}}
)\,\varphi (x)\varphi (y) = 
\dot{\varphi }(\vec{x} \,, x_{0})\varphi (\vec{y} \,, x_{0})- 
\varphi (\vec{x} \,, x_{0})\dot{\varphi }(\vec{y} \,, x_{0})
\,, 
\\ && \hspace*{1cm}
\lim_{x_{0} \rightarrow y_{0}^{+}}\,
(\frac{\partial }{\partial x_{0}} - \frac{\partial }{\partial y_{0}}
)^{2}\,\varphi (x)\varphi (y) = 
\nonumber 
\\ && 
(\nabla ^{2} - M^{2})\,\varphi (\vec{x}\,, x_{0})\varphi (\vec{y}\,, x_{0})
+ \varphi (\vec{x} \,, x_{0})
(\nabla ^{2} - M^{2})\,\varphi (\vec{y} \,, x_{0})
\nonumber \\ && \hspace*{-1cm}
-\, 2\,\dot{\varphi } (\vec{x} \,, x_{0})\dot{\varphi } (\vec{y} \,, x_{0})
- \lambda (\varphi \chi ^{2})(\vec{x} \,, x_{0})\varphi (\vec{y}\,, x_{0})
- \lambda \varphi (\vec{x}\,, x_{0})(\varphi \chi ^{2})(\vec{y} \,, x_{0})
\,.
\end{eqnarray}
These terms have different structure for the even and odd number of
derivatives, and it is convenient to separately deal with
the even and the odd powers of the $F(\omega )$ expansion, with
the exception of the first term $F(0)$.
One thus arrives at the following
operator identity in our $\varphi ^{2}\chi ^{2}$ model;
\begin{eqnarray}
&& \hspace*{1cm}
\lim_{x_{0} \rightarrow y_{0}^{+}}\,
F\left(-\, \frac{i}{2}
(\frac{\partial }{\partial x_{0}} - \frac{\partial }{\partial y_{0}}
)\right)
\varphi (x)\varphi (y) =
\nonumber \\ && \hspace*{0.5cm} 
\left[ \,
F(0)\,\varphi (x)\varphi (y) 
- 
F_{-}(\Delta )\,
\frac{i}{2}\,(\dot{\varphi }(x)\varphi (y) - \varphi (x)\dot{\varphi }(y))
+  F_{+}(\Delta )
\right.
\nonumber \\ && \hspace*{-0.5cm}
\left. \cdot \frac{1}{2}\, 
\left( \dot{\varphi }(x)\dot{\varphi }(y) -
\frac{1}{2} (\nabla ^{2}\varphi (x)\varphi (y) + \varphi (x)
\nabla ^{2}\varphi (y)) + (M^{2} + \lambda \chi ^{2})
\varphi (x)\varphi (y) \,\right)
\,\right]_{x_{0} = y_{0}}
\,, \nonumber \\ && \hspace*{2cm}
\label{operator derivative rel} 
\\ && \hspace*{1cm} 
\Delta \equiv \sqrt{\,-\nabla ^{2} + M^{2} + \lambda \chi ^{2}\,}
\,, 
\\ && \hspace*{0.5cm} 
F_{+}(\omega ) \equiv \frac{F(\omega ) + F(-\omega ) - 2F(0)}{2\omega ^{2}}
\,, \hspace{0.5cm} 
F_{-}(\omega ) \equiv \frac{F(\omega ) - F(-\omega ) }{2\omega } \,.
\end{eqnarray}

Here we assumed, for simplicity and without losing practical utility for 
our problem,
that $\chi ^{2}$ is independent of the spatial coordinate.
For our purpose we may take 
\( \:
\langle \chi ^{2} \rangle = T^{2}/12
\: \)
for this constant value, to give the temperature dependent $\varphi $
mass $M^{2} + \lambda T^{2}/12$
in thermal medium.  We shall have more to say on the temperature
dependent mass later.
Furthermore, in the homogeneous thermal medium of our interest one may take
the  spatial Fourier component, 
\begin{eqnarray}
&&
\int\,d^{3}x\,e^{i\vec{k}\cdot \vec{x}}\,\langle \varphi (\vec{x} \,, x_{0})
\varphi (\vec{0} \,, y_{0}) \rangle
= \langle \tilde{\varphi }^{\dag }(\vec{k}\,, x_{0} )\tilde{\varphi }
(\vec{k} \,, y_{0}) \rangle  \,,
\end{eqnarray}
thereby replacing $\nabla ^{2}$
by the relevant momentum, to get
\begin{equation}
\Delta = \omega _{k}(T) = \sqrt{\,\vec{k}^{2} + M^{2}
+ \frac{\lambda }{12}\,T^{2}\,} \,.
\end{equation}
For a notational simplicity we often omit the temperature dependence
of the $\varphi $ mass, simply using the notation $\omega _{k}$
for $\omega _{k}(T)$.

Using the isotropy of medium, one has for the odd power terms,
\begin{eqnarray}
&& \hspace*{0.5cm} 
\lim_{x_{0} \rightarrow y_{0}^{+}}\,\frac{-i}{2}
(\frac{\partial }{\partial x_{0}} - \frac{\partial }{\partial y_{0}})
\,\langle \tilde{\varphi }^{\dag }(\vec{k} \,, x_{0})
\tilde{\varphi }(\vec{k} \,, y_{0}) \rangle
\nonumber \\ &&
= \,
\lim_{x_{0} \rightarrow y_{0}^{+}}\,\frac{-i}{2}
(\frac{\partial }{\partial x_{0}} - \frac{\partial }{\partial y_{0}})
\,\frac{1}{2}\, \langle \tilde{\varphi }^{\dag }(\vec{k} \,, x_{0})
\tilde{\varphi }(\vec{k} \,, y_{0})  +
\tilde{\varphi }(\vec{k} \,, x_{0})\tilde{\varphi }^{\dag }
(\vec{k} \,, y_{0})\rangle
\nonumber \\ && \hspace*{-0.5cm}
= \, \frac{-\,i}{4}\,
\langle \,\left[ \frac{d\tilde{\varphi }^{\dag }(\vec{k} \,, x_{0})}{dt}
\,, \tilde{\varphi }(\vec{k}\,, x_{0})\right]
+ \left[ \frac{d\tilde{\varphi }(\vec{k} \,, x_{0})}{dt}
\,, \tilde{\varphi }^{\dag }(\vec{k}\,, x_{0})\right]\, \rangle
= -\,\frac{1}{2}
\,,
\end{eqnarray}
using the equal time canonical commutator,
\begin{equation}
\left[\, \frac{d\tilde{\varphi }(\vec{k} \,, t)}{dt} \,, 
\tilde{\varphi }(-\vec{k'} \,, t) \,\right] 
= -\,i\,\delta ^{3}(\vec{k} - \vec{k'})
\,.
\end{equation}
Even power terms, on the other hand, contain the modified occupation number,
\begin{equation}
f(\vec{k} \,, t) + \frac{1}{2}= 
\frac{1}{2\omega _{k}(T)}\, 
\langle \,\frac{d\tilde{\varphi }^{\dag }(\vec{k}\,, t)}{dt}
\frac{d\tilde{\varphi }(\vec{k}\,, t)}{dt}
+ \omega _{k}^{2}(T)\,
\tilde{\varphi }^{\dag }(\vec{k}\,, t)\tilde{\varphi }(\vec{k}\,, t)
\, \rangle
\,.
\end{equation}

Using this operator relation for
\begin{eqnarray}
&&
F(k_{0}') = \frac{2\,r_{\chi }(k_{0} + k_{0}' \,, \vec{k} + \vec{k'})}
{1 - e^{-\beta (k_{0} + k_{0}')}}
 \,,
\end{eqnarray}
one obtains the spectral weight for the Hartree model;
the main result of this section,
eqs.(\ref{discontinuity formula}) $-$ (\ref{v definition}).
Even and odd power terms have been lumped according to
\begin{eqnarray}
&&
\frac{F(\omega _{k'}) + F(-\,\omega _{k'})}{2\omega _{k'}}
\,\left( f(k' \,, t) + \frac{1}{2} \right)
- \frac{F(\omega _{k'}) - F(-\,\omega _{k'})}{2\omega _{k'}}\,\frac{1}{2}
\nonumber 
\\ &&
\hspace*{1cm} = \,
\frac{1}{2\omega _{k'}}\,
\left( \, F(\omega _{k'})\,f(k' \,, t) + F(-\,\omega _{k'})\,
(1 + f(k'\,, t))\,\right)  \,,
\end{eqnarray}
except the first power term $F(0)$ giving the third term in 
the right side of eq.(\ref{discontinuity formula}).
We shall later show that the combination of $f$ and $1 + f$ correctly
describes the destructive and the creative processes, including
the stimulated emission effect for bosons.

It is important to realize that the spectral function in the reduced
Hartree model, $r(k\,, t)$ of eq.(\ref{discontinuity formula}),
cannot be written in terms of the distribution function 
\( \:
f(\vec{k} \,, t)= \langle a^{\dag }(\vec{k} \,, t)a(\vec{k} \,, t) \rangle
\: \)
of $\varphi $ particle alone, and contains another combination of the
momentum and the coordinate,
\( \:
v(\vec{k}) =
\langle  p_{k}^{2}/(2\omega _{k}) - \omega _{k}\,q_{k}^{2}/2 \rangle \,,
\: \)
which vanishes for a system of a set of independent harmonic oscillators.
To this extent the particle picture does not completely hold in
a full quantum theory such as ours.
In the next section we shall derive a self-consistent set of equations
that determine the particle distribution function under the
assumption of a slow variation of the particle number density.
The $v$ term becomes a higher order effect in $\lambda ^{2}$.
This way one can derive a useful quantum kinetic equation.
It is convenient, with this in mind, to separate the contribution to
the spectral function without the $v$ term of (\ref{v definition}),
\begin{eqnarray}
r_{0}(k \,, t) &=&
2\,\int\,\frac{d^{3}k'}{(2\pi )^{3}2\omega _{k'}}\,
\left( \,
\frac{r_{\chi }(k_{0} + \omega _{k'} \,, \vec{k} + \vec{k}')}
{1 - e^{-\beta (k_{0} + \omega _{k'})}}\,f(\vec{k'} \,, t)
\right.
\nonumber \\ 
&+& \left.
\frac{r_{\chi }(k_{0} - \omega _{k'} \,, \vec{k} - \vec{k}')}
{1 - e^{-\beta (k_{0} - \omega _{k'})}}\,(1 + f(\vec{k'}\,, t)\,) 
\,\right) 
\,. \label{leading discontinuity formula} 
\end{eqnarray}

As a minor note, we point out that 
the relation $f(\vec{k}' \,, t) = f(|\vec{k}'| \,, t)$ following from
the isotropy was assumed in the derivation above.
We also ignored a possible tadpole type of contribution, which
will be discussed shortly.

The same operator equation, eq.(\ref{operator derivative rel}),
when applied to 
\begin{equation}
F(k_{0}) = e^{ik_{0}(x_{0} - y_{0})}e^{-i\vec{k}\cdot (\vec{x} - \vec{y})}
\,, 
\end{equation}
gives the full propagator in terms of the distribution function,
\begin{eqnarray}
-\,i\,G(x \,, y) &=& \int\,\frac{d^{3}k}{(2\pi )^{3}\,2\omega _{k}}\,
e^{-\,i\vec{k}\cdot (\vec{x} - \vec{y})}
\left( \,
e^{i\omega _{k}(x_{0} - y_{0})}\,f(\vec{k} \,, t) 
\right. \nonumber 
\\ 
&+& \left.
e^{-i\omega _{k}(x_{0} - y_{0})}\,(\,1 + f\,(\vec{k} \,, t))
- 2\,v(\vec{k}\,, t)
\,\right)
\,, \label{expansion of propagator} 
\end{eqnarray}
where 
\( \:
\omega _{k} = \sqrt{\,\vec{k}^{2} + M^{2} + \lambda \chi ^{2}\,}
\: \)
with a constant $\chi ^{2}$, and 
$t = (x_{0} + y_{0})/2$.
Again, the full propagator contains $v$ in addition to
the distribution function.

In the previous discussion  we ignored a possibility of
a tadpole type of contraction.
For instance, in the correlator of
\begin{eqnarray}
&& \hspace*{-2cm} 
\langle\, \varphi (1)\varphi (2)\,\int\,dx\,dy\,
(\,\varphi ^{2}(x) - \varphi '\,^{2}(x)\,)\,(\,\alpha (x - y)\,
\varphi ^{2}(y) - \alpha ^{*}(x - y)\,\varphi '\,^{2}(y)\,)\, \rangle
\,,
\end{eqnarray}
one may contract both field operators at $1$ and $2$ with those of
$\varphi^{2}$ at the same spacetime point $x$. This would lead to
a new kernel of the form,
\( \:
\:\propto  \:
\delta (x - y)\,\langle \varphi ^{2}(0) \rangle \,.
\: \)
More specifically, one allows a local term of the form,
\begin{eqnarray}
&&
\delta \beta_{4} (x - y) = \delta (x - y)\,
\langle \varphi ^{2}(0) \rangle\,\int_{x_{0} > y_{0}}\,dy\,
\alpha_{I}(x - y)
 \,.
\end{eqnarray}
It is then easy to show that
\begin{eqnarray}
\delta \beta _{4}(x ) = -\,i\lambda ^{2}\,
\delta (x)\,4\langle\varphi ^{2}(0) \rangle
\,\int_{2m}^{\infty }\,dk_{0}\,\frac{r_{\chi }(k_{0}\,,
\vec{0})}{k_{0}}\,\coth \frac{\beta k_{0}}{2} \,,
\label{tadpole 1} 
\end{eqnarray}
with $m$ the $\chi $ mass.
There is another tadpole term to even lower order $\lambda $ from
\begin{eqnarray}
&&
i\,\frac{\lambda }{2}\,
\langle  \varphi (1)\varphi (2)\,\int\,dx\,\varphi ^{2}(x)\chi ^{2}(x) 
\,\rangle \,, 
\end{eqnarray}
which gives
\begin{eqnarray}
\delta \beta _{2} (x ) = i\,\delta (x )\,\lambda \,
\langle \chi^{2}(0) \rangle 
= i\,\delta (x )\,\lambda \,
\int\,\frac{d^{3}k}{(2\pi )^{3}2\omega _{k}}\,\coth \frac{\beta \omega_{k}}{2}
\,.
\label{tadpole 2} 
\end{eqnarray}
These tadpoles, (\ref{tadpole 1}) and (\ref{tadpole 2}), are shown in Fig.1.

Both of these $\delta \beta _{i}(x)$ here 
are local ($\propto \delta (x)$) and are purely imaginary.
These contain the mass and the coupling constant renormalization,
the $\varphi ^{4}$ coupling.
Thus, the infinite part of these is eliminated by their respective 
counter terms.
The remaining, temperature dependent finite terms give physical effects of
finite temperature correction to the mass and the coupling constant.
For discussion of order $\lambda ^{2}$ effects in the kinetic equation,
only the temperature dependent mass contributes
in the Hartree model, changing the renormalized mass to
\begin{equation}
M^{2} \:\rightarrow  \: M^{2}(T) = M^{2} + \frac{\lambda }{12}\,T^{2} \,.
\label{temperature dep mass} 
\end{equation}
The coefficient $\frac{\lambda }{12}$ was calculated from the vacuum
subtracted 
\( \:
\lambda \,
\langle \chi ^{2}(0) \rangle - \langle \chi ^{2}(0) \rangle_{T = 0} 
\: \)
above, and precisely coincides with
the result of finite temperature field theory \cite{textbook on finite temp}.
The temperature dependent $\varphi ^{4}$
coupling of order $\lambda ^{2}$ gives
a higher order term 
to our subsequent kinetic equation, hence may effectively be ignored.

The temperature dependent mass has the structure of the tadpole term, 
\begin{equation}
-i\,\frac{\lambda }{2}\,\langle \chi ^{2} \rangle\,
(\varphi ^{2}(x) - \varphi '\,^{2}(x))
\end{equation}
in the influence functional. In the usual operator formalism at
zero temperature this type of term $\langle \chi ^{2} \rangle$
is absent by the normal ordering, but in thermal
environment $\langle \chi ^{2} \rangle$ has a temperature
dependence, hence has a physical meaning.
After renormalization one may write the temeperature dependent mass term 
as
\( \:
\frac{1}{2}\, M^{2}(T)\,\varphi ^{2}
\: \)
in the Lagrangian density.
With this new term, one should replace the previous 
$\alpha (x)$, eq.(\ref{2-point kernel}) by
\begin{equation}
\alpha (x) = \lambda ^{2}\,
\left( \,{\rm tr}\;T[\chi ^{2}(x)\,\chi ^{2}(0)]\,\rho _{\beta }
- ({\rm tr}\;\chi ^{2}\,\rho _{\beta })^{2}\,\right) \,.
\label{modified 2-point kernel by tadpole} 
\end{equation}
With this new contribution included, one can forget about the tadpole term
provided that one uses the temperature dependent mass, $M(T)$, instead of
the renormalized mass parameter.
We shall come back to further aspects of the renormalization
in Appendix B
after we introduce the slow variation approximation in the next section.

The two-body spectral weight $r_{\chi }(k)$ for $\alpha (x)$
is calculable from the analytically
continued imaginary-time thermal Green's function \cite{fetter-walecka}, 
\cite{hjmy-97}. We first note the oddness,
\( \:
r_{\chi }(-\omega \,, \vec{k}) = -\,r_{\chi }(\omega \,, \vec{k}) \,, 
\: \)
hence
\( \:
r_{\chi } (- k) = -\,r_{\chi }(k)
\: \)
combined with the isotropy of space.
The spectral weight is given by 
a discontinuity along the real axis of the energy $\omega = k_{0}$
corresponding to two $\chi $ particle states in thermal medium.
Since the kinematics is modified in thermal medium from that
in vacuum, the relevant expressions are different,
depending on relation between $k_{0}$ and $|\vec{k}|$.
For both $k_{0} > \sqrt{\vec{k}^{2} + 4m^{2}}$ and
$|\vec{k}| > k_{0} > 0$ \cite{jmy-98-1},
\begin{eqnarray}
&& \hspace*{-1.5cm}
r_{\chi }(k _{0} \,, k) = 
\frac{\lambda ^{2}}{16\pi ^{2}}\,\left( \,
\sqrt{1 - \frac{4m^{2}}{k_{0}^{2} - k^{2}}}\,\theta (k_{0}
- \sqrt{k^{2} + 4m^{2}}) + \frac{2}{\beta k}\,
\ln \frac{1 - e^{-\beta \omega _{+}}}{1 - e^{-\beta |\omega _{-}|}}\,\right)
\,, \label{two-body spectral} 
\\ && \hspace*{1cm} 
\omega _{\pm } = \frac{k_{0} }{2} \pm \frac{k}{2}\,
\sqrt{1 - \frac{4m^{2}}{k_{0}^{2} - k^{2}}} \,,
\end{eqnarray}
where $k = |\vec{k}|$.
The formula is better understood \cite{weldon} if one writes this
separately in
the respective kinematic regions, using the thermal distribution function
$f_{{\rm th}}(\omega ) = 1/(e^{\beta \omega } - 1)$; \\
for $\omega  > \sqrt{\vec{k}^{2} + 4m^{2}}$ 
\begin{eqnarray}
&& \hspace*{-0.5cm}
r_{\chi }(\omega \,, k) = 
\frac{\lambda ^{2}}{16\pi ^{2}\,k}\,\int_{\omega _{-}}^{\omega _{+}}\,
dE\,\left( \,(\,1 + f_{{\rm th}}(E)\,)\,(\,1 + f_{{\rm th}}(\omega - E)\,)
- f_{{\rm th}}(E)\,f_{{\rm th}}(\omega - E)\,\right)
\,, \nonumber \\ && \label{2-body spectral above th} 
\end{eqnarray}
whereas for $|\vec{k}| > k_{0} > 0$
\begin{eqnarray}
&& \hspace*{-0.5cm}
r_{\chi }(\omega \,, k) = 
\frac{\lambda ^{2}}{8\pi ^{2}\,k}\,\int_{-\,\omega _{-}}^{\infty }\,
dE\,\left( \,f_{{\rm th}}(\omega )\,(\,1 + f_{{\rm th}}(\omega + E)\,)
- (\,1 + f_{{\rm th}}(E)\,)\,f_{{\rm th}}(\omega + E)\,\right)
\,. \nonumber \\ &&
\label{2-body spectral below th} 
\end{eqnarray}
Thus, the indivisual discontinuity has a one-to-one
correspondence to a physical process such
as something $\leftrightarrow \chi \chi $ and something 
$+\, \chi \leftrightarrow \chi $.
An example of the two-body spectral function $r_{\chi }$ is given
in Fig.2.

The spectral weight $r_{0}(\omega \,, \vec{p} \,, t)$
in the Hartree model given by 
(\ref{leading discontinuity formula}) takes a particularly
simple form if one uses the energy conservation for the the thermal factor,
\( \:
f^{-1}_{{\rm th}}(\omega ) + 1 = e^{\beta \omega } \,.
\: \)
For instance, near the mass shell, $\omega \approx  \pm\, \omega _{p}$,
the relevant processes are the annihilation and the scattering process
for $\omega \approx \omega _{p}$ and their inverse processes for
$\omega = -\,\omega _{p}$;
explicitly
\begin{eqnarray}
&& \hspace*{1cm}
r_{0}(\omega _{p} \,, \vec{p}) = 2\lambda ^{2}
\,\int\,dp'\,\int\,dk_{1}\,\int\,dk_{2}
\nonumber \\ && 
\cdot 
\left( \, (\,1 + f_{{\rm th}}(k_{1})\,)\,(\,1 + f_{{\rm th}}(k_{2})\,)\,
f(p')\,\delta (p + p' - k_{1} - k_{2})
\right.
\nonumber \\ && 
\left. +\, 
2f_{{\rm th}}(k_{1})\,(\,1 + f_{{\rm th}}(k_{2})\,)\,(\,1 + f(p')\,)
\delta (p + k_{1} - p' - k_{2})\,\right)
\,, \label{on-shell spectral +} 
\\ && \hspace*{1cm}
r_{0}(-\,\omega _{p} \,, \vec{p}) = 2\lambda ^{2}
\,\int\,dp'\,\int\,dk_{1}\,\int\,dk_{2}
\nonumber \\ &&
\cdot 
\left( \, f_{{\rm th}}(k_{1})\,f_{{\rm th}}(k_{2})\,(\,1 + f(p')\,)
\,\delta (k_{1} + k_{2} - p - p')
\right.
\nonumber \\ && 
\left. +\, 2f_{{\rm th}}(k_{2})\,(\,1 + f_{{\rm th}}(k_{1})\,)\,f(p')\,
\delta ( p' + k_{2} - p - k_{1})\,\right)
\,. \label{on-shell spectral -} 
\end{eqnarray}
Four processes appearing in these equations are depicted in Fig.3.
Here
\begin{equation}
f_{{\rm th}}(k) = \frac{1}{e^{\beta \omega _{k}} - 1} \,, \hspace{0.5cm} 
\omega_{k} = \sqrt{k^{2} + m^{2}} \,.
\end{equation}
We used an abbreviated notation for the phase space integral;
\begin{eqnarray}
&&
\int\,dk\,(\cdots ) = \int\,\frac{d^{3}k}{(2\pi )^{3}2\omega _{k}}\,
(\cdots) \,, {\rm etc.}
\,, \label{abbreviated phase space} 
\end{eqnarray}
and 
\( \:
\delta (p + p' - k_{1} - k_{2}) = (2\pi )^{3}\,
\delta^{4} (p + p' - k_{1} - k_{2}) \,, 
\: \)
etc.
When multiplied by $f(p)$ and $1 + f(p)$, these give the on-shell
destruction and
the production rate of the momentum mode $p$ with a suitable normalization.

\hspace*{0.5cm} 
\section{Slow variation and self-consistent
equation}

\vspace{0.5cm} 
\hspace*{0.5cm} 
Once the Hartree approximation is defined with the kernel $\beta (x\,, y)$
in the influence functional given, one may proceed as follows.
We have in mind a physical situation in
which one can clearly separate two time scales;
the microscopic time scale for the quantum behavior governed by
the Hamiltonian, 
and a macroscopic time scale for the change of the $\varphi $
particle distribution function $f(\vec{k} \,, t)$.
The separation naturally occurs for a small system
within a dilute thermal medium at very low temperatures, where
the system itself is not far from equilibrium. Under this circumstance 
one may consider a slowly varying $f(\vec{k} \,, t)$ 
in the time range, $t \gg $ microscopic relaxation time.
We thus take, in considering the short time variation, the 
time $t$ in the distribution function as a fixed constant.
It is then straightforward to work
out consequences of the short time dynamics,
since the truncated Hartree model is formally
equivalent to the harmonic oscillator
model, known to be completely solvable \cite{jmy-98-1}, \cite{jmy-97}. 
Since the relaxation rate towards the equilibrium depends on
the distribution function at that time, this consideration leads to
a self-consistent equation for the the quasi-stationary
distribution function.

Separation of the two time scales to distinguish
the slow and the fast process also appears in other approaches
\cite{real-time th green}, \cite{closed time path} when 
the quantum Boltzmann equation is derived in different contexts.

We shall briefly summarize the main point of how the exactly solvable
model is used in our Hartree model.
One first recalls that one can discuss each Fourier $\vec{k}$ mode separately.
The Fourier mode kernel $\beta (\vec{k} \,, t)$ in the influence
functional is related to the kernel given 
at the relative coordinate $\vec{x}$ by
\begin{eqnarray}
\beta (\vec{x} \,, x_{0}) = 
\int\,\frac{d^{3}k}{(2\pi )^{3}}\,\beta (\vec{k} \,, x_{0})\,
e^{i\vec{k}\cdot \vec{x}} \,.
\label{kernel fourier} 
\end{eqnarray}
For the sake of clarity we sometimes omit the mode index $\vec{k}$
unless confusion occurs.
One notes that the path integral for the $\varphi $ system
in the influence functional has an exponent of the form,
\begin{eqnarray}
&& \hspace*{1cm} 
\frac{i}{2}\,\int_{t_{i}}^{t_{f}}\,d\tau \left( 
\dot{\xi }(\tau ) \dot{X}(\tau ) - \omega^{2}(T)\,\xi(\tau ) X (\tau ) 
\right) 
\nonumber \\ &&
- \,
\int_{t_{i}}^{t_{f}}\,d\tau \,\int_{t_{i}}^{\tau }\,ds\,
\left( \,\xi (\tau )\,\beta_{R} (\tau - s)\,\xi (s)
+ i\,\xi (\tau )\,\beta _{I}(\tau - s)\,X(s)\,\right) \,.
\label{exponent of influence functional} 
\end{eqnarray}
Here $\omega ^{2}(T) = k^{2} + M^{2}(T)$ includes 
the temperature dependent mass given by eq.(\ref{temperature dep mass}).

The time $x_{0}$ in eq.(\ref{kernel fourier}) 
is a common CM time in the original $\beta (x\,, y)$.
Since there is only one common time, one has the time translation
invariance for $\beta $, which was absent in the original $\beta (x\,, y)$.

A remarkable feature of this exponent (\ref{exponent of influence functional})
is that it is linear in the variable
$X = \varphi + \varphi '$, 
hence its path integration yields a trivial delta function.
This gives an equation for the semiclassical path for the 
$\xi ( = \varphi - \varphi ')$ variable;
\begin{eqnarray}
&&
\frac{d^{2}\xi }{d\tau^{2}} + \omega^{2}(T)\,\xi (\tau )
+ 2\,\int_{\tau }^{t}\,ds\,\xi (s)\beta _{I}(s - \tau ) = 0 \,.
\label{integro-differential eq for g} 
\end{eqnarray}
We set hereafter
\( \:
t_{i} = 0
\: \)
and $t_{f} = t$.
This integro-differential equation can be solved by the standard
technique of the Laplace transform, as described in \cite{jmy-98-1}.
Its solution $\xi _{{\rm cl}}$ is given by
\begin{eqnarray}
&& \hspace*{1cm} 
\xi_{{\rm cl}}(\tau ) = \xi _{i}\,y(\tau ) + \xi _{f}\,z(\tau ) \,, 
\\ &&
y(\tau ) = \frac{g(t - \tau )}{g(t)} \,, \hspace{0.5cm} 
z(\tau ) = \dot{g}(t - \tau ) - \frac{g(t - \tau )\dot{g}(t)}{g(t)}
\,, 
\\ && 
g(\tau ) = \frac{1}{2\pi i}\,\int_{p_{0} - i\infty }^{p_{0} + i\infty }\,
dp\,\frac{e^{p\tau }}{p^{2} + \omega ^{2}(T)
+ 2\,\tilde{\beta }(p)} \,, 
\\ && \hspace*{1cm}
\tilde{\beta }(p) = 
\int_{0}^{\infty }\,d\tau \,e^{-p\tau }\,\beta _{I}(\tau ) \,,
\end{eqnarray}
with $p_{0} > 0$.
Note that $\dot{g}(0) = 1$.

Applied to our specific model, 
\begin{eqnarray}
&&
\tilde{\beta }(p \,, \vec{k}) = -\,\int_{0}^{\infty }\,d\omega \,
\frac{\omega \,r_{-}(\omega \,, \vec{k}) }
{p^{2} + \omega ^{2}} \,,
\\ &&
r_{\pm }(\omega \,, \vec{k}) = r(\omega \,, \vec{k}) \pm 
r(-\,\omega \,, \vec{k}) \,,
\end{eqnarray}
for the momentum mode $\vec{k}$.
The basic function $g(\tau )$ for this mode is then
\begin{eqnarray}
&&
g(\vec{k} \,, \tau ) = \frac{1}{2\pi }\,\int_{-\infty + i0^{+}}
^{\infty + i0^{+}}\,dz\,e^{-iz\tau }\,F(z\,, \vec{k}) \,, 
\label{basic function g} 
\\ &&
-\,F(z\,, \vec{k})^{-1} = z^{2} - \omega _{k}^{2}(T) - 2\,
\int_{0}^{\infty }\,d\omega \,\frac{\omega \,r_{-}(\omega \,, \vec{k})}
{z^{2} - \omega ^{2}} \,.
\label{analytic self-energy f} 
\end{eqnarray}
We deformed the contour of integration as
$p \rightarrow -\,iz$ in the Laplace inverted formula.

Both $g(\tau )$ and its time derivative $\dot{g}(\tau )$ obey
a related equation to (\ref{integro-differential eq for g});
\begin{equation}
\frac{d^{2}x}{d\tau ^{2}} + \omega ^{2}(T)\,x(\tau ) +
2\,\int_{0}^{\tau }\,ds\,\beta _{I}(\tau - s)\,x(s) = 0 \,, 
\end{equation}
with different boundary conditions,
\( \:
g(0) = 0\,, \hspace{0.3cm} \dot{g}(0) = 1 \,.
\: \)

Important quantities for the influence functional in the Hartree model are
two real functions $\beta _{i}$;
\begin{eqnarray}
&&
\beta _{R}(\vec{k}\,, t) = \int_{0}^{\infty }\,d\omega \,
r_{+}(\omega \,, \vec{k})\,\cos (\omega t) \,, 
\\ &&
\beta _{I}(\vec{k} \,, t) = -\,\int_{0}^{\infty }\,d\omega \,
r_{-}(\omega \,, \vec{k})\,\sin (\omega t) \,.
\end{eqnarray}
Different combinations $r_{\pm }$ appear, because these integrals
originally defined in $-\,\infty < \omega < \infty $
have definite parities; $\beta _{R}\, (\,\beta _{I}\,)$ is even (odd) in
$t$.
These $\beta_{i} $ 
depend on $f(\vec{k})$ of the $\varphi $ particle distribution
via $r_{\pm }(\omega \,, \vec{k})$.

For comparison we write the corresponding quantity $\beta ^{(d)}$
in the unstable particle decay described by the Lagrangian density, 
$\lambda \,\varphi \chi ^{2}$ ($\lambda $ having a mass dimension
in this case);
\begin{eqnarray}
&&
\beta ^{(d)} _{R}(\vec{k} \,, t) = \int_{0}^{\infty }\,d\omega \,
\coth \frac{\beta \omega }{2}\,
r_{\chi }(\omega \,, \vec{k})\,\cos (\omega t)  \,, 
\\ &&
\beta ^{(d)} _{I}(\vec{k}\,, t) = -\,\int_{0}^{\infty }\,d\omega \,
r_{\chi }(\omega \,, \vec{k})\,\sin (\omega t) \,,
\end{eqnarray}
with $r_{\chi }$ given by eq.(\ref{two-body spectral}).
Thus, in this case $r_{\pm }^{(d)} $ that defines $\beta _{R \,, I}^{(d)}$
satisfies
\( \:
r_{+}^{(d)} (\omega \,, \vec{k}) = \coth \frac{\beta \omega }{2}\,
r_{-}^{(d)} (\omega \,, \vec{k}) \,.
\: \)

On the other hand,
it turns out that in the present annihilation-scattering problem
a similar relation holds only on the mass shell and only when one takes
the thermal distribution $f_{{\rm th}}$ for $f(\vec{k})$;
\begin{equation}
r_{+}(\omega _{k} \,, \vec{k})_{{\rm th}} 
= \coth \frac{\beta \omega _{k}}{2}\,
r_{-}(\omega _{k} \,, \vec{k})_{{\rm th}} \,,
\end{equation}
which further reduces to the detailed balance relation,
\begin{equation}
r(\omega _{k} \,, \vec{k})_{{\rm th}}\,f_{{\rm th}}(\vec{k})
= 
r(-\,\omega _{k} \,, \vec{k})_{{\rm th}}\,
\left( \,1 + f_{{\rm th}}(\vec{k})\,\right) \,.
\end{equation}
A deep reason why this relation does not hold for the more general 
non-equilibrium case in our annihilation-scattering problem
is that $r_{\pm }(\omega \,, \vec{k})$
are functionals of the non-equilibrium $\varphi $ 
distribution and in general,
$f(\vec{k}) \neq f_{{\rm th}}(\vec{k})$.
In the simple case of the unstable particle decay 
into two thermal particles the corresponding spectrum is independent of
the distribution function of the decaying particle.
We further note that distinction of two types of the spectrum
$r_{\pm }$ is crucial to obtain the correct form of the Boltzmann
equation in the next section.

Under a certain condition the function $F(z\,, \vec{k})$ is analytic
except on the real axis where there is a branch cut singularity
as shown in Fig.4. 
As seen by taking the imaginary part of 
the defining equation (\ref{analytic self-energy f}),
this analyticity holds under the condition,
\begin{equation}
r_{-}(\omega \,, \vec{k}) = 
r(\omega \,, \vec{k}) - r(-\,\omega \,, \vec{k}) > 0 \,.
\end{equation}
This condition for the analytic property of $F(z \,, \vec{k})$, 
when evaluated on the mass shell $\omega = \omega _{k}$, 
physically means that the destructive process of $\varphi $ 
given by the first term $r(\omega _{k} \,, \vec{k})$ dominates
over the production process given by the second term 
$r(-\,\omega _{k} \,, \vec{k})$, 
a situation we are practically interested in.
Furthermore, 
in the weak coupling limit the above condition needs to be obeyed
only near $\omega = \omega _{k}$, 
since the off-shell contribution in the weak coupling
limit is negligible in determining the analyticity.
The analytic extention of $F(z\,, \vec{k})$ has two simple
poles in the second Riemann sheet, approximately at
\begin{eqnarray}
&&
z = \pm \,\omega _{k}(T) - i\,\frac{\pi \,r_{-}(\omega _{k} \,, \vec{k})}
{2\omega _{k}}  \,.
\end{eqnarray}
The poles in the second sheet
are close to the real axis in the weak coupling limit,
$\lambda ^{2} \rightarrow 0$.
When pole terms dominate in the integral,
the exponential decay law follows;
\begin{eqnarray}
&&
g(\vec{k} \,, \tau ) \approx \frac{\sin \omega _{k}\tau }{\omega _{k}}\,
e^{-\,\Gamma _{k}\tau /2} \,, 
\\ &&
\Gamma _{k} = \frac{\pi \,r_{-}(\omega _{k}\,, \vec{k})}{\omega _{k}} 
= \frac{\pi }{\omega _{k}}\,\left( \,r(\omega _{k} \,, \vec{k})
- r(-\omega _{k} \,, \vec{k}) \,\right)
\,.
\end{eqnarray}
As is clear in (\ref{on-shell spectral +}), (\ref{on-shell spectral -}),
the decay rate $\Gamma _{k}$ has both annihilation and
scattering contribution along with their inverse processes.

On the other hand, 
when $r_{-}(\omega _{k} \,, \vec{k}) < 0$, these poles appear
in the first Riemann sheet. 
The physical situation here is the dominance of the production process,
which is not of our immediate concern.
It however appears that this case can be dealt with analogously
to the above case of $r_{-} > 0$.

We now turn to derivation of the self-consistent equation for the distribution
function.
Recall first for each Fourier mode $\vec{k}$,
\begin{eqnarray}
f(\vec{k} \,, t) = \langle\, a^{\dag }(\vec{k}\,, t)
a(\vec{k}\,, t)\,\rangle \,.
\label{def of occupation-n} 
\end{eqnarray}
Here the creation ($a^{\dag }$) and the annihilation ($a$)
operators of $\varphi $ particle are Heisenberg
operators such as $e^{iHt}\,a\,e^{-iHt}$ with $H$ the total Hamiltonian
including the system-environment interaction.
In the harmonic oscillator basis which is an essential element of
the plane wave decomposition of the field operator, this is 
equal to
\begin{equation}
f(\vec{k} \,, t) = \langle \,\frac{1}{2\omega _{k}}\,p_{k}(t)
p_{-k}(t)
+ \frac{\omega_{k}}{2}\,q_{k}(t)q_{-k}(t) - \frac{1}{2}\, \rangle \,, 
\label{original def of occ-n} 
\end{equation}
where the coordinate and the momentum, $q_{k} \,, p_{k}$, are
identified using the field decomposition into the plane wave;
in the previous notation,
\begin{eqnarray}
&&
q_{k}(t) \equiv \tilde{\varphi }(\vec{k} \,, t)
\,, \hspace{0.5cm} 
p_{k}(t) \equiv \frac{d\tilde{\varphi }(\vec{k} \,, t)}{dt}
\,,
\\ &&
\tilde{\varphi }(\vec{k} \,, t) = \int\,d^{3}x\,\varphi (\vec{x} \,, t)
e^{-i\vec{k}\cdot \vec{x}} \,.
\end{eqnarray}
The subtracted factor $\frac{1}{2}$ in 
(\ref{original def of occ-n}) is the well known contribution from
vacuum fluctuation. 
For simplicity we subsequently use the notation, for example $q^{2}_{k}$ 
to mean
\( \:
q_{k}q_{-k} \,.
\: \)

We mention here an ambiguity for the choice of the reference
energy $\omega _{k}$ to define the occupation number.
We used here the "free" part of the oscillator energy
$\omega _{k} = \sqrt{\vec{k}^{2} + M^{2}}$.
In the presence of the interaction with thermal environment
the use of this unobservable energy is however dubious.
Indeed, we confirmed, as will be explained in Appendix B,
that this choice would lead to an unacceptable result of
the absence of the equilibrium distribution at very low temperatures.
As will be discussed later, the proper choice of the reference
energy turns out to be the renormalized energy including $O[\lambda ^{2}]$
correction of the finite self-energy shift in medium,
eq(\ref{renormalized mass at finite t}).

We need to sharpen the concept of various time scales
in computing the correlator.
Consider the correlator at different times such as
\( \:
\langle q(t_{0} + t/2)q(t_{0} - t/2) \rangle \,, 
\: \)
and suppose that $t_{0} \gg t$, all the times measured from
some specified initial time.
We then let $t_{0} \gg $ several $\times $ relaxation time,
and vary the relative time
$t$ in the range from 0 to several $\times $ the
relaxation time.
Under this circumstance we approximate
\( \:
\langle q(t_{0} + t/2)q(t_{0} - t/2) \rangle 
\approx \langle q^{2}(t_{0}) \rangle \,.
\: \)
We thus need the coincident limit at large times ($\gg $ relaxation time).

Some means to compute the correlator at the coincident time such as
\( \:
\langle q^{2}(t) \rangle
\: \)
becomes necessary.
This or a more general multi-time correlator such as \\
\( \:
\langle q(t_{1})q(t_{2})\cdots q(t_{n}) \rangle 
\: \)
can be computed with the machinery of the generating functional.
Instead of being much involved in technical details, we shall
give only a general idea here and relegate all technical points of the
generating functional to Appendix A.
An alternative method of computation is to use the exact operator solution
for the harmonic system, as is done in \cite{jmy-97}.

Calculation of the Green's function in the path integral approach is
performed by introducing a coupling of external source terms of
the form,
\begin{equation}
j(\tau )q(\tau ) + l(\tau )p(\tau ) -
j'(\tau )q'(\tau ) - l'(\tau )p'(\tau )
= \frac{1}{2}\, 
(\,S_{j}\xi + D_{j}X + S_{l}p_{\xi } + D_{l}p_{X} \,) \,,
\end{equation}
where
\( \:
S_{j}= j + j' \,, \; D_{j} = j - j'\,, \;
p_{X \,, \, \xi } = p \pm p' \,, 
\: \)
etc.
Functional differentiation with respect to the sources,
$S_{i} \,, D_{i}$ then
gives various combination of correlators, which in turn gives
necessary two point correlators,
\( \:
\langle q(1)q(2) \rangle \,, \; \langle p(1)p(2) \rangle \,, 
\; \langle q(1)p(2) \rangle \,.
\: \)

The result is
\begin{eqnarray}
&&
\langle\, q_{k}^{2}(t)\, \rangle = \int_{0}^{\infty }\,d\omega 
\,r_{+}(\omega \,, \vec{k})\,|h(\omega \,, \vec{k} \,, t)|^{2} 
\nonumber \\ &&
\hspace*{1cm} 
+\,
g^{2}(\vec{k} \,, t)\,\overline{p_{i}^{2}} + 
\dot{g}^{2}(\vec{k} \,,t)\,\overline{q_{i}^{2}} + 
g(\vec{k} \,,t)\dot{g}(\vec{k} \,,t)\,
\overline{p_{i}q_{i} + q_{i}p_{i}} \,, 
\label{correlator q^2} 
\\ && 
\langle \,p_{k}^{2}(t)\, \rangle = \int_{0}^{\infty }\,d\omega 
\,r_{+}(\omega \,, \vec{k})\,|k(\omega \,,\vec{k}\,,  t)|^{2} 
\nonumber 
\\ &&
\hspace*{1cm} 
+\, 
\dot{g}^{2}(\vec{k} \,,t)\,\overline{p_{i}^{2}} + 
\stackrel{..}{g}^{2}(\vec{k} \,,t)\, \overline{q_{i}^{2}} + 
\dot{g}(\vec{k} \,,t)\stackrel{..}{g}(\vec{k} \,,t)\,
\overline{p_{i}q_{i} + q_{i}p_{i}} \,, 
\label{correlator p^2} 
\\ &&
\frac{1}{2}\, \langle \,q_{k}(t)p_{k}(t) + p_{k}(t)q_{k}(t)\, \rangle
= 
\int_{0}^{\infty }\,d\omega \,r_{+}(\omega \,, \vec{k})
\,\Re \left( \,h(\omega \,, \vec{k}\,, t)k^{*}(\omega \,,\vec{k}\,,  t) 
\,\right)
\nonumber \\ && \hspace*{-1cm} 
+\,
\dot{g}(\vec{k} \,,t)g(\vec{k} \,,t)\,
\overline{p_{i}^{2}} + \dot{g}(\vec{k} \,,t)
\stackrel{..}{g}(\vec{k} \,,t)\,
\overline{q_{i}^{2}} 
+ (\,\dot{g}^{2}(\vec{k} \,,t) + 
g(\vec{k} \,,t)\stackrel{..}{g}(\vec{k} \,,t)\,)
\,\frac{1}{2}\, \overline{p_{i}q_{i} + q_{i}p_{i}} \,, 
\label{correlator qp} \nonumber \\ &&
\\ && 
\hspace*{1.5cm} 
h(\omega \,, \vec{k} \,, t) = \int_{0}^{t}\,d\tau 
\,g(\vec{k}\,, \tau )\,e^{-i\omega \tau } \,, 
\hspace{0.5cm} 
\label{basic function h} 
\\ && \hspace*{1.5cm} 
k(\omega \,, \vec{k} \,, t) = \int_{0}^{t}\,d\tau 
\,\dot{g}(\vec{k}\,, \tau )\,e^{-i\omega \tau } \,.
\label{basic function k} 
\end{eqnarray}
The coincident time limit in these formulas may be understood
only with the condition, $|{\rm relative \;time}| \ll t$ , thus
the relative time can be as large as the relaxation time.
Quantities such as $\overline{q_{i}^{2}} $ are the values 
averaged over the ensemble at a specified initial time.
We have chosen  the initial ensemble such that
\( \:
\overline{q_{i}} = 0 
\: \)
and 
\( \:
\overline{p_{i}} = 0 \,.
\: \)
Note that the relevant kernel for the correlator is 
$\beta _{R}(\vec{k} \,, t)$, 
as seen from (\ref{qq correlator at different t}), hence the
corresponding spectral combination is $r_{+}(\omega \,, \vec{k})$ 
appearing in the $\omega $ integral here instead of $r_{-}$ .

We now coarse-grain the short time dynamics to derive the self-consistent
equation for the equilibrium occupation number.
We first note that except at very early and very late times 
the simple exponential decay law for $g(\vec{k} \,, t)$
is an excellent approximation.
This intermediate epoch of the exponential decay law
is ideal to obtain the coarse-grained behavior.
A sacrifice resulting from the replacement by the intermediate
exponential law is that the very early quantum behavior is lost.
We lose nothing, however, in the late time behavior, because the power law
behavior present 
at much later times is not realized due to a large multiplicity
of actual reactions; indeed, the relaxation rate $\Gamma _{k}$ 
which will be precisely defined later depends on the time 
weakly via the slowly varying distribution function $f(\vec{k} \,, t)$.

We  observe that in the formulas for the dynamical variables, 
eq.(\ref{correlator q^2}) $-$ (\ref{correlator qp}), the initial
value dependence disappears with the exponentially 
decaying $g(\vec{k} \,, t)$.
We then use the limiting behavior of $h(\omega \,, \vec{k} \,, t)$
and $k(\omega \,, \vec{k} \,, t)$ as $t \rightarrow \infty $, 
to actually mean $t \gg $ relaxation time.
From the expression of $g(\vec{k} \,, t)$, 
eq.(\ref{basic function g}), and
the definition of these functions
in (\ref{basic function h}), (\ref{basic function k})
we find that
\begin{eqnarray}
&&
|h(\omega \,, \vec{k} \,, \infty )|^{2} \approx 
\frac{1}{(\omega ^{2} - \tilde{\omega }_{k}^{2}(T) )^{2} + (\pi \,
r_{-}(\omega \,, \vec{k}))^{2}} \,, 
\label{limit h} 
\\ &&
|k(\omega \,, \vec{k} \,, \infty )|^{2} \approx 
\frac{\omega ^{2}}{(\omega ^{2} - \tilde{\omega }_{k}^{2}(T))^{2} + (\pi \,
r_{-}(\omega \,, \vec{k}))^{2}} \,,
\label{limit k} 
\\ &&
\tilde{\omega }_{k}^{2}(T) = \vec{k}^{2} + M^{2} + \frac{\lambda }{12}\,
T^{2} + \Pi (\omega _{k} \,, \vec{k}) \,,
\\ &&
\Pi (\omega \,, \vec{k}) = -\,{\cal P}\,\int_{-\infty }^{\infty }\,d\omega' \,
\frac{r_{-}(\omega' \,, \vec{k})}{\omega' - \omega } \,.
\end{eqnarray}
A possible infinity in the proper self-energy
$\Pi (\omega _{k} \,, \vec{k})$ is removed by
a mass counter term such that the combination
\( \:
M^{2} + \Pi (\omega _{k} \,, \vec{k})
\: \)
is made finite by renormalization;
after the wave function renormalization as explained in 
Appendix B,
\begin{eqnarray}
&&
\tilde{\omega }_{k}^{2}(T) \rightarrow (\omega _{k}^{R})^{2}
= \vec{k}^{2} + M_{R}^{2} + \frac{\lambda }{12}\,T^{2}
+ \delta \Pi (\omega _{k} \,, \vec{k}) \,, 
\label{renormalized mass at finite t} 
\end{eqnarray}
where $M_{R}$ is the finite renormalized mass and
$\delta \Pi (\omega _{k} \,, \vec{k})$ is defined in 
eq.(\ref{subtracted finite self-energy}).

The self-consistent equation for the stationary value is then
\begin{eqnarray}
&&
f_{{\rm eq} }(\vec{k}) + \frac{1}{2} = 
\int_{0}^{\infty }\,d\omega \,(\,\frac{\omega _{k}}{2} +
\frac{\omega ^{2}}{2\omega _{k}}\,)\,
\frac{r_{+}(\omega \,, \vec{k})}{(\omega ^{2} - 
\tilde{\omega }_{k}^{2}(T))^{2}
+ (\pi r_{-}(\omega \,, \vec{k}))^{2}} \,, 
\label{stationary self-c eq1} 
\\ &&
v_{{\rm eq} }(\vec{k}) = \int_{0}^{\infty }\,d\omega \,
\frac{\omega ^{2} - \omega _{k}^{2}}{2\omega _{k}}\,
\frac{r_{+}(\omega \,, \vec{k})}{(\omega ^{2} - 
\tilde{\omega }_{k}^{2}(T))^{2}
+ (\pi r_{-}(\omega \,, \vec{k}))^{2}} \,, 
\label{stationary self-c eq2} 
\\ &&
\hspace*{1cm} r_{\pm }(\omega \,, \vec{k}) =
r(\omega \,, \vec{k}) \pm r(- \omega \,, \vec{k}) \,, 
\label{stationary self-c eq3} 
\\ &&
r(\omega \,, \vec{k}) =
2\,\int\,\frac{d^{3}k'}{(2\pi )^{3}2\omega _{k'}}\,
\left( \,
\frac{r_{\chi }(\omega  + \omega _{k'} \,, \vec{k} + \vec{k}')}
{1 - e^{-\beta (\omega  + \omega _{k'})}}\,f_{{\rm eq} }(\vec{k'} )
\right.
\nonumber \\ && \hspace*{0.5cm} 
\left.
+ \,
\frac{r_{\chi }(\omega  - \omega _{k'} \,, \vec{k} - \vec{k}')}
{1 - e^{-\beta (\omega  - \omega _{k'})}}\,(1 + f_{{\rm eq} }(\vec{k'})\,) 
- \,
\frac{2\,r_{\chi }(\omega  \,, \vec{k} + \vec{k'})}{1 - e^{-\beta \omega }}\,
v_{{\rm eq} }(\vec{k'})
\,\right) \,. \label{stationary self-c eq4} 
\end{eqnarray}
In this self-consistency equation
the two-body kernel $r_{\chi }$ is a given function.
The function $v_{{\rm eq} }(\vec{k})$ is the stationary
value of $v(\vec{k} \,, t)$ given by
\begin{eqnarray}
&&
v(\vec{k} \,, t) = \langle \,\frac{1}{2\omega _{k}}\,p_{k}^{2}(t)
- \frac{\omega _{k}}{2}\,q_{k}^{2}(t) \, \rangle \,,
\end{eqnarray}
and may be considered as a functional of $f_{{\rm eq} }(\vec{k})$
when one first solves eq.(\ref{stationary self-c eq2}) for $v_{{\rm eq} }$.

The set of self-consistent equations, eqs.(\ref{stationary self-c eq1})
$-$ (\ref{stationary self-c eq4}), along with the definition of
$r_{\chi }$ (\ref{two-body spectral}), does not depend on
the initial $\varphi $ state.
We assume that there exists a unique solution to the self-consistent
equation, when a proper renormalization is performed.
We also anticipate and later deomonstrate that
the high temperature limit of this equation gives the familiar
distribution function $1/(e^{\beta \omega _{k}} - 1)$.

A simplified computation is possible in the weak coupling
limit. In the limit of $\lambda \rightarrow 0$, one can ignore
the $\omega $ dependence of $r_{-}(\omega \,, \vec{k})$ in
the denominator of the 
Breit-Wigner function since it is narrowly peaked around 
$\omega = \omega _{k}$.
By separating the narrow width contribution, one has 
\begin{eqnarray}
&& \hspace*{0.5cm} 
f_{{\rm eq} }(\vec{k}) = \frac{r(-\,\omega _{k}\,, \vec{k})}
{r_{-}(\omega _{k} \,, \vec{k})} + \delta \tilde{f}_{{\rm eq} }(\vec{k}) \,, 
\\ && 
\delta  \tilde{f}_{{\rm eq} }(\vec{k}) = \frac{1}{4\omega _{k}}\,
\int_{-\infty }^{\infty }\,d\omega \,
\,\frac{r_{+}(\omega \,, \vec{k}) - r_{+}(\omega _{k}\,,\vec{k})}
{(\omega - \tilde{\omega }_{k}(T) )^{2} + \Gamma _{k}^{2}/4}
\,, \label{off-shell stationary} 
\end{eqnarray}
with 
\( \:
\Gamma _{k} = \pi\, r_{-}(\omega _{k} \,, k)/\omega _{k} \,.
\: \)
We have used $r_{+}(-\omega \,, \vec{k}) = r_{+}(\omega \,, \vec{k})$.
The tilded $\delta \tilde{f}_{{\rm eq} }$ 
contains terms to be renormalized away by
subtraction, in contrast to the finite $\delta f_{{\rm eq} }$ later 
defined after renormalization.
Note that both $\delta \tilde{f}_{{\rm eq} }$ and $v_{{\rm eq} }$ are
of the same coupling order, $O[\lambda ^{2}]$,
since the on-shell term of $O[\lambda ^{0}]$ is absent.
It is then easy to see that the quantity $v_{{\rm eq} }$ gives
an even higher order $O[\lambda ^{4}]$ correction to $r_{\pm }$,
and one can forget about $v_{{\rm eq} }$ altogether to
$O[\lambda ^{2}]$.

The self-consistent equation in the present form is not particularly
illuminating, since cancellation occurs among scattering
terms in the on-shell part
when the distribution function is integrated over momenta,
namely in the quantity,
\begin{equation}
\int\,\frac{d^{3}k}{(2\pi )^{3}}\,\Gamma _{k}\,\left( \,
f_{{\rm eq} }(\vec{k}) - \frac{r(-\,\omega _{k} \,, \vec{k})}
{r_{-}(\omega _{k} \,, \vec{k})}\,\right) \,.
\end{equation}
In the next section we shall directly work out the stationary
number density integrated over momenta.
In the rest of this section we shall focus on the off-shell part
$\delta \tilde{f}_{{\rm eq} }$.

To further simplify the off-shell contribution, we 
use a sum rule that results from a consistency condition;
from the equality,
\( \:
\dot{g}(0) = 1 \,, 
\: \)
\begin{eqnarray}
&&
2\,\int_{0}^{\infty }\,d\omega \,
\frac{\omega \,r_{-}(\omega \,, \vec{k})}
{(\omega ^{2} - \tilde{\omega } _{k}(T)^{2})^{2} 
+ (\pi r_{-}(\omega \,, \vec{k}))^{2}}
= 1 \,.
\label{consistency integral} 
\end{eqnarray}
Using
\begin{eqnarray}
&& \hspace*{1.5cm} 
\int_{-\infty }^{\infty }\,d\omega \,\frac{1}
{(\omega  - \tilde{\omega } _{k}(T))^{2} 
+ \Gamma _{k}^{2}/4}
= \frac{2\pi }{\Gamma _{k}} \,,
\\ && \hspace*{-1cm}
\int_{0}^{\infty }\,d\omega \,
\frac{\omega \,r_{-}(\omega \,, \vec{k})}
{(\omega ^{2} - \tilde{\omega } _{k}(T)^{2})^{2} 
+ (\pi r_{-}(\omega_{k} \,, \vec{k}))^{2}}
\approx \frac{1}{4\tilde{\omega } _{k}(T)} \,
\int_{-\infty }^{\infty }\,d\omega \,\frac{r_{-}(\omega \,, \vec{k})}
{(\omega  - \tilde{\omega } _{k}(T))^{2} 
+ \Gamma _{k}^{2}/4}
\,, \nonumber \\ &&
\end{eqnarray}
with 
\( \:
\Gamma _{k} = \pi r_{-}(\omega _{k} \,, \vec{k})
/\tilde{\omega }_{k}(T) \approx \pi r_{-}(\omega _{k} \,, \vec{k})
/\omega _{k} \,, 
\: \)
the consistency integral (\ref{consistency integral}) may be rewritten as
\begin{equation}
\int_{-\infty }^{\infty }\,d\omega \,
\frac{r_{-}(\omega \,, \vec{k}) - r_{-}(\omega _{k} \,, \vec{k})}
{(\omega - \tilde{\omega } _{k}(T))^{2} + \Gamma _{k}^{2}/4}  = 0 \,.
\label{consistency integral 2} 
\end{equation}
We have neglected the $\omega $ dependence of $r_{-}(\omega \,, \vec{k})$
in the denominator in deriving this equation.

Since 
\( \:
r_{\pm }(\omega ) = r(\omega ) \pm r(-\omega ) \,, 
\: \)
the off-shell equilibrium distribution becomes
\begin{eqnarray}
&&
\delta \tilde{f}_{{\rm eq} } (\vec{k}) = \frac{1}{2\omega _{k}}\,
\int_{-\infty }^{\infty }\,d\omega \,
\frac{r(-\,\omega \,, \vec{k}) - r(-\,\omega _{k} \,, \vec{k})}
{(\omega - \tilde{\omega }_{k}(T))^{2} + \Gamma _{k}^{2}/4}
\,, \label{finite t-dependent off-shell f} 
\\ && \hspace*{0.5cm} 
r(-\,\omega \,, \vec{k}) = 2\,\int\,\frac{d^{3}k'}{(2\pi )^{3}2\omega _{k'}}
\,\left( \,
\frac{r_{\chi }(|\omega - \omega _{k'}| \,, \vec{k} - \vec{k'})}
{e^{\beta |\omega - \omega _{k'}|} - 1}\,f(\vec{k'} \,, t)
\right.
\nonumber \\ && \hspace*{-1cm}
+ \,
\frac{r_{\chi }(|\omega + \omega _{k'}| \,, \vec{k} + \vec{k'})}
{e^{\beta |\omega + \omega _{k'}|} - 1}\,
(\,1 + f(\vec{k'} \,, t)\,) 
+ \theta (\omega _{k'} - \omega )\,
r_{\chi }(|\omega - \omega _{k'}| \,, \vec{k} - \vec{k'})\,
f(\vec{k'} \,, t)
\nonumber \\ && \hspace*{1cm} 
\left.
+\, \theta (-\,\omega - \omega _{k'})\,
r_{\chi }(|\omega + \omega _{k'}| \,, \vec{k} + \vec{k'})\,
(\,1 + f(\vec{k'} \,, t)\,) \,\right)
 \,.
\end{eqnarray}
In the spectral function $r(-\,\omega \,, \vec{k})$ here,
one can disregard the term involving $v_{{\rm eq} }$ in eq.
(\ref{stationary self-c eq4}),
since it is of higher order, $O[\lambda ^{4}]$.

A region of $\omega $ integral in the formula 
(\ref{off-shell stationary}) for $\delta \tilde{f}_{{\rm eq} }$ does not
contribute; this is the region of
$|\omega - \omega _{k}| < \omega _{c}$, where
$\omega _{c}$ is the energy scale for which $r(\omega\,, \vec{k})$ 
appreciably varies and
in the weak coupling limit $\omega _{c} \gg \Gamma _{k}$.
This means that the resulting $\delta \tilde{f}_{{\rm eq} }(\vec{k})$ is
insensitive to the actual value of $\Gamma _{k}$.
We may use this fact to replace $\Gamma _{k}$ by a small quantity
$\tilde{\Gamma }_{k}$ which is taken independent 
of the distribution function $f(\vec{k})$;
\begin{eqnarray}
&&
\delta \tilde{f}_{{\rm eq} }(\vec{k}) \approx 
\frac{1}{2\omega _{k}}\,\int_{-\infty }^{\infty }\,
\frac{r(-\, \omega \,, \vec{k}) - r(-\,\omega _{k} \,, \vec{k})}
{(\omega - \tilde{\omega }_{k}(T))^{2} + \tilde{\Gamma }_{k}^{2}/4}  \,.
\end{eqnarray}
We shall specify $\tilde{\Gamma }_{k}$ shortly
in (\ref{universal width}) by a quantity even independent of
the momentum $\vec{k}$.
A great virture of this formula is that the off-shell part 
$\delta \tilde{f}_{{\rm eq} }$ of the
self-consistency formula  becomes a linear functional equation
of $f_{{\rm eq} }(\vec{k})$, and the perturbative treatment becomes
transparant.
This form of the equilibrium distribution is most convenient and
is later used frequently.

The universal width factor $\tilde{\Gamma }_{k}$ independent of
$f$ may be defined by using the value of the width $\Gamma _{k}$ 
calculated at $f = 0$. 
Since 
\( \:
\Gamma _{k} = \pi r_{-}(\omega _{k} \,, \vec{k})/\omega _{k}
\: \)
is the on-shell value, 
it is dominated by the scattering contribution,
\( \:
\Gamma _{k} = \pi r_{s}(\omega _{k} \,, \vec{k})/\omega _{k} \,.
\: \)
Thus, it numerically
has the thermal number density factor $n_{{\rm th}} \approx T^{3}$
times the scattering cross section of order $\lambda ^{2}/M^{2}$. 
More explicitly, $\tilde{\Gamma }_{k}$ becomes independent of
the mode $\vec{k}$, hence is denoted by $\tilde{\Gamma }$ with
\begin{eqnarray}
&&
\tilde{\Gamma } = \frac{\zeta (3)\,\lambda ^{2}T^{3}}
{4\pi ^{3}\,M^{2}} \,. \label{universal width} 
\end{eqnarray}

In Appendix B we identify the renormalization term in the proper
self-energy $\Pi (\omega \,, \vec{k})$.
This gives constant counter terms to be added to the distribution function;
the renormalized finite occupation number is
\begin{eqnarray}
&& 
f_{{\rm ren}}(\vec{k}) =  f_{{\rm eq}}(\vec{k}) 
- \frac{A + B\,\vec{k}^{2}}{4\omega _{k}^{2}} 
\,.
\end{eqnarray}
The infinite counter term $A \,, B$ cancells the corresponding
infinity of the $O[\lambda ^{2}]$ correction in the distribution
function.

In Appendix D we give a detailed account of our computation
of 
\begin{equation}
\delta f_{{\rm eq} }(\vec{k}) = f_{{\rm ren}}(\vec{k}) -
\frac{r(-\,\omega _{k} \,, \vec{k})}{r_{-}(\omega _{k} \,, \vec{k})}
\,.
\end{equation}
To the leading order of
$T/M$ the result of this computation is summarized as a sum
of a $f$ dependent and a $f$ independent term,
\begin{eqnarray}
&& \hspace*{1cm} 
\delta f_{{\rm eq} }(\vec{k}) = f_{f}(\vec{k}) + f_{2}^{0}(\vec{k})
\,. \label{off-shell f final 1} 
\end{eqnarray}
As will be explained in Appendix D, the $f-$independent term has
a dominant contribution from the inverse process $\chi \chi 
\rightarrow \varphi \varphi $, and is computed as
\begin{eqnarray}
&&
f_{2}^{0}(\vec{k}) = 
\frac{1}{2\omega _{k}}\,\int_{-\infty }^{\infty }\,d\omega \,
\frac{1}{(\omega - \omega _{k})^{2} + \Gamma _{k}^{2}/4}\,
2\,\int\,\frac{d^{3}k'}{(2\pi )^{3}2\omega _{k'}}\,
\nonumber \\ && \hspace*{2cm} \cdot 
\left(\, \frac{r_{\chi }(|\omega + \omega _{k'}| \,, \vec{k} + \vec{k'})}
{e^{\beta |\omega + \omega _{k'}| } - 1} - (\,\omega \rightarrow \omega _{k}\,)
\,\right)
\label{integral formula f-independent f} 
\\ &&
\approx 
\frac{2}{\omega _{k}}\,\int\,\frac{d^{3}k'}{(2\pi )^{3}2\omega _{k'}}\,
\frac{1}{(\omega _{k} + \omega_{k'})^{2}}\,\int_{0}^{\infty }\,d\omega \,
\frac{r_{\chi }(\omega \,, \vec{k} + \vec{k'})}{e^{\beta \omega } - 1}
\nonumber 
\\ && \hspace*{-1cm}
\approx 
\frac{\zeta (2)\lambda ^{2}}
{16\pi ^{4}}\,\frac{T^{2}}{k\omega _{k}}\,
\int_{0}^{\infty }\,dq\,\frac{q}{2q + \zeta (2)T}\,
\frac{1}{e^{q/2T} - 1}\,\left( \frac{1}{\omega _{k} + \omega _{k - q}}
- \frac{1}{\omega _{k} + \omega _{k + q}}\right) 
\,, \label{off-shell f final 2} 
\end{eqnarray}
The $f-$dependent term $f_{f}$ is given in Appendix D, but actually
is not necessary to write
the time evolution equation for the number density.
The use of the low temperature formula for $r_{\pm }$ for
computation of $\delta f_{{\rm eq} }$ even at high temperatures is
justified, since at higher temperatures the on-shell contribution
dominates and the term $\delta f_{{\rm eq} }$ becomes irrelevant
in $f_{{\rm eq} }(\vec{k})$ which is dominated by the familiar $f_{{\rm th}}$.

An example of the $\omega-$integrand (\ref{integral formula f-independent f})
is shown in Fig.5, where the total, the Planck, and the rest of the 
integrand are separately given.

Physical processes that contribute to the important piece $f_{2}^{0}$
are predominantly inverse annihilation 
$\chi \chi \rightarrow \varphi \varphi $, and 1 to 3 process,
$\chi \rightarrow \chi \varphi \varphi $, which gives a smaller fraction
of the total number density.
On the other hand, the term $f_{f}$ arises from the inverse scattering
process. We shall show in the next section that
the scattering-related 
term $f_{f}(\vec{k})$ is subdominant compared to $f_{2}^{0}(\vec{k})$
in determining the equilibrium number density.

Although it is technically complicated to calculate the equilibrium
distribution function $f_{{\rm eq}}$, one may regard this calculation
as a self-consistent approximation for the quantity,
\begin{equation}
f_{{\rm eq}}(\vec{k}) = \frac{{\rm tr}\; (\,a_{k}^{\dag }a_{k}\,
e^{-\,\beta H_{{\rm tot}}}\,)}{{\rm tr}\;e^{-\,\beta H_{{\rm tot}}}}
\,, 
\end{equation}
where $H_{{\rm tot}}$ is the total Hamiltonian including interaction
between the $\varphi $ system and the $\chi $ environment.
We explicitly demonstrated this relation for the exactly solvable
harmonic oscillator model in \cite{jmy-97}, whose result is
used for our self-consistent solution.
Thus, the Gibbs formula is valid, while the ideal gas form of
the distribution $1/(e^{\beta \omega _{k}} - 1)$ is changed.

\hspace*{0.5cm} 
\section{Quantum kinetic equation
}

\vspace{0.5cm} 
\hspace*{0.5cm} 
We go back to the time dependent expectation values, given in
eq.(\ref{correlator q^2}) $-$ (\ref{correlator qp}).
It is sometimes convenient to write a formula for the time derivative;
\begin{eqnarray}
&&
\hspace*{1cm} 
\frac{df(\vec{k} \,, t)}{dt} = -\,\Gamma (\vec{k}\,, t)
\nonumber 
\\ && 
\cdot 
\left( \,f(\vec{k} \,, t) - \int_{0}^{\infty }\,d\omega \,
r_{+}(\omega \,, \vec{k})
\,(\,
\frac{\omega_{k}}{2}|h(\omega \,, \vec{k} \,, t)|^{2} 
+ \frac{1}{2\omega _{k}}\,
|k(\omega \,, \vec{k} \,, t)|^{2}\,)
\,\right)
\nonumber \\ && \hspace*{0.5cm} 
+ \,
\int_{0}^{\infty }\,d\omega \,
\frac{d}{dt}\,r_{+}(\omega \,, \vec{k})\,(\,
\frac{\omega_{k}}{2}|h(\omega \,, \vec{k} \,, t)|^{2} 
+ \frac{1}{2\omega _{k}}\,
|k(\omega \,, \vec{k} \,, t)|^{2}\,)
\,, \label{non-Markovian kinetic eq} 
\\ && \hspace*{1cm} 
\Gamma (\vec{k} \,, t) \equiv 
\nonumber \\ &&
\hspace*{-1cm}
-\,\frac{d}{dt}\,\ln \left( 
(\frac{\omega_{k} }{2}\,g^{2} + \frac{\dot{g}^{2}}{2\omega _{k}})\,
\overline{p_{i}^{2}} + (\frac{\omega _{k}}{2}\,\dot{g}^{2} +
\frac{\stackrel{..}{g}^{2}}{2\omega _{k}})\,\overline{q_{i}^{2}}
+ \frac{\dot{g}}{2}\,(\omega _{k}\,g + \frac{\stackrel{..}{g}}
{\omega _{k}})\,\overline{q_{i}p_{i}+p_{i}q_{i}}
\,\right) \,. \nonumber \\ &&
\label{annihilation rate gamma} 
\end{eqnarray}
Regarded as a time evolution equation, this equation has 
a remarkable property;
the initial memory effect is confined to, and isolated by,
the rate $\Gamma (\vec{k} \,, t)$.
All other quantities in this equation are written in terms of
those at the same local time, including the distribution function
$f(\vec{k} \,, t)$ itself.

We take a Markovian limit of this equation.
The idea is as follows.
The equation above, although exact in the slow variation limit of the Hartree
model, has the initial memory effect.
To retain the memory effect to late times may not make much sense
when one wants to discuss an average behavior of the time evolution 
for a collective body of particles in a complex environment.
After all, it is impossible to follow the time evolution for all
particles in the ensemble. It may be possible to specify the initial data 
such as the ensemble-averaged 
$\overline{q_{i}^{2}}$ for all momentum modes at some specific time,
and the data may have a simple regulated form 
if one assumes a simple initial distribution function for $\varphi $.
But, remember that in the problem of our interest there are processes
occuring rapidly such as the scattering, distinct from the other 
simultaneously occuring slow process such as the annihilation,
which is  of our main interest.
Under this circumstance different particles undergo different
history of evolution, and after a while it
is almost impossible to keep track of the updated initial data.

One really does not care much about the details of the complicated
time history. We only care about a global and slow change for
a few key quantities.
Fortunately, in many physical situations one has a clear separation
of at least two time scales; one due to an elementary quantum
behavior and the other for the bulk behavior.
With this in mind
it is much more sensible to erase the initial memory effect and
to coarse-grain the time evolution averaging out 
the fast oscillatory behavior.
A nice feature of the Hartree model is that 
if one eliminates the initial time dependence in
$\Gamma (\vec{k} \,, t)$, a simple Markovian description of
the transport phenomenon becomes possible.

In the weak coupling limit the exponential law 
\( \:
g(\vec{k} \,, t) \approx \sin (\omega _{k}\,t)\,e^{-\,\Gamma _{k}t/2}
/\omega _{k} 
\: \)
gives a constant decay rate,
\begin{eqnarray}
\Gamma_{k} = 
\frac{\pi }{\omega_{k} }\,r_{-}(\omega _{k} \,, \vec{k})
= \frac{\pi }{\omega_{k} }\,\left( \,r(\omega _{k} \,, \vec{k})
- r(-\,\omega _{k} \,, \vec{k})\,\right)
\,, \label{on-shell annihilation gamma} 
\end{eqnarray}
after the time average. 
Except at very early times the initial memory effect is almost
completely erased.
Via the spectral function this $\Gamma _{k}$
depends on the $\varphi $ distribution function.
As seen from eqs.(\ref{on-shell spectral +}), (\ref{on-shell spectral -}),
this combination $r_{-}$ is the rate of decrease of heavy particles
due to the annihilation, the scattering and their inverse processes.

Our Markovian evolution equation is then
\begin{eqnarray}
&& 
\frac{df(\vec{k} \,, t)}{dt} = -\,\Gamma_{k}
\,\left(\, f(\vec{k} \,, t) - f_{{\rm eq}}(\vec{k}) \,\right) \,, 
\label{kinetic equation} 
\\ && 
f_{{\rm eq}}(\vec{k}) = \frac{r(-\,\omega _{k}\,, \vec{k})}
{r_{-}(\omega _{k} \,, \vec{k})} + \delta f_{{\rm eq}}(\vec{k}) 
\,.
\end{eqnarray}
Here the off-shell contribution $\delta f_{{\rm eq}}(\vec{k})$ is
given by (\ref{off-shell f final 1}) $-$ 
(\ref{off-shell f final 2}).
Although not written explicitly, the distribution function
used to compute $\Gamma _{k}$ and $f_{{\rm eq}}(\vec{k})$
should be the instantaneous one,
$f(\vec{k} \,, t)$, with the same common time $t$ as in the left side.
The Markovian evolution equation for the momentum distribution
function thus derived is the most fundamental result in the present
work.

We note that the same Markovian approximation, when applied to the model
of unstable particle decay, gives the kinetic equation identical to
eq.(\ref{kinetic equation}), except that $\Gamma _{k}$ in that case 
is the decay rate on the mass shell and that $r_{\pm }(\omega \,, \vec{k})$
in $f_{{\rm eq}}(\vec{k})$
should be replaced by the appropriate spectral weight for the decay
process such as $r_{\chi }(\omega \,, \vec{k})$;
more precisely
\( \:
r_{+} \rightarrow r_{\chi } \coth \frac{\beta \omega }{2} \,, 
\hspace{0.3cm}
r_{-} \rightarrow r_{\chi } \,.
\: \)
We give in Appendix C the expression and its actual value 
of the stationary distribution
$f_{{\rm eq}}(\vec{k})$ for the boson decay model. 

We shall recapitulate the problem related to $v(\vec{k})$.
The Markovian equation derived above does contain the spectral
combination $r_{\pm }$, which is written using the distribution function
$f(\vec{k} \,, t)$ and and also $v(\vec{k} \,, t)$. 
This new function describes a deviation
from the simple particle picture in the field theory;
in terms of the harmonic coordinate and its conjugate momentum,
\begin{equation}
v(\vec{k} \,, t) = \langle \,\frac{1}{2\omega _{k}}\,p_{k}^{2}(t)
- \frac{\omega _{k}}{2}\,q_{k}^{2}(t)\, \rangle \,.
\label{v def} 
\end{equation}
This combination  of dynamical variables obeys another time
evolution equation different from the distribution function.
The coarse-graining, under the same slow variation approximation
applied previously, gives
\begin{eqnarray}
&&
v(\vec{k} \,, t) = \int_{0}^{\infty }
\,d\omega \,\frac{\omega ^{2} - \omega _{k}^{2}}
{2\omega _{k}}\,\frac{r_{+}(\omega \,, \vec{k})}
{(\omega ^{2} - \omega _{k}^{2})^{2} + (\pi r_{-}(\omega \,, \vec{k}))^{2}}
\,.
\label{v-integral} 
\end{eqnarray}
The initial memory term for the combination (\ref{v def})
has the fast oscillatory behavior, which
vanishes on the average. This behavior is different from that of the
occupation number which has a slowly decaying component,
non-vanishing after the time average.
In this sense it is best to use the expression (\ref{v-integral})
for $v(\vec{k} \,, t)$
and the defining equation of $r_{\pm }$ containing both $f(\vec{k} \,, t)$
and $v(\vec{k} \,, t)$, 
as a functional equation to determine $v(\vec{k} \,, t)$
in terms of $f(\vec{k} \,, t)$.
This way we arrive at a closed form of time evolution equation for
the distribution function.
In the Markovian time evolution equation the quantity $v(\vec{k}\,, t)$
appears via the spectral function $r(\pm \,\omega \,, \vec{k})$,
which we now show to be of negligible higher order.

In the weak coupling limit of our main concern,
both $r_{\pm }$ has an overall small coupling, and
the integral (\ref{v-integral}), that excludes contribution from
the Breit-Wigner region of $\omega \approx \omega _{k}$ by the factor
$\omega ^{2} - \omega _{k}^{2}$, gives  $v = O[\lambda ^{2}]$, 
hence gives $O[\lambda ^{4}]$ contribution to the spectral
function $r$, as seen from (\ref{discontinuity formula}).
We shall thus take $v$ vanishing from now on.

With the vanishing $v$, 
the usual Boltzmann equation follows when one approximates the
energy integral for $f_{{\rm eq}}$ by the pole term, namely
\( \:
r(-\omega _{k} \,, \vec{k})/r_{-}(\omega _{k} \,, \vec{k}) \,.
\: \)
This narrow width approximation gives the evolution equation,
\begin{eqnarray}
&&
\frac{df(\vec{k} \,, t)}{dt} = 
-\,\Gamma_{k}
\left( \, f(\vec{k}\,, t) -
\frac{r(-\,\omega_{k} \,, \vec{k})}{r(\omega _{k} \,, \vec{k}) -
r(-\,\omega _{k} \,, \vec{k})} 
\,\right)
\,.
\end{eqnarray}
One may use the formula (\ref{on-shell annihilation gamma}) 
for $\Gamma _{k}$ and write this as
\begin{equation}
\frac{df(\vec{k} \,, t)}{dt} = - \frac{\pi }{\omega _{k}}\,
\left( \,r(\omega _{k} \,, \vec{k})\,f(\vec{k}\,, t)
- r(-\,\omega _{k}\,, \vec{k})\,(1 + f(\vec{k}\,, t))\,\right) \,.
\end{equation}
Again, the instantaneous $\varphi $ distribution function is
taken in computing $r(\pm\, \omega _{k} \,, \vec{k})$ of 
the right hand side.
In view of the on-shell relation (\ref{on-shell spectral +}),
(\ref{on-shell spectral -}) for $r(\pm\, \omega _{k}\,, \vec{k})$,
this is equivalent to the Boltzmann equation for the present
annihilation-scattering problem in thermal medium;
\begin{eqnarray}
&& \hspace*{1cm} 
\frac{df(\vec{k} \,, t)}{dt} = \lambda ^{2}\,
\int\,dk'\,\int\,dk_{1}\,\int\,dk_{2}\,
\nonumber \\ &&
\left( \,
(\,1 + f_{{\rm th}}(k_{1})\,)\,(\,1 + f_{{\rm th}}(k_{2})\,)\,
f(k')\,f(k)\,\delta (k + k' - k_{1} - k_{2})
\right.
\nonumber \\ && 
+\, 
2f_{{\rm th}}(k_{1})\,(\,1 + f_{{\rm th}}(k_{2})\,)\,(\,1 + f(k')
\,)\,f(k)\,\delta (k + k_{1} - k' - k_{2})
\nonumber 
\\ && 
- \,f_{{\rm th}}(k_{1})\,f_{{\rm th}}(k_{2})\,(\,1 + f(k')\,)
(\,1 + f(k)\,)\,\delta (k + k' - k_{1} - k_{2})
\nonumber \\ && 
\left. -\,2f_{{\rm th}}(k_{2})\,(\,1 + f_{{\rm th}}(k_{1})\,)\,f(k')\,
(\,1 + f(k)\,)\,\delta (k + k_{1} - k' - k_{2})
\,\right)\,,
\label{boltzmann eq} 
\end{eqnarray}
where all relevant processes are explicitly written.
We thus find that our Markovian Hartree model gives a foundation
to  derivation of the Boltzmann equation under the narrow
resonance approximation.

Note that the thermally averaged rate in the right  side of
the Boltzmann equation (\ref{boltzmann eq}) has separate contributions of
the annihilation and its inverse process characterized by 
$\delta (k + k' - k_{1} - k_{2})$, and the scattering plus its inverse by
$\delta (k + k_{1} - k' - k_{2})$. 
There is no contribution from 1 to 3, and 3 to 1 processes
such as $\varphi \leftrightarrow \varphi \chi \chi $ in the
Boltzmann equation.
When one integrates over the particle momentum $\vec{k}$ to
discuss the time evolution of the number density,
the entire scattering contribution drops out.
This is reasonable, because the scattering process does conserve the
particle number, hence the scattering process
does not cause the change of the $\varphi $ particle number.
However, the momentum distribution function changes by the scattering.
Moreover, it is not immediately clear how the off-shell scattering term 
contributes within our crossing symmetric approach.
We thus keep the scattering term for the time being, as it stands.

The equilibrium distribution function is defined by setting
$df/dt = 0$. We assume as before that this has the unique solution,
which in the case of the Boltzmann equation
must agree with that of the thermal and the chemical equilibrium,
\( \:
f(k) = 1/(e^{\beta \omega _{k}} - 1) \,.
\: \)
The concept of partial equilibrium may however be of some use.
For instance, if the scattering process takes place much more frequently
than the annihilation process, it may be useful to suppose that
the energy exchange is equilibrated, but the species number change is
not fast enough. 
In this case the thermal equilibrium, and not
the chemical equilibrium, is reached with
\begin{equation}
f(\vec{k}) = \frac{1}{e^{\beta (\omega _{k} - \mu )} - 1} 
\equiv f_{{\rm th}}^{\mu } (k) \,, 
\label{distribution with chemical potential} 
\end{equation}
where $\mu $ is the chemical potential.
Under this thermal equilibrium of a finite chemical potential
all scattering-related terms in
eq.(\ref{boltzmann eq}) cancels each other since
\begin{eqnarray}
&& \hspace*{0.5cm} 
\left[ r_{s}(\omega _{k}\,, k)\right]_{f = f_{{\rm th}}^{\mu }}\,
f_{{\rm th}}^{\mu }(k) - 
\left[ r_{s}(- \,\omega _{k}\,, k)\right]_{f = f_{{\rm th}}^{\mu }}\,
(\,1 + f_{{\rm th}}^{\mu }(k)\,) = 0  \,,
\label{balance of scattering} 
\end{eqnarray}
and only the annihilation-related terms are to be retained.
Here
\begin{eqnarray}
&&
\hspace*{0.5cm}
r_{s}(\omega_{k} \,, \vec{k}) = 2\lambda ^{2}\,
\int\,dk'\,\int\,dk_{1}\,\int\,dk_{2}\,
\nonumber \\ && \hspace*{1cm} 
\cdot f_{{\rm th}}(k_{1})\,(\,1 + f_{{\rm th}}(k_{2})\,)\,
(\,1 + f(k')\,)
\,\delta (k + k_{1} - k' - k_{2})
\,, \label{on-shell scat rate} 
\\ && \hspace*{0.5cm}
r_{s}(- \omega _{k}\,, \vec{k}) = 4\lambda ^{2}\,\int\,dk'\,\int\,dk_{1}\,
\int\,dk_{2}\,
\nonumber \\ &&
\hspace*{1cm} 
\cdot f_{{\rm th}}(k_{2})(\,1 + f_{{\rm th}}(k_{1})\,)\,
f(k')\,\delta (k + k_{1} - k' - k_{2}) \,.
\end{eqnarray}
In this case the Boltzmann equation effectively describes the time evolution
for the chemical potential.
Note however that the thermal equilibrium distribution of a finite
chemical potential is realized, assuming that the on-shell
Boltzmann equation is a valid description of our problem without
the off-shell contribution.

As will be made clear shortly, it is not easy to obtain a readily
calculable form of the
equilibrium distribution function at low temperatures in
the annihilation-scattering problem, and we shall directly work out
the integrated number density.
On the other hand, for the unstable particle decay we have an approximate 
analytic result, eqs.(\ref{occupation-n for decay 1}) and
(\ref{low-T interpolation of f}), for the distribution function,
which is shown in Fig.13. At low $T < 0.1\,M$ deviation from the
Planck form becomes large, but the distribution is not described
by the form (\ref{distribution with chemical potential}) with
a finite chemical potential.

More generally, it is convenient to separate the on-shell term and
write the rest of contribution, as is done in the preceeding section;
\begin{eqnarray}
&& \hspace*{0.5cm} 
f_{{\rm eq}}(\vec{k}) = 
\frac{r(-\,\omega_{k} \,, \vec{k})}{r(\omega _{k} \,, \vec{k}) -
r(-\,\omega _{k} \,, \vec{k})} + \delta f_{{\rm eq}}(\vec{k})
\,,
\label{off-shell occupation} 
\\ && \hspace*{1cm}
\delta f_{{\rm eq}}(\vec{k}) = f_{f}(\vec{k}) + f_{2}^{0}(\vec{k})
\,,
\end{eqnarray}
with
\( \:
\tilde{\Gamma } = \frac{\zeta (3)\,\lambda ^{2}}{4\pi ^{3}}\,\frac{T^{3}}
{M^{2}} 
\,.
\: \)
The function $f_{2}^{0}$ is given in eq.(\ref{off-shell f final 2}).
Physically, the off-shell contribution $\delta f_{{\rm eq}}(\vec{k})$
in this formula  consists of two terms;
the first $f$ dependent one $f_{f}$ 
due to the inverse scattering process,
and the second $f$ independent one $f_{2}^{0}$
due to the inverse annihilation process, 
\( \:
\chi  \chi \rightarrow  \varphi \varphi  \,,
\: \)
along with a small contribution from 1 to 3 process, 
\( \:
\chi  \leftrightarrow \chi \varphi \varphi  \,.
\: \)

For discussion of the time evolution,
we shall be content with the integrated number density,
\begin{equation}
n(t) = \int\,\frac{d^{3}k}{(2\pi )^{3}}\,f(\vec{k} \,, t) 
\,, 
\end{equation}
and its evolution.
In the discussion of the relic abundance of WIMP this integrated
quantity is of prime interest in cosmology.
As already noted, the evolution equation for the number density
is simplified considering cancellation of the scattering terms
in the on-shell part,
\begin{eqnarray}
&& 
\frac{dn}{dt} = -\,\int\,\frac{d^{3}k}{(2\pi )^{3}}\,
\left( \,
\Gamma _{k}^{{\rm ann}}\,f(\vec{k}\,, t)
- \Gamma _{k}^{{\rm inv}}\,f_{{\rm eq}}(\vec{k} \,, t)  
\,\right) \,,
\end{eqnarray}
where $\Gamma _{k}^{{\rm ann}}\,(\Gamma _{k}^{{\rm inv}})$ is 
the rate keeping the annihilation (inverse annihilation and
inverse scattering) term.
The approximate form of the evolution equation for the number density
at low temperatures ($ T \ll M$) is then given by
\begin{eqnarray}
&& 
\frac{dn}{dt} = -\,
\int\,\frac{d^{3}k}{(2\pi )^{3}}\,\frac{\pi }{\omega _{k}}
\left( \,
r_{a}(\omega _{k} \,, k)\,f(k) - \Gamma _{k}^{{\rm inv}}\,
\delta f_{{\rm eq}}(k) 
\,\right)  \,,
\\ && \hspace*{0.5cm} 
r_{a}(\omega_{k} \,, \vec{k}) = 2\lambda ^{2}\,
\int\,dk'\,\int\,dk_{1}\,\int\,dk_{2}\,
\nonumber \\ && \hspace*{1cm} 
\cdot (\,1 + f_{{\rm th}}(k_{1})\,)\,(\,1 + f_{{\rm th}}(k_{2})\,)\,f(k')
\,\delta (k +  k' - k_{1} - k_{2})
\,. \label{on-shell ann rate} 
\end{eqnarray}

The equilibrium number density is obtained by setting 
$dn/dt = 0$, hence is given by
\begin{eqnarray}
&&
{\rm RHS} = -\,\int\,\frac{d^{3}k}{(2\pi )^{3}\,2\omega _{k}}\,
\left( \,
r_{a}(\omega _{k} \,, k)\,f(k) - \Gamma _{k}^{{\rm inv}}\,
\delta f_{{\rm eq}}(k) 
\,\right) = 0  \,.
\end{eqnarray}
A more explicit form of this equation is $2\lambda ^{2}$ times
\begin{eqnarray}
&& \hspace*{-1cm}
\int\,dk\,\int\,dk'\,\int\,dk_{1}\,\int\,dk_{2}\,
\left[ \,
(\,1 + f_{{\rm th}}(k_{1})\,)\,(\,1 + f_{{\rm th}}(k_{2})\,)\,f(k')
\,f(k)\,\delta (k +  k' - k_{1} - k_{2})
\right.
\nonumber \\ &&
\left.
-\,
f_{{\rm th}}(k_{1})\,(\,1 + f_{{\rm th}}(k_{2})\,)
\,\delta f_{{\rm eq}}(k)\,(\,1 + f(k')\,)\,\delta (k + k_{1} - k' - k_{2})
\,\right] = 0 \,, 
\label{equilibrium condition} 
\end{eqnarray}
where we used the abbreviated notation for the phase space integral
(\ref{abbreviated phase space}).
Thus, $\Gamma _{k}^{{\rm inv}}$ is given by the second term
of eq.(\ref{equilibrium condition}).

Some sort of averaged cross sections are here in this equation
when $\lambda ^{2} $ is multiplied,
and it would be useful to crudely estimate these. Let us assume that
$\varphi $ particles are non-relativistic and
ignore the effect of stimulated emission. 
One has in the first term 
an averaged annihilation cross section $\overline{\sigma _{a}\,v}$
times the $\varphi $ number density squared $n_{\varphi }^{2}$,
where $v$ is the relative velocity between two annihilating particles,
and $\sigma _{a}v$ is the invariant cross section.
Since the annihilation cross section slowly changes with energy
at low temperatures, one may take the zero energy limit of the
cross section, 
\begin{equation}
\sigma _{a}v_{a} \approx \frac{\lambda ^{2}}{16\pi \,M^{2}} \,.
\end{equation}

In the second term 
there are two contributions corresponding to the two $f$ in 
$\delta f_{{\rm eq}}$, $f_{f}$ and $f_{2}^{0}$.
One of these $f_{f}$ is multiplied by the scattering cross section,
\begin{equation}
\sigma_{s}v_{s} \approx  \frac{\lambda ^{2}}{4\pi \,M^{2}} \,,
\end{equation}
while the second $f_{2}^{0}$ is multiplied by $\sigma _{a}v_{a}$.
The first one is given by $\sigma _{s}v_{s}$ times the
quantity of order
\( \:
n_{{\rm th}}\cdot  n_{1}^{f} \,, 
\: \)
with 
\( \:
n_{{\rm th}} = \frac{\zeta (3)}{\pi ^{2}}\,T^{3} \,,
\: \)
hence, using (\ref{inverse scat to n}),
\begin{eqnarray}
&&
n_{{\rm th}}\,n_{1}^{f} \approx \frac{\zeta (3)\lambda ^{2}}{32\pi ^{6}}
\,(\frac{T}{M})^{2}\,T^{3}\,n_{\varphi } 
\,.
\end{eqnarray}
The second one is given by $\sigma _{a}v_{a}$ times
\begin{eqnarray}
&&
n_{\rm th}\,n_{2}^{0} \approx 
\frac{c\,\zeta (3)\lambda ^{2}}{192\pi ^{5}}\,
\frac{T^{7}}{M} \,,
\end{eqnarray}
with $c \approx 0.27$, eq.(\ref{decay ph integral}).
Thus, once the $\varphi $ number density becomes $\ll O[T^{3}]$,
the second contribution is dominant by $O[MT^{2}/n_{\varphi }]$.
Equating this $\sigma _{a}v_{a}\,n_{\rm th}n_{2}^{0}$ to the
annihilation rate $\sigma _{a}v_{a}\,n_{\varphi }^{2}$ gives
\begin{equation}
n_{\varphi } \approx \sqrt{n_{{\rm th}}n_{2}^{0}} \approx 
\frac{1}{\pi ^{2}}\,\sqrt{\frac{c\,\zeta (3)}{192\pi }}\,\lambda \,
\sqrt{\frac{T}{M}}\,T^{3} 
\approx 0.0023\,\times 
\lambda \,\sqrt{\frac{T}{M}}\,T^{3}
\,.
\end{equation}
We thus derived the equilibrium number density roughly of order
\begin{equation}
10^{-3}\times \lambda \,\sqrt{\frac{T}{M}}\,T^{3}
\end{equation}
at low temperatures which may become much larger than of order
\( \:
(MT)^{3/2}\,e^{-M/T}
\: \)
determined from the Maxwell-Boltzmann distribution of zero chemical
potential.

This argument shows that with 
\( \:
n_{2}^{0} = O[\lambda ^{2}\,T^{4}/M]
\: \)
the $f-$dependent off-shell contributions $f_{i}^{f} $ of order
$n_{i}^{f} = O[\lambda ^{2}\,(T/M)^{\alpha }\,n_{\varphi }]$
are subdominant unless
\begin{equation}
\lambda (\frac{T}{M})^{\alpha - \frac{1}{2}} \geq 1 \,.
\end{equation}
Even the possibly largest case obtained numerically, 
eq.(\ref{f-dependent n numerical}), gives $\alpha \approx 1.35$,
thus confirming that the dominant off-shell contribution is
$n_{2}^{0}$, eq.(\ref{f-independent equil-n}).

\hspace*{0.5cm} 
\section{
Application to cosmology: relic abundance
}

\vspace{0.5cm} 
\hspace*{0.5cm} 
The cosmic expansion has a drastic effect on the annihilation
process of heavy stable particles.
The temperature of cosmic environment particles such as the $\chi $ particle
in our toy model decreases with the scale factor,
\( \:
T \approx 1/a(t) \,,
\: \)
which in turn results in less frequent reaction.
This gives rise to a phenomenon called the freeze-out or the decoupling
of the process.

The freeze-out is described introducing a term of the cosmic expansion
in the evolution equation.
For the evolution of the distribution function $f(\vec{k} \,, t)$,
the time derivative operator is modified to
\begin{equation}
\frac{\partial }{\partial t} - \frac{\dot{a}}{a}\,k\,
\frac{\partial }{\partial k} \,.
\end{equation}
Here $\dot{a}/a $ is the Hubble parameter $H$.
When one integrates over the phase space to get the number density,
the left hand side of the evolution equation becomes
\begin{equation}
\frac{dn}{dt} + 3H\,n = \cdots \,,
\end{equation}
assuming that the distribution function is sufficiently damped in the high
momentum limit.
The evolution equation is further simplifed by introducing
the relative yield $Y$,
\begin{equation}
Y \equiv \frac{n}{T^{3}} \,,
\end{equation}
since
\begin{equation}
\frac{dY}{dt} = \frac{d}{dt}\,\frac{n}{T^{3}} =
T^{-3}\,(\,\frac{dn}{dt} + 3H\,n\,)
\,.
\end{equation}
This holds owing to the temperature-scale relation,
\( \:
T \propto 1/a \,,
\: \)
hence $H = -\,\dot{T}/T$.

The approximate Markovian kinetic equation in the expanding universe
is then
\begin{eqnarray}
&&
\frac{dY}{dt} = -\,2\pi \,\frac{2\lambda ^{2}}{T^{3}}\,
\int\,dk\,\int\,dk'\,\int\,dk_{1}\,\int\,dk_{2}\,
\nonumber \\ && \hspace*{-0.5cm}
\left( \,
(\,1 + f_{{\rm th}}(k_{1})\,)\,(\,1 + f_{{\rm th}}(k_{2})\,)\,
\left( \,f(k')\,f(k) - f_{{\rm MB}}(k')\,f_{{\rm MB}}(k)\,\right)
\,\delta (k +  k' - k_{1} - k_{2})
\right.
\nonumber \\ && \hspace*{0.5cm} 
\left.
- \,
f_{{\rm th}}(k_{1})\,(\,1 + f_{{\rm th}}(k_{2})\,)
\,(\,1 + f(k')\,)\,\delta f_{{\rm eq}}(k)\,\delta (k + k_{1} - k' - k_{2})
\,\right)
\,,
\end{eqnarray}
where we introduced the Maxwell-Boltzmann distribution for
the zero chemical potential,
\begin{equation}
f_{{\rm MB}}(k) = e^{-M/T}\,\exp (-\,\frac{k^{2}}{2MT}) \,.
\end{equation}

A very crude estimate of the freeze-out temperature goes as follows.
One equates the equilibrium annihilation rate 
\( \:
\approx \sigma _{a}v_{a}\cdot n_{\varphi }
\: \)
to the Hubble rate,
\begin{equation}
H = d\,\frac{T^{2}}{m_{{\rm pl}}} \,, \hspace{0.5cm} 
d = \sqrt{\frac{4\pi ^{3}\,N}{45}} \approx 1.66\,\sqrt{N} \,, 
\end{equation}
where $N$ is the number of particle species contributing to
the cosmic energy density. 
This argument, when applied for $n_{\varphi } \gg n_{{\rm MB}} $, 
gives the freeze-out temperature,
\begin{eqnarray}
&&
T_{f} \approx 700 \times N^{1/3}\,
(\frac{M}{m_{{\rm pl}}})^{2/3}\,\frac{M}{\lambda ^{2}}
\approx 1.3 \times 10^{5}\,(\frac{e^{2}}{\lambda })^{2}\,N^{1/3}\,
(\frac{M}{m_{{\rm pl}}})^{2/3}\,M
\nonumber 
\\ &&
\hspace*{0.5cm} 
\sim 50\,keV\,(\frac{e^{2}}{\lambda })^{2}\,
N^{1/3}\,(\frac{M}{100\,GeV})^{5/3}
\,.
\end{eqnarray}
We used the unit of $\lambda $, anticipating a strength of order
the electromagnetic
interaction $\lambda \approx e^{2} = 4\pi \alpha $.
The use of the off-shell formula for $n_{\varphi }$ is usually
justified for $\lambda = O[e^{2}]$, since at low temperatures 
$n_{\varphi } > n_{{\rm MB}}$.

The freeze-out yield is defined by
\( \:
Y_{f} = (n_{\varphi }/T^{3})_{T = T_{f}} \,.
\: \)
We find for the range of parameters,
\( \:
\lambda > 9.3 \times 10^{-5}\,(M/GeV)^{0.32} \,, \;
10^{-3}\,GeV < M < 1\,TeV
\: \)
\begin{eqnarray}
&& 
Y_{f} \approx 0.06 \times N^{1/6}\,(\frac{M}{m_{{\rm pl}}})^{1/3} 
\approx 2.4 \times 10^{-8}\,N^{1/6}\,(M/GeV)^{1/3}
\,.
\end{eqnarray}
Remarkably, this quantity is insensitive to the coupling
constant $\lambda $.
Prior to the freeze-out epoch, this value $Y$ does vary, but
only gradually, since 
\( \:
n_{\varphi } \propto T^{3.5} \,.
\: \)
The freeze-out yield $Y_{f} $ is almost invariant
in the rest of cosmic expansion, as will also be discussed in
analytic estimate below.
The present relic mass density is then estimated from
\begin{equation}
\rho _{0} = M\,Y_{f}\,T_{0}^{3} \,, 
\end{equation}
with $T_{0}$ the present microwave temperature of $\approx 3 \,K$.
Numerically,
\begin{equation}
\rho _{0} \approx 4.1 \times 10^{4} \,N^{1/6}\,(\frac{M}{GeV})^{4/3}
\,eV\,cm^{-3} \,.
\end{equation}
The closure mass density of order 
$2\times 10^{-29}\,g\,cm^{-3} \approx 10^{4}\,eV\,cm^{-3}$ requires
that $M \leq O[1\,GeV]$, assuming $N = 43/4$. 
Thus, WIMP in the mass range far above
$1\,GeV$ is excluded for the S-wave boson-pair annihilation model.

Extension to the annihilation in a higher angular momentum state is of
great interest, since LSP in SUSY models pair-annihilates
with a large P-wave contribution \cite{lsp}.
This P-wave annihilation for LSP is related to the Majorana nature
of LSP. 
It is beyond the scope of the present work to accurately
calculate the annihilation
rate in supersymmetric models, taking into account the Majorana
nature and all contributing Feynman diagrams.
It is however not too difficult to qualitatively estimate the P-wave
annihilation rate, by simply taking into account the momentum, hence
temperature dependence of various rates, $\langle k^{2} \rangle
\propto T$.

A more detailed behavior of the $\varphi $ number density may be
worked out by examining
\begin{eqnarray}
&&
\frac{dn}{dt} + Hn = -\,\langle \sigma _{a}v_{a} \rangle\,(\,n^{2} - 
\delta \,(\frac{T}{M})^{p + 1}\,T^{6} - n_{{\rm MB}}^{2}\,) \,, 
\\ && 
\langle \sigma _{a}v_{a} \rangle= \frac{\lambda ^{2}}{16\pi M^{2}}\,
(\frac{T}{M})^{p} \,, 
\hspace{0.5cm} 
\delta  = 0.27\times \frac{\zeta (3)\,\lambda ^{2}}{192\pi ^{5}} \,,
\\ && \hspace*{1cm} 
n_{{\rm MB}} = (\frac{MT}{2\pi })^{3/2}\,e^{-\,M/T} \,.
\end{eqnarray}
Our toy model gives the S-wave annihilation with $p = 0$, while
$p = 1$ for the P-wave annihilation.
We have extended the S-wave annihilation to the case of higher angular
momentum without changing the effective coupling constant $\delta $
relevant to our toy model.
Equivalently, using the inverse temperature and the yield,
\begin{eqnarray}
&&
\frac{dY}{dx} = -\,\frac{\eta }{x^{p + 2}}\,(\,Y^{2} - Y_{{\rm eq}}^{2}\,)  \,,
\label{y evolution} 
\\ &&
Y_{{\rm eq}}^{2} = 
\frac{\delta }{x^{p + 1}} + (\frac{x}{2\pi })^{3}\,e^{-\,2x}
\,, 
\\ && 
x = \frac{M}{T} \,, \hspace{0.5cm} 
\eta = \frac{\lambda ^{2}\,m_{{\rm pl}}}{16\pi \,dM} \,.
\end{eqnarray}
The parameter $\eta $ is roughly the (on-shell) annihilation rate
$\sigma _{a}v_{a}\,T^{3}$ divided by the Hubble rate at the temperature
equal to the particle mass, $T = M$.

We plot in Fig.6 $-$ Fig.8
a typical solution to the time evolution equation, (\ref{y evolution});
Fig.6 and Fig.7 for the S-wave annihilation and Fig.8 for the P-wave
annihilation. 

The analytic estimate of the freeze-out temperature and the 
freeze-out yield \cite{heavy particle decoupling}, which well reproduces 
the numerical estimate above, is as follows.
One may consider with a good precision that the yield follows the
equilibrium abundance $Y_{{\rm eq}}$ until the freeze-out temperature.
This temperature $T_{f}$ is given by
\begin{equation}
\frac{dY_{{\rm eq}}}{dx_{f}} = -\,\frac{\eta }{x_{f}^{p + 2}}\,
Y_{{\rm eq}}^{2}
\,, 
\end{equation}
since after this epoch the inverse process is frozon and 
the yield follows
\begin{equation}
\frac{dY}{dx} = -\,\frac{\eta }{x^{p + 2}}\,Y^{2} \,.
\end{equation}
Integration of this equation gives the final yield,
\begin{equation}
Y(x) = 
\frac{Y_{f}}{1 - Y_{f}\,\frac{\eta}{p + 1} \,(\,x^{-p-1} - x_{f}^{-p-1}\,)} 
\,, 
\end{equation}
which agrees with
\begin{equation}
Y \approx 
\frac{Y_{f}}{1 + Y_{f}\,\frac{\eta}{p + 1} \,x_{f}^{-p-1}} 
\end{equation}
as $T \rightarrow 0$.
Usually $Y_{f}$ is very small along with $x_{f}^{-1} \ll 1$, 
and in this case $Y \approx Y_{f}$ after the freeze-out.

In Fig.9 $-$ Fig.11 we show the present WIMP mass density in
the parameter space $(M\,, \lambda )$.
The off-shell dominance region is shown in Fig.9, along with the
closure mass density, while contours of smaller mass densities
are shown in Fig.10.
In Fig.11 the P-wave case is shown.
The excluded region due to the overclosure is larger for the P-wave
annihilation than for the S-wave, with the same set of $(M\,, \lambda )$.

For a given temperature $T$, there exists an upper bound on the heavy
particle mass $M_{{\rm max}}$ that can be produced in the equilibrium
abundance $n_{{\rm eq}}$. In the Boltzmann equation approach this bound
is roughly of order $T$, but our new contribution $\delta n_{{\rm eq}}$
substantially changes this value. 
There are two considerations to be taken here.
The first one is the condition on energetics;
produced energy $<$ thermal environment energy.
From 
\( \:
M\,n_{{\rm eq}} < c\,T^{4}
\: \)
with $c$ a constant of order unity,
\begin{equation}
M < M_{{\rm max}} \,, \hspace{0.5cm} 
M_{{\rm max}} \approx 2\times 10^{5}\,c^{2}\,\frac{T}{\lambda ^{2}} \,.
\end{equation}
This bound is not stringent, since 
\( \:
M_{{\rm max}}/T \approx 10^{5}/\lambda ^{2}
\: \)
can be quite large.

The second, a more important constraint comes from the relaxation
time. For the inverse process 
\( \:
\chi \chi \rightarrow \varphi \varphi 
\: \)
to occur frequently, its rate $\Gamma _{{\rm inv}}$ must be larger
than the Hubble rate $H$. With
\begin{equation}
\Gamma _{{\rm inv}} \sim \langle \sigma _{a}v \rangle\,n_{{\rm eq}}
\approx 3 \times 10^{-4}\,\lambda ^{3}\,(\frac{T}{M})^{7/2}\,M \,, 
\end{equation}
this consideration gives
\begin{equation}
M_{{\rm max}} \approx 3\times 10^{-2}\,\lambda ^{6/5}\,N^{-1/5}\,
(\frac{m_{{\rm pl}}}{T})^{2/5}\,T \,.
\end{equation}
As $T \rightarrow 0$, 
\( \:
M_{{\rm max}}  \rightarrow 0
\: \)
like $T^{3/5}$, but $M_{{\rm max}}/T$ can become very large.

\vspace{1cm}
\begin{center}
{\bf Acknowledgment}
\end{center}

This work has been supported in part by the Grand-in-Aid for Science
Research from the Ministry of Education, Science and Culture of Japan,
No. 08640341. The work of Sh. Matsumoto is partially
supported by the Japan Society of the Promotion of Science.

\newpage
\appendix
\section{
Generating functional in the influence functional method
}

\vspace{0.5cm} 
\hspace*{0.5cm} 
We explain the technique of the generating functional applied to
the influence functional, taking the example of the exactly solvable
model \cite{caldeira-leggett 83}.
When one wants to apply this method to the Hartree model of
our annihilation-scattering problem, one should use relevant
spectrum $r_{\pm }$ instead of $r(\omega )$ below.

We first
introduce the source terms coupled to both the harmonic coordinate
and the conjugate momentum as
\begin{equation}
j(\tau )q(\tau ) + l(\tau )p(\tau ) -
j'(\tau )q'(\tau ) - l'(\tau )p'(\tau )
= \frac{1}{2}\, 
(\,S_{j}\xi + D_{j}X + S_{l}p_{\xi } + D_{l}p_{X} \,) \,,
\end{equation}
where a convenient combination for the influence functional method
is introduced
\begin{equation}
S_{j}= j + j' \,, \; D_{j} = j - j'\,, \;
S_{l}= l + l' \,, \; D_{l} = l - l'\,, \;
p_{X\,,\,  \xi } = p \pm p' \,.
\end{equation}
In subsequent formulas we often omit the momentum index $\vec{k}$
for the sake of simplicity and discuss each Fourier mode separately.
Funtional differentiation of the resulting density matrix $\rho ^{(j\,, l)}$
with respect to these sources, when evaluated at the vanishing source, 
gives various combinations of correlators; for instance,
\begin{eqnarray}
&&
\langle q(\tau_{1} )q(\tau _{2}) \rangle 
=
-\,\left[\,\frac{\delta ^{2}}{\delta j(\tau _{1})\delta j(\tau _{2}) }\,
{\rm tr}\,\rho ^{(j\,, l)}\,\right]_{j = 0\,, l=0} = \\
&& 
-\,\left[ \,
\frac{\delta ^{2}}{\delta S_{j}(\tau _{1})\delta S _{j}(\tau _{2})}
+ \frac{\delta ^{2}}{\delta S_{j}(\tau _{1})\delta D_{j}(\tau _{2})}
+ \frac{\delta ^{2}}{\delta D_{j}(\tau _{1})\delta S_{j}(\tau _{2})}
\right.
\nonumber \\ &&
\hspace*{1cm} \left.
+\, \frac{\delta ^{2}}{\delta D_{j}(\tau _{1})\delta D_{j}(\tau _{2})}
\,{\rm tr}\;\rho ^{(j\,, l)}\,\right]_{j = 0 \,, l=0}
\,, 
\\ &&
\langle p(\tau_{1} )p(\tau _{2}) \rangle 
=
-\,\left[\,\frac{\delta ^{2}}{\delta l(\tau _{1})\delta l(\tau _{2}) }\,
{\rm tr}\,\rho ^{(j\,, l)}\,\right]_{j = 0\,, l=0} \,.
\end{eqnarray}

Computation of the new density matrix $\rho ^{(j\,, l)}$ 
under the action of the source is similar to
the case without the source terms, because introduction of
the source does not change the Gaussian nature of the Hartree model.
The semiclassical $\xi $ equation and the the effective action
is thus given by extending the analysis sketched in the text;
\begin{eqnarray}
\xi _{{\rm cl}}(\tau ) &=&
-\,\dot{\xi } _{f}g(t - \tau ) + \xi _{f}\dot{g}(t - \tau )
+ \int_{\tau }^{t}\,ds\,g(s - \tau )\,(\,D_{j} - \dot{D}_{l}\,)(s) \,, 
\\ && \hspace*{2cm}
J^{(j\,, l)} =
\frac{e^{iS_{{\rm cl}}^{(j\,, l)}}}{2\pi g} \,, 
\label{j function} 
\\ 
iS_{{\rm cl}}^{(j\,, l)} &=&
-\,\int_{0}^{t}\,d\tau \,\int_{0}^{\tau }\,ds\,\xi _{{\rm cl}}(\tau )
\beta _{R}(\tau - s)\xi _{{\rm cl}}(s) \nonumber 
\\
&+&
\frac{i}{2}\,\int_{0}^{t}\,d\tau \,\left( \,S_{j}(\tau )\xi _{{\rm cl}}(\tau )
+ S_{l}(\tau )\dot{\xi }_{{\rm cl}}(\tau )\,\right)
+ \frac{i}{2}\,\left( \,X_{f}\dot{\xi }_{f} - X_{i}\dot{\xi }_{i}\,\right)
\,.
\end{eqnarray}
Here the change of the variable,
$\dot{\xi }_{f} \rightarrow \xi _{i}$, is computed using
\begin{equation}
\xi _{i} = -\,\dot{\xi }_{f}g(t) + \xi _{f}\dot{g}(t) +
\int_{0}^{t}\,ds\,g(s)(D_{j} - \dot{D}_{l})(s) \,,
\end{equation}
with
\begin{eqnarray}
&&
g(t) = \frac{1}{2\pi }\,\int_{-\infty }^{\infty }\,d\omega \,
e^{-\,i\omega t}\,F(\omega + i0^{+})
\,,
\\ &&
\beta _{R}(t) + i\beta _{I}(t)
= \int_{-\infty }^{\infty }\,d\omega \,r(\omega )\,e^{-i\omega t} \,,
\\ &&
-\,F(z)^{-1} = z^{2} - \omega ^{2}(T) - 2\,
\int_{0}^{\infty }\,d\omega \,\frac{\omega \,r_{-}(\omega )}
{z^{2} - \omega ^{2}} \,.
\end{eqnarray}

For calculation of the correlator one convolutes the $J^{(j\,, l)}$ 
function (\ref{j function})
above with an initial density matrix of $\varphi $ system
$\rho _{i}(X_{i} \,, \xi _{i})$ and traces out the final $X_{f}$ and
$\xi _{f}$ variables. Thus, it is convenient to take the trace with
regard to the final variable, to get
\begin{eqnarray}
&& \hspace*{2cm}
{\rm tr}\; J^{(j\,, l)} =
\nonumber \\ && \hspace*{-1cm}
\delta \left(\, \xi _{i} - 
\int_{0}^{t}\,d\tau \,g(\,D_{j} - \dot{D}_{l}\,)\,\right)\,
\exp \left[ \,- 
\int_{0}^{t}\,d\tau \,\int_{0}^{\tau }\,ds\,
\xi _{0}(\tau )\beta _{R}(\tau - s)\xi _{0}(s) \right.
\nonumber \\ && \hspace*{-2cm} \left.
+ \frac{i}{2}\,\int_{0}^{t}\,d\tau \,\left( \,S_{j}(\tau )\xi _{0}(\tau )
+ S_{l}(\tau )\dot{\xi }_{0}(\tau ) \, \right)
+ \frac{i}{2}
X_{i}\,\int_{0}^{t}\,d\tau \,\dot{g}(\tau )(\,D_{j}(\tau ) - \dot{D}_{l}
(\tau )\,)\,\right] \,,
\\ && \hspace*{1cm} 
\xi _{0} =
\int_{\tau }^{t}\,ds\,(\,D_{j} - \dot{D}_{l}\,)(s)\,g(s - \tau ) \,.
\end{eqnarray}

As an example of the denisty matrix with the external source attached,
we may work out the case for initial thermal state of temperature
$T_{0} = 1/\beta _{0}$, to obtain
\begin{eqnarray}
&& 
{\rm tr}\,\rho ^{(j\,, l)} =
\exp \left[ \, - 
\int_{0}^{t}\,d\tau \,\int_{0}^{\tau }\,ds\,
\xi _{0}(\tau )\beta _{R}(\tau - s)\xi _{0}(s) \right. \nonumber 
\\ 
&& \hspace*{0.5cm} 
+ \,
\frac{i}{2}\int_{0}^{t}\,d\tau \,\int_{0}^{\tau }ds\,
(\,D_{j} - \dot{D}_{l}\,)(\tau )g(\tau - s)S_{j}(s)
\nonumber \\
&& \hspace*{0.5cm} 
- \,
\frac{i}{2}\int_{0}^{t}\,d\tau \,\int_{0}^{\tau }ds\,
(\,D_{j} - \dot{D}_{l}\,)(\tau )\dot{g}(\tau - s)S_{l}(s)
\nonumber \\
&& \hspace*{-1cm}
- \,
\left.
\frac{1}{4\omega _{0}}\coth (\frac{\beta _{0}\omega _{0}}{2})
\left[ \left( \omega _{0}\int_{0}^{t}ds\,g(s)
(D_{j} - \dot{D}_{l})(s)\right)^{2} + 
\left( \int_{0}^{t}ds\,\dot{g}(s)
(D_{j} - \dot{D}_{l})(s)\right)^{2}\right]\,\right] \,.
\nonumber \\ &&
\end{eqnarray}

General solution for arbitrary initial uncorrelated states
can be derived with the aid of a conjugate Wigner transform;
we define with the initial system density matrix
\( \:
\rho _{i}(X = q + q' \,, \xi = q - q') \,, 
\: \)
\begin{eqnarray}
f_{p\xi } (p \,, \xi ) = \int_{-\infty }^{\infty }\,dx\,
e^{ipX/2}\,\rho _{i}(X \,, \xi ) \,.
\end{eqnarray}
Tracing out the final variables leads to
\begin{eqnarray}
&&
{\rm tr}\;\rho ^{(j \,, l)} = \frac{1}{2}\, 
\int_{-\infty }^{\infty }\,dX\,\int_{-\infty }^{\infty }\,d\xi \,
\rho _{i}(x\,, \xi )\,{\rm tr}\;J^{(j\,, l)} =
\\
&&
\hspace*{-1cm}
\exp \left[ \,- 
\int_{0}^{t}\,d\tau \,\int_{0}^{\tau }\,ds\,
\xi _{0}(\tau )\beta _{R}(\tau - s)\xi _{0}(s) 
+ \frac{i}{2}\,\int_{0}^{t}\,d\tau \,\left( \,S_{j}(\tau )\xi _{0}(\tau )
+ S_{l}(\tau )\dot{\xi }_{0}(\tau ) \, \right)\,\right] \nonumber 
\\
&&
\cdot f_{p \xi }\left( \,
\int_{0}^{t}\,d\tau \,\dot{g}(\tau )(\,D_{j}(\tau ) - \dot{D}_{l}
(\tau )\,) \,, 
\int_{0}^{t}\,d\tau \,g(\tau )(\,D_{j}(\tau ) - \dot{D}_{l}
(\tau )\,)\,\right) \,.
\end{eqnarray}

Functional differentiation with respect to $j(\tau ) \,, l(\tau )$
gives the correlator such as
\begin{eqnarray}
&&
\langle q(\tau _{1}) q(\tau _{2})\rangle =
-\,\frac{i}{2}\,g(\tau _{1} - \tau _{2}) 
+ \int_{0}^{\tau _{1}}\,d\tau \,\int_{0}^{\tau _{2}}\,ds\,
g(\tau _{1} - \tau )\beta _{R}(\tau - s)g(\tau _{2} - s) 
\nonumber \\
&& \hspace*{1cm} 
-\, g(\tau _{1})g(\tau _{2})\,
\left(\,\frac{\partial  ^{2}f_{p\xi }}{\partial  \xi ^{2}}\,\right)_{00}
- \dot{g}(\tau _{1})\dot{g}(\tau _{2})\,
\left(\,\frac{\partial ^{2}f_{p\xi }}{\partial p^{2}}\,\right)_{00}
\nonumber 
\\ &&
\hspace*{2cm} 
-\, \left( \,g(\tau _{1})\dot{g}(\tau _{2}) + \dot{g}(\tau _{1})g(\tau _{2})
\,\right)
\left(\,\frac{\partial ^{2}f_{p\xi }}
{\partial  \xi \partial  p}\,\right)_{00} \,.
\label{qq correlator at different t} 
\end{eqnarray}
Here the suffix 00 is understood to mean
\( \:
p = 0 \,, \xi = 0 \,.
\: \)
Needless to say, the derivatives are related to the averages of
dynamical variables;
\begin{eqnarray}
&& \hspace*{-0.5cm}
\left(\,\frac{\partial  ^{2}f_{p\xi }}{\partial  \xi ^{2}}\,\right)_{00}
= -\,\overline{p_{i}^{2}} \,, \hspace{0.5cm} 
\left(\,\frac{\partial ^{2}f_{p\xi }}{\partial p^{2}}\,\right)_{00}
= -\,\overline{q_{i}^{2}}  \,, \hspace{0.5cm} 
\left(\,\frac{\partial ^{2}f_{p\xi }}
{\partial  \xi \partial  p}\,\right)_{00}
= -\,\frac{1}{2}\, \overline{p_{i}q_{i} + q_{i}p_{i}} \,.
\nonumber \\ &&
\end{eqnarray}
We assumed in deriving this formula that
\begin{equation}
\left( \frac{\partial f_{p\xi }}{\partial \xi }\right)_{00}
= 0 \,, \hspace{0.5cm} 
\left( \frac{\partial f_{p\xi }}{\partial p }\right)_{00} = 0 \,.
\end{equation}

The initial memory effect appears in two ways for the exact result
of the correlator $\langle q(\tau _{1})q(\tau _{2}) \rangle$;
first, via the initial state dependence, $f_{p\xi }$, and
secondly, via the explicit lower limit of the time integration,
$\tau = 0$ taken to be the initial time.
The memory effect thus violates the time translation invariance
under $\tau _{i} \rightarrow \tau _{i} + \delta $.

The coincident limit of the correlator is computed, using
\( \:
g(0) = 0 \,, 
\: \)
and the relation,
\begin{equation}
\int_{0}^{t}\,d\tau \,\int_{0}^{t}\,ds\,
g(t - \tau )\,\beta _{R}(\tau - s)\,g(t - s) =
\int_{0}^{\infty }\,d\omega \,r_{+}(\omega )\,|h(\omega \,,t)|^{2}
\,.
\end{equation}
The result is eq.(\ref{correlator q^2}) and similar ones for
the other quantities, eq.(\ref{correlator p^2}), (\ref{correlator qp})
in the text.

We finally give some other examples of 
the conjugate Wigner function $f_{p\xi }$;
\begin{enumerate}
\item 
thermal state of temperature $T_{0} = 1/\beta_{0} $
\begin{equation}
f_{p\xi }(p\,, \xi ) = \exp [\,- \frac{1}{4}\coth (\frac{\beta _{0}\omega _{0}}
{2})\,(\,\frac{p^{2}}{\omega _{0}} + \omega _{0}\xi ^{2}\,)\,] \,, 
\end{equation}
\item 
moving packet of momentum $p_{0}$, with a spread
$\Delta p \approx  \sqrt{\alpha }$
\begin{eqnarray}
f_{p\xi }(p\,, \xi ) = \exp [\,- \frac{1}{4}\,(\,\frac{p^{2}}{\alpha }
+ \alpha \xi ^{2}\,) + ip_{0}\xi ] \,.
\end{eqnarray}
\end{enumerate}
In these cases the correlators such as 
(\ref{qq correlator at different t}) can be worked out explicitly.

\vspace{0.5cm} 
\section{
Renormalization of distribution function
}

\vspace{0.5cm} 
\hspace*{0.5cm} 
One may use the equal time limit of the full propagator for
the purpose of renormalization; in particular,
\begin{eqnarray}
&&
\lim_{x_{0} \rightarrow y_{0}^{+}}\,
\left( \frac{\partial }{\partial x_{0}} - 
\frac{\partial }{\partial y_{0}}\right)\,
\tilde{G}(x_{0} \,, y_{0} \,; \vec{k}) = 
Z  \,,
\label{z sum rule} 
\\ &&
\lim_{x_{0} \rightarrow y_{0}^{+}}\,\frac{i}{4}\,
\left( \frac{\partial }{\partial x_{0}} - 
\frac{\partial }{\partial y_{0}}\right)^{2}\,
\tilde{G}(x_{0} \,, y_{0} \,; \vec{k}) = 
\langle {\cal H}(\vec{k} \,, t) \rangle
\,, \label{hamiltonian density} 
\end{eqnarray}
where $Z$ is the wave function renormalization factor and
${\cal H}(\vec{k} \,, t) $ is the Hamiltonian for the momentum $\vec{k}$ mode.
Using the expansion (\ref{expansion of propagator}), we find 
\begin{eqnarray}
&& \hspace*{-1cm}
\lim_{x_{0} \rightarrow y_{0}^{+}}\,
\left( \frac{\partial }{\partial x_{0}} - 
\frac{\partial }{\partial y_{0}}\right)\,G(x\,, y)
= \int\,\frac{d^{3}k}{(2\pi )^{3}}\,e^{-i\vec{k}\cdot (\vec{x} - \vec{y})}
\,\left( 1 + f(\vec{k} \,, t) - f(\vec{k} \,, t)\right)
\nonumber 
\\ &&
\hspace*{1cm} 
=\, \delta ^{3}(\vec{x} - \vec{y})
\,,
\end{eqnarray}
in contradiction to eq.(\ref{z sum rule}).
Thus, the original expansion (\ref{expansion of propagator})
must be modifid to allow a counter term in
\( \:
-\,iG(x\,, y) \,.
\: \)
The term independent of $f$ should thus be replaced as
\begin{eqnarray}
&&
e^{i\omega _{k}(x_{0} - y_{0})}\,f(\vec{k} \,, t) +
e^{-i\omega _{k}(x_{0} - y_{0})}\,(\,1 + f\,(\vec{k} \,, t))
\: \rightarrow  \:
\nonumber \\ && \hspace*{-0.5cm}
e^{i\omega _{k}(x_{0} - y_{0})}\,\left( \frac{1 - Z}{2} + f(\vec{k} \,, t)
\right) +
e^{-i\omega _{k}(x_{0} - y_{0})}\,\left( \frac{1 + Z}{2} + f\,(\vec{k} \,, t)
\right)
\,. \label{wave function renorm} 
\end{eqnarray}

Another useful relation for the renormalization is the Fourier inversion
formula,
\begin{eqnarray}
&&
\hspace*{0.5cm} 
\frac{1}{2\omega _{k}}\,\left( 1 + 2f(\vec{k} \,, x_{0}) -
2v(\vec{k} \,, x_{0}) \right)
\nonumber \\ && \hspace*{-0.5cm}
= \,\lim_{x_{0} \rightarrow y_{0}^{+}}\,\int\,d^{3}x\,
e^{i\vec{k}\cdot \vec{x}}\,\langle \varphi (\vec{x} \,, x_{0})
\varphi (\vec{0} \,, y_{0}) \rangle = -\,i\,
\int_{-\infty }^{\infty }\,\frac{dk_{0}}{2\pi }\,\tilde{G}(k_{0} \,, \vec{k}
\,; x_{0}) \,. \label{inverted f formula} 
\end{eqnarray}
The correlator 
\( \:
G(x\,, y) = i\langle \varphi (x)\varphi (y) \rangle
\: \)
in thermal equilibrium
is what is called the real-time thermal Green's function in the 
literature \cite{fetter-walecka}, and its Fourier transform
$\tilde{G}$ here is related to the analytic function 
(\ref{analytic self-energy f}) by
\begin{equation}
-\,\tilde{G}(k_{0} \,, \vec{k}) = 
\frac{1}{1 - e^{-\beta k_{0}}}\,F(k_{0} + i0^{+} \,, \vec{k}) + 
\frac{1}{1 - e^{\beta k_{0}}}\,F(k_{0} - i0^{+} \,, \vec{k}) \,.
\end{equation}
This relation holds when $\varphi $ particles are in thermal
equilibrium with $\chi $. We shall first derive the renormalization
condition assuming thermal equilibrium, and then
extend its result to the non-equilibrium circumstance.

Subtraction term for the distribution function consists of an infinite
quantity and its associated finite term.
In the spirit of the on-shell renormalization in field theory
it is important to identify the pole contribution with possible
infinities of the wave function and the mass correction included.
The pole part of the propagator is extracted from the terms of
$F^{-1}(k_{0} \,, \vec{k})$ to the quadratic order in $k_{0}$;
prior to the renormalization,
\begin{eqnarray}
&& \hspace*{-1cm}
i\,F_{{\rm pole}}(k_{0} \pm i0^{+} \,, \vec{k}) = 
\frac{-\,i}{k_{0}^{2} - \vec{k}^{2} - M^{2}(T) - 
\Pi (k_{0} \,, \vec{k}) \pm i\pi r_{-}(\omega _{k} \,, \vec{k})\,
\epsilon (k_{0})}
 \,,
\\ &&
\Pi (\omega \,, \vec{k}) = -\,{\cal P}\,\int_{-\infty }^{\infty }\,d\omega' \,
\frac{r_{-}(\omega '\,, \vec{k})}{\omega ' - \omega } =
-\,{\cal P}\,\int_{0}^{\infty }\,d\omega' \,
\frac{2\omega '\,r_{-}(\omega ' \,, \vec{k})}{\omega '\,^{2} - \omega ^{2}}
\,, 
\end{eqnarray}
with
\( \:
M^{2}(T) = M^{2} + \frac{\lambda }{12}T^{2}
\: \)
including the $O[\lambda ]$ temperature dependent mass.

The mass and the wave function renormalization is done perturbatively
for the proper self-energy $\Pi (k_{0} \,, \vec{k})$.
Expanding in powers of $k_{0}^{2} - M^{2} $ and $ \vec{k}^{2}$, 
one identifies the renormalized temperature dependent mass as
\begin{equation}
M_{R}^{2}(T) = M^{2}(T) + \Pi (M\,, \vec{0}) \,, 
\end{equation}
and the wave function renormalization factor as
\begin{equation}
Z^{-1} = 1 - \frac{1}{2}\,\left( \frac{\partial^{2} \Pi (k_{0} \,, \vec{0})}
{\partial k_{0}^{2}}\right)_{k_{0} = M} \,.
\end{equation}
Thus, defining the subtracted finite part $\delta \Pi $ by
\begin{eqnarray}
&&
\delta \Pi (k_{0} \,, \vec{k}) = \Pi (k_{0} \,, \vec{k})
-  \Pi (M \,, \vec{0}) + (\,Z^{-1} - 1\,)\,(\,k_{0}^{2} - M^{2}\,) \,,
\label{subtracted finite self-energy} 
\end{eqnarray}
one has
\begin{eqnarray}
&&
i\,F_{{\rm pole}}(k_{0} \pm i0^{+} \,, \vec{k}) = 
\frac{-\,iZ}{k_{0}^{2} - (\omega _{k}^{R})^{2} 
\pm i\pi r_{-}(\omega _{k} \,, \vec{k})\,\epsilon (k_{0})}
\,,
\\ && \hspace*{1cm} 
\omega _{k}^{R} = \sqrt{\, \vec{k}^{2} + M_{R}^{2}(T) +
\delta \Pi (\omega _{k} \,, \vec{k})\,} \,.
\label{renormalized mass at finite t 2} 
\end{eqnarray}
This $F_{{\rm pole}}^{-1}(k_{0} \,, \vec{k})$ is an optimal Gaussian
approximation for the low energy dynamics, to the order $O[\lambda ^{2}]$.
One may regard $\delta \Pi (\omega _{k} \,, \vec{k})$ as a finite
energy shift due to the interaction with environment.
When one uses 
\( \:
\omega _{k} = \sqrt{\vec{k}^{2} + M_{R}^{2}} \,, 
\: \)
for the energy at the pole,
as is done in the on-shell Boltzmann equation,
one has to compensate for the difference 
\begin{equation}
\omega _{k}^{R} - \omega _{k}
\approx \frac{\delta \Pi (\omega _{k} \,, \vec{k})}{2\omega _{k}}
\,, \hspace{0.5cm} 
\omega _{k} = \sqrt{\vec{k}^{2} + M_{R}^{2}} \,, 
\end{equation}
in the distribution function as a kind of finite energy renormalization.

Relativistic covariance requires 
the equality of the temperature independent part;
\begin{equation}
Z^{-1} - 1 = \frac{1}{2}\, \frac{\partial ^{2} \Pi (M \,, \vec{k})}
{\partial k_{i}^{2}} \,,
\label{covariance requirement for infinity} 
\end{equation}
with $k_{i}$ a spatial component of $\vec{k}$.
We shall check in Appendix D that the infinite part of this relation
holds.
The finite, temperature dependent part however needs not to satisfy
the covariance relation,
\begin{equation}
\frac{\partial ^{2} \Pi (k_{0} \,, \vec{k})}{\partial k_{0}^{2}}
= -\,\frac{\partial ^{2} \Pi (k_{0} \,, \vec{k})}{\partial k_{i}^{2}}
\,,
\end{equation}
due to the presence of the preferred frame in which the temperature
is uniquely defined.

One may then compute (\ref{inverted f formula}) in the narrow width
limit (or in the weak coupling limit).
With $v = 0$, the renormalized pole term $f_{{\rm pole}}^{R}$
is given by
\begin{eqnarray}
&&
\frac{1}{2\omega _{k}}\,\left( 1 + 2f_{{\rm pole}}(\omega _{k})\right)
= \frac{Z}{2\omega _{k}^{R}}\,\left( 1 + 2f_{{\rm pole}}^{R}(\omega _{k}^{R})
\right) \,. 
\end{eqnarray}
We thus find that
\begin{equation}
f_{{\rm pole}}^{R}(\omega_{k}^{R}) =
\frac{\omega _{k}^{R} - Z\omega _{k}}{2Z\omega _{k}}
+ \frac{\omega _{k}^{R}}{Z\omega _{k}}\,f_{{\rm pole}}(\omega _{k}) \,, 
\end{equation}
from which to $O[\lambda ^{2}]$
\begin{eqnarray}
&& \hspace*{1cm} 
\delta f_{{\rm ren}}(\vec{k}) \equiv 
f_{{\rm pole}}^{R}(\omega _{k}^{R}) - f_{{\rm pole}}(\omega _{k}^{R})
\nonumber 
\\ &&
=\, 
\left( -\,\delta Z + \frac{
\delta \Pi (\omega _{k} \,, \vec{k})}{2\omega _{k}^{2}}\right)\,
\left( \,\frac{1}{2} + f_{{\rm pole}} \,\right) -
\frac{
\delta \Pi (\omega _{k} \,, \vec{k})}{2\omega _{k}}
\,\frac{df_{{\rm pole}}}{d\omega _{k}} \,, 
\end{eqnarray}
with $\delta Z = Z - 1$.
Since the occupation number $f(\vec{k})$ is defined 
in reference to the energy of
the harmonic oscillator of mode $\vec{k}$,
\( \:
\langle\, p_{k}^{2}/(2\omega _{k}) + \omega _{k}q_{k}^{2}/2 \,\rangle
\: \)
the finite part of the proper self-energy $\delta \Pi (\omega _{k}
\,, \vec{k})$ appears in this formula as a change of the reference
energy
\( \:
\omega _{k} \rightarrow \omega _{k}^{R} \,.
\: \)

In the low temperature limit the distribution function $f_{{\rm pole}}$
is very small, and one approximately has 
\begin{equation}
\delta f_{{\rm ren}}(\vec{k}) \approx \frac{1}{2}\, 
\left( -\,\delta Z + \frac{\delta \Pi (\omega _{k} \,, \vec{k})}
{2\omega _{k}^{2}} \right) \,.
\end{equation}

We extend this renormalization term to the case of the non-equilibrium
state, to get
\begin{equation}
\delta f_{{\rm ren}}(\vec{k}) = \frac{1}{2}\,\left( \,-\,\delta Z +
\frac{\delta  \Pi(\omega _{k} \,, \vec{k})}{2\omega _{k}^{2}}\,\right)\,
\,,
\end{equation}
to be used at low temperatures.
The renormalized distribution function is then
\begin{equation}
f_{{\rm ren}}(\vec{k}) = f(\vec{k}) + \delta f_{{\rm ren}}(\vec{k})
\approx f(\vec{k}) - \frac{\delta Z}{2} 
+ \frac{\delta \Pi (\omega _{k}\,, \vec{k})}{4\omega _{k}^{2}} \,.
\end{equation}

The last term of this formula is related to the finite energy shift
of the harmonic oscillator for the mode $\vec{k}$.
Its effect is absorbed by changing the reference energy when one
defines the occupation number for this mode. 
In another word, 
this last term disappears
by the replacement,
\( \:
\omega _{k} \rightarrow \omega _{k}^{R} \,,
\: \)
with $\omega _{k}^{R}$ given by (\ref{renormalized mass at finite t 2}).
Thus, if one modifies the occupation number according to
\begin{eqnarray}
&&
f_{{\rm new}}(\vec{k}) \equiv \langle\, \frac{p_{k}^{2}}{2\omega _{k}^{R}} + 
\frac{\omega _{k}^{R}}{2}\,q_{k}^{2} \,\rangle - \frac{1}{2} \,, 
\label{new def of occupation n} 
\end{eqnarray}
then the renormalized occupation number becomes
\begin{equation}
f_{{\rm ren}}(\vec{k}) = f(\vec{k}) - 
\frac{A + B\,\vec{k}^{2}}{4\omega _{k}^{2}} \,, \hspace{0.5cm} 
B = \delta Z \,.
\end{equation}
Here $A$ and $B$ are infinite counter terms independent of the
momentum $\vec{k}$.

Thie new definition appears very reasonable, because
the $\varphi $ system in isolation from the $\chi $ environment
is an ideal setting which has nothing to do with actual
observation.
The new reference energy $\omega _{k}^{R}$ includes the interaction
with the environment, and in this sense is directly related to
observation.
We studied the problem more closely
than this, by retaining the energy shift term
$\delta \Pi(\omega _{k} \,, \vec{k})/4\omega _{k}^{2} $.
We found a disease with this term;
at very low temperatures the equilibrium distribution does not
exist.
This happens by having a dominantly negative term for the
stationary distribution $f_{{\rm eq}}$.
Since this is clearly unacceptable, 
we shall use the definition (\ref{new def of occupation n})
and its associated renormalization of the proper self-energy to
define our occupation number.

\vspace{0.5cm} 
\section{
Kinetic equation for unstable particle decay
}

\vspace{0.5cm} 
\hspace*{0.5cm} 
We consider two-body decay of a boson $\varphi $ described by a Lagrangian
density,
\begin{equation}
{\cal L}_{{\rm int}} = -\,\frac{\mu }{2}\,\varphi \chi ^{2} \,, 
\end{equation}
where $\mu $ is a coupling constant of mass dimension.
The environment particle $\chi $ is taken to make up a thermal
bath of temperature $T = 1/\beta $.
For simplicity, we assume that the mass of $\chi $ particle vanishes.

The fundamental quantity for the quantum kinetic equation
of the unstable particle decay is the spectral weight given by
\begin{eqnarray}
&&
r_{\chi }(k) = \frac{\mu ^{2}}{16\pi ^{2}}\,\epsilon (k_{0})\,
\left( \,\frac{1}{2}\, \theta (|k_{0}| - k) + \frac{1}{\beta k}\,
\ln \frac{1 - e^{-\beta \omega _{+}}}{1 - e^{-\beta |\omega _{-}|}}\,\right)
\,,
\\ && \hspace*{1cm} 
\omega _{\pm } = \frac{1}{2}\, (|k_{0}| \pm k) \,,
\end{eqnarray}
where $k = |\vec{k}|$.
This spectral function corresponds to $\varphi \leftrightarrow \chi \chi $
for $k _{0} > k$ and $\varphi \chi \leftrightarrow \chi $ for
$0 < k_{0} < k$, as depicted in Fig.12.
Note that we do not assume the on-shell kinematic condition
\( \:
k_{0}^{2} - k^{2} = M^{2} \,, 
\: \)
thus the process $\varphi \chi \leftrightarrow \chi $ becomes possible.
For $k_{0} < 0$ 
the spectral weight is extended according to
\( \:
r_{\chi }(-\,k) = -\,r_{\chi }(k) \,.
\: \)
The Feynman-Vernon kernel 
\begin{eqnarray}
&&
\alpha (x - y) = (\frac{\mu }{2})^{2}\,
{\rm tr}\;\left( \,T[\tilde{\chi }^{2}(x)\,\tilde{\chi }^{2}(y)\,
\rho _{\beta }]\,\right)
\end{eqnarray}
given by 
eq.(\ref{fv kernel for 2-body}) contributes to the influence functional
of this problem;
\begin{equation}
{\cal F}[\varphi \,, \varphi '] =
\exp [\,-\,\int_{x_{0} > y_{0}}\,dx\,dy\,(\,\varphi (x) - \varphi'(x)\,)
\,\left( \,\alpha (x - y)\varphi (y) - \alpha ^{*}(x - y)\varphi'(y)
\,\right)\,] \,.
\end{equation}
Neglect of higher powers of $\varphi $ with the kernel of the type,
\( \:
\langle \tilde{\chi }^{2}\tilde{\chi }^{2} \cdots \tilde{\chi }^{2} \rangle
\,, 
\: \)
defines the Hartree approximation in this decay model.

The Markovian kinetic equation can be worked out in the same way
as in the annihilation-scattering problem in the text, and
it is
\begin{eqnarray}
&& \hspace*{1cm}
\frac{df(\vec{k} \,, t)}{dt} = -\,\Gamma_{k}
\,\left(\, f(\vec{k} \,, t) - f_{{\rm eq}}(\vec{k}) \,\right) \,, 
\\ && 
f_{{\rm eq}}(\vec{k}) = \int_{0}^{\infty }\,d\omega \,
(\frac{\omega _{k}}{2} + \frac{\omega ^{2}}{2\omega _{k}})\,
\coth \frac{\beta \omega }{2}\,\frac{r_{\chi }(\omega \,, \vec{k})}
{(\omega ^{2} - \omega _{k}^{2})^{2} + 
(\pi r_{\chi }(\omega \,, \vec{k}))^{2}} 
\nonumber \\ &&
\hspace*{2cm} 
-\, (\,T = 0 \; {\rm contribution}\,)
\label{energy integral for decay f} 
\\ && \hspace*{0.5cm} 
\approx 
\frac{1}{2\omega _{k}}\,\int_{0}^{\infty }\,d\omega \,
\frac{r_{\chi }(\omega \,, \vec{k})}{e^{\beta \omega } - 1}\,
\left( \,\frac{1}{(\omega - \omega _{k})^{2} + \frac{\Gamma _{k}^{2}}{4}}
+ \frac{1}{(\omega + \omega _{k})^{2} + \frac{\Gamma _{k}^{2}}{4}}\,\right)
\,.
\end{eqnarray}
Here we ignored a minor correction, the temperature dependence of mass,
and took an advantage of the weak coupling limit in writing the last form
of this equation.
The rate $\Gamma _{k} = \pi \,r_{\chi }(\omega _{k} \,, \vec{k})/\omega _{k}$
is the decay rate of unstable $\varphi $ particle with time dilatation
effect included.

As demonstrated in \cite{jmy-97}, the equilibrium distribution
function can be interpreted by a Gibbs formula,
\begin{equation}
f_{{\rm eq}}(\vec{k}) = \frac{{\rm tr}\;
a_{k}^{\dag }a_{k}\,e^{-\beta H_{{\rm tot}}}}
{{\rm tr}\;e^{-\beta H_{{\rm tot}}}} \,, 
\end{equation}
where $H_{{\rm tot}}$ is the total Hamiltonian including interaction
between the $\varphi $ system and the $\chi $ environment.

Let us work out the stationary distribution $f_{{\rm eq}}(\vec{k})$
in more detail.
The narrow width approximation to the energy integral 
(\ref{energy integral for decay f}) gives the
the usual Planck distribution, and the rest of contribution 
at low temperatures is approximately given as
\begin{eqnarray}
&& \hspace*{1cm} 
f_{{\rm eq}}(\vec{k}) \approx \frac{1}{e^{\beta \omega _{k}} - 1}
+ \delta f_{{\rm eq}}(\vec{k})
\label{occupation-n for decay 1} 
\,,
\\ && \hspace*{1cm} 
\delta f_{{\rm eq}}(\vec{k}) \approx 
\frac{1}{\omega _{k}^{3}}\,\int_{0}^{\infty }\,d\omega \,
\frac{r_{\chi }(\omega \,, \vec{k})}{e^{\beta \omega } - 1}
\nonumber 
\\ &&
= \frac{\mu ^{2}}{16\pi ^{2}\,\omega _{k}^{3}}\,\int_{0}^{\infty }\,d\omega \,
\frac{1}{e^{\beta \omega } - 1}\,
\left( \,\frac{1}{2}\, \theta (\omega - k) +
\frac{1}{\beta k}\,\ln \frac{1 - e^{-\beta (\omega + k)/2}}
{1 - e^{-\beta |\omega - k|/2}} \,\right) \,.
\end{eqnarray}
The last $\omega $ integral can be worked out analytically both in the
limit of $k\ll T$ and $k \gg T$. For $k \ll T$ the dominant part of the 
integral extends both from the region $\omega < k$ and to $\omega > k$, 
while for $k \gg T$ it comes only from $\omega < k$,
giving
\begin{equation}
\frac{2T}{e^{k/2T} - 1} \,, \hspace{0.5cm} 
\frac{\zeta (2)\,T^{2}}{k}\,\frac{1}{e^{k/2T} - 1} \,, 
\end{equation}
respectively.
Here $\zeta (2) = \frac{\pi ^{2}}{6}$ is the Riemann's zeta function.
Our interpolating formula is a smooth match of these two limiting
functions,
\begin{equation}
\delta f_{{\rm eq}}(\vec{k}) = \frac{\mu ^{2}}{16\pi ^{2}\,\omega _{k}^{3}}\,
\frac{2\zeta (2)\,T^{2}}{2k + \zeta (2)\,T}\,\frac{1}{e^{k/2T} - 1} \,.
\label{low-T interpolation of f} 
\end{equation}

We numerically compare this interpolating formula with the result of
exact integration at $T = M/20$ in Fig.13, where we plot
the quantity $k^{2}f_{{\rm eq}}(k)/(2\pi ^{2})$ using 
(\ref{energy integral for decay f}).
The agreement of our approximate formula $f_{{\rm eq}}(\vec{k})$ 
(\ref{low-T interpolation of f}) and
the exact numerical integration is not excellent, but
is adequate for evaluation of the total number density.
Thus, at low temperatures the stationary distribution for the unstable
particle is approximately given by our 
$\delta f_{{\rm eq}}(k)$, eq.(\ref{low-T interpolation of f}).

Integration over the phase space gives the number density
at low temperatures of $T \ll M$ (parent mass);
\begin{eqnarray}
&&
\delta n_{{\rm eq}} = \frac{1}{2\pi ^{2}}\,\int_{0}^{\infty }\,dk\,
k^{2}\delta f_{{\rm eq}}(k) = \frac{\zeta (2)\,\mu ^{2}}
{2\pi ^{4}\,M^{3}}\,c\,T^{4} \,,
\\ &&
c = \int_{0}^{\infty }\,dx\,\frac{x^{2}}{4x + \zeta (2)}\,\frac{1}{e^{x} - 1}
\approx 0.27 \,. 
\label{decay ph integral} 
\end{eqnarray}
Thus, 
\begin{equation}
n_{{\rm eq}}(T) \approx (\frac{MT}{2\pi })^{3/2}\,e^{-\,M/T} +
0.23\,\Gamma (\frac{T}{M})^{2}\,T^{2} \,,
\end{equation}
where $\Gamma = \mu ^{2}/(32\pi M)$ is the decay rate for 
$\varphi \rightarrow \chi \chi $ in the $\varphi $ rest frame.
See Fig.14 for the number density.
Emergence of the temperature power term has been found recently 
\cite{jmy-98-1}, \cite{jmy-97},
and its relevance to the GUT baryogenesis has been discussed in 
\cite{jmy-98-2}.

\vspace{0.5cm} 
\section{
Computation of off-shell distribution function
}

\vspace{0.5cm} 
\hspace*{0.5cm} 
We separate both $\delta \tilde{f}_{{\rm eq}}$ 
into three pieces and write
\begin{eqnarray}
&& \hspace*{0.5cm}
\delta \tilde{f}_{{\rm eq}} + \delta f_{{\rm ren}} = f_{1} + f_{2} + f_{3}
- \frac{A + B\,\vec{k}^{2}}{4\omega _{k}^{2}} \,, 
\\ && 
f_{i}(\vec{k}) = \frac{1}{2\omega _{k}}\,\int_{-\infty }^{\infty }\,d\omega \,
\frac{r_{i}(\omega \,, \vec{k}) - r_{i}(\omega _{k} \,, \vec{k})
}
{(\omega - \omega _{k})^{2} + \Gamma _{k}^{2}/4} \,, 
\hspace{0.5cm} (i = 1 \,, 2) \,,
\\ && 
r_{1}(\omega \,, \vec{k}) = 
2\,\int\,\frac{d^{3}k'}{(2\pi )^{3}2\omega _{k'}}\,
\frac{r_{\chi }(|\omega - \omega _{k'}| \,, \vec{k} - \vec{k'})}
{e^{\beta |\omega - \omega _{k'}|} - 1}\,
f(\vec{k'} )
\,, 
\\ && 
r_{2}(\omega \,, \vec{k}) = 
2\,\int\,\frac{d^{3}k'}{(2\pi )^{3}2\omega _{k'}}\,
\frac{r_{\chi }(|\omega + \omega _{k'}| \,, \vec{k} + \vec{k'})}
{e^{\beta |\omega + \omega _{k'}|} - 1}\,
(\,f(\vec{k'}) + 1\,) \,.
\end{eqnarray}
We shall give the explicit form of $f_{3}(\vec{k})$ later, which
has the temperature dependence only via the two-body spectral
function $r_{\chi }$.
For each piece we shall give an integrated number density
to show their respective importance at low temperatures.

The first piece $f_{1}$ has the main contribution
from the region, $|\omega - \omega _{k'}| \leq T$, due to the exponential
suppression outside this region. One may use the expansion formula,
\begin{equation}
\frac{r_{\chi }(|\omega - \omega _{k'}| \,, \vec{k} - \vec{k'})}
{e^{\beta |\omega - \omega _{k'}|} - 1} \approx 
T\,\left( \frac{dr_{\chi }(x \,, \vec{k} - \vec{k'})}{dx}
\right)_{x = 0} \,, 
\end{equation}
which gives the corresponding spectral function,
\begin{eqnarray}
&&
r_{1}(-\,\omega \,, \vec{k}) \approx 
\frac{\lambda ^{2}T}{4\pi ^{2}}\,\int\,
\frac{d^{3}k'}{(2\pi )^{3}2\omega _{k'}}\,\frac{\theta (|\vec{k} - \vec{k'}|
- |\omega - \omega _{k'}|)}{e^{\beta |\vec{k} - \vec{k'}|/2} - 1}\,
\frac{f(\vec{k'})}{|\vec{k} - \vec{k'}|}  \,.
\end{eqnarray}
This contributes to the stationary distribution function off the mass shell;
\begin{eqnarray}
&& \hspace*{-1cm}
f_{1}(\vec{k}) \approx 
\frac{\lambda ^{2}\,T^{2}}{32\pi ^{4}\,k\omega _{k}}\,
\int_{0}^{\infty }\,dk'\,\frac{k'}{\omega _{k'}}\,f(\vec{k'})
\,\int_{-\infty }^{\infty }\,d\omega \,
\frac{D(\omega \,, k\,, k') - D(\omega _{k} \,, k \,, k')}
{(\omega - \omega _{k})^{2} + \Gamma ^{2}/ 4} 
\,, \label{dominant off-shell d-function 1} 
\\ && \hspace*{1cm} 
D(\omega \,, k \,, k') = \theta (|k - k'| - |\omega - \omega _{k'}|)\,
\ln \frac{1 - e^{-\beta (k + k')/2}}{1 - e^{-\beta |k - k'|/2}}
\nonumber \\ && \hspace*{1.5cm} 
+\, \theta (|\omega - \omega _{k'}| - |k - k'|)\,
\ln \frac{1 - e^{-\beta (k + k')/2}}{1 - e^{-\beta |\omega - \omega _{k'}|/2}}
\,.
\end{eqnarray}
Since the $k'$ integral is dominated in the region
around $k$ of width of order $T$, the integral roughly gives
\begin{equation}
f_{1} (\vec{k}) \approx 
O[1]\times 
\frac{\lambda ^{2}\,T^{2}}{32\pi ^{4}\,\omega _{k}^{2}}\,
f(\vec{k})  \,.
\end{equation}
Integrated over momenta, this gives the number density
\begin{eqnarray}
&&
n_{1}^{f} \approx O[1]\times 
\frac{\lambda ^{2}\,T^{2}}{32\pi ^{4}\,\omega _{k}^{2}}\,n_{\varphi } \,, 
\label{inverse scat to n} 
\end{eqnarray}
with $n_{\varphi }$ the $\varphi $ number density.
One may interpret this result, by saying that a $\varphi $ fraction
of order $\lambda ^{2}T^{2}/M^{2}$ is created by scattering off
the mass shell.

Since the magnitude of the first term $n_{1}^{f}$ is important to
estimate the relic abundance, we performed
direct numerical computation for this term 
without resorting to the above approximation.
The number density due to this term is
\begin{eqnarray}
&&
\delta n_{1}^{f} = \frac{\lambda ^{2}\,T^{2}}{128\pi ^{6}}\,
\int_{0}^{\infty }\,\frac{k'\,dk'}{\omega _{k'}}\,f(k')\,
C(k')  \,,
\\ && \hspace*{-1cm}
C(k') = \int_{0}^{\infty }\,dx\,\int_{\omega _{k' - x} - \omega _{k'}}
^{\omega _{k' + x} - \omega _{k'}}\,d\alpha \,
\int_{0}^{\infty }\,d\omega \,
\frac{A(\omega + \alpha ) + A(-\,\omega + \alpha ) - 2A(\alpha )}
{\omega ^{2}} \,, 
\\ &&
A(\omega ) = \frac{1}{e^{\omega } - 1}\,\ln \frac{1 - e^{-|\omega + x|/2}}
{1 - e^{-|\omega - x|/2}} \,, 
\end{eqnarray}
in an energy unit measured by the temperature $T$.
The narrow width limit $\Gamma \rightarrow 0$ can trivially
be taken in this form
of the integral. We numerically confirmed that $C(k')$ is alomost
in proportion to $k'$ such that the last $k'$ integral approximately
gives the $\varphi $ number density $n_{\varphi }$, and moreover 
the constant factor is roughly 
\( \:
5.5 \times (T/M)^{0.35}
\: \)
within an accuracy of 10 \%,
if the temprature range is $M/100 < T < M/10$. 
This gives 
\begin{equation}
\delta n_{1}^{f} \approx 5.5 \times \,\frac{\lambda ^{2}}{64\pi ^{4}}\,
(\frac{T}{M})^{1.35}\,n_{\varphi } 
\approx 0.89\times 10^{-5}\,\lambda ^{2}\,(\frac{T}{M})^{1.35}\,
n_{\varphi } \,.
\label{f-dependent n numerical} 
\end{equation}
This result has a slightly slower decrease in temperature than the
analytic estimate of $\propto T^{2}\,n_{\varphi }$.
It nevertheless gives a subdominant contribution for $n_{{\rm eq}}$
at low temperatures.

The second piece $f_{2}$ is dominant in the region around
\( \:
\omega \approx -\,\omega _{k'} \,; 
\: \)
\begin{eqnarray}
&& \hspace*{-1cm}
f_{2}(\vec{k}) \approx 
\frac{2}{\omega _{k}}\,\int\,\frac{d^{3}k'}{(2\pi )^{3}2\omega _{k'}}\,
\frac{1 }{(\omega _{k'} + \omega _{k})^{2}}\,
\int_{0}^{\infty }\,d\omega \,\frac{r_{\chi }(\omega \,, \vec{k} + \vec{k'})}
{e^{\beta \omega } - 1}\,
\left( \,f(\vec{k'}) + 1 \,\right)
\\ &&
\hspace*{1cm} 
= f_{2}^{f}(\vec{k}) + f_{2}^{0}(\vec{k}) 
\,. 
\end{eqnarray}
We separated the second contribution into a sum of $f-$dependent
and $f-$independent integrals.
The last $\omega $ integral here is explicitly done in Appendix C
in a related model, giving with $q = |\vec{k} + \vec{k'}|$
\begin{eqnarray}
&&
\int_{0}^{\infty }\,d\omega \,\frac{r_{\chi }(\omega \,, q)}
{e^{\beta \omega } - 1} 
\approx \frac{4\zeta (2)\,T^{2}}{2q + \zeta (2)\,T}\,
\frac{1}{e^{q/2T} - 1} 
\,.
\end{eqnarray}
This gives, for the $f-$dependent part,
\begin{eqnarray}
&& \hspace*{-1.5cm}
f_{2}^{f}(\vec{k}) \approx \frac{\zeta (2)\,\lambda ^{2}}{16\pi ^{4}}\,
\frac{T^{2}}{k\,\omega _{k}}\,\int_{0}^{\infty }\,dk'\,\frac{k'}{\omega _{k'}}
\,\frac{f(k')}{(\omega _{k} + \omega _{k'})^{2}}\,
\int_{|k - k'|}^{k + k'}\,dq\,\frac{q}{2q + \zeta (2)T}\frac{1}{e^{q/2T} - 1}
\,. \nonumber \\ && 
\end{eqnarray}
Considering that the $q$ integral is suppressed at $q \gg T$, one obtains
for the integrated number density,
\begin{equation}
n_{2}^{f} \approx \frac{c\,\zeta (2)\lambda ^{2}}{4\pi ^{4}}\,
(\frac{T}{M})^{4}\,n_{\varphi } \,,
\end{equation}
for the $f$ dependent part.
Here $c \approx 0.27$ from eq.(\ref{decay ph integral}) of Appendix C.
This is smaller by a factor of order $(T/M)^{2}$ 
than the first piece $n_{1}^{f}$.

On the other hand,
the $f$ independent part gives for the distribution function and
the number density,
\begin{eqnarray}
&& 
\hspace*{-1cm}
f_{2}^{0}(\vec{k}) = \frac{\zeta (2)\lambda ^{2}}
{16\pi ^{4}}\,\frac{T^{2}}{k\omega _{k}}\,
\int_{0}^{\infty }\,dq\,\frac{q}{2q + \zeta (2)T}\,
\frac{1}{e^{q/2T} - 1}\,\left( \frac{1}{\omega _{k} + \omega _{k - q}}
- \frac{1}{\omega _{k} + \omega _{k + q}}\right) 
\,, \nonumber \\ && \label{dominant inverse f} 
\\ && \hspace*{0.5cm} 
n_{2}^{0} = \int_{0}^{\infty }\,\frac{k^{2}\,dk}{2\pi ^{2}}\,f_{2}^{0}
(k) 
\approx 
\frac{\zeta (2)c\,c'\,\lambda ^{2}}{8\pi ^{6}}\,\frac{T^{4}}{M}
\,, \label{f-independent equil-n} 
\\ &&
\hspace*{0.5cm} 
c' = \int_{1}^{\infty }\,dx\,\frac{\sqrt{x^{2} - 1}}{x^{3}} 
= \frac{\pi }{4} \,.
\end{eqnarray}
The total number density thus derived is roughly of order
\( \:
\lambda ^{2}\,T^{4}/M \,.
\: \)

Physical processes that contribute to the second piece $f_{2}^{0}$
are predominantly inverse annihilation 
$\chi \chi \rightarrow \varphi \varphi $, and 1 to 3 process,
$\chi \rightarrow \chi \varphi \varphi $, which gives a small fraction
of the number density.

The third piece $f_{3}(\vec{k})$ has the following spectral function,
\begin{eqnarray}
&& 
r_{3}(\omega \,, \vec{k}) = 2\,
\int\,\frac{d^{3}k'}{(2\pi )^{3}2\omega _{k'}}\,
\left( \,
\theta (\omega _{k'} - \omega )\,r_{\chi }(|\omega - \omega _{k'}|\,, 
\vec{k} - \vec{k'})\,f(\vec{k'}) 
\right.
\nonumber \\ && \hspace*{1.5cm} 
\left.
+\,
\theta (-\,\omega _{k'} - \omega )\,r_{\chi }(|\omega + \omega _{k'}|\,, 
\vec{k} + \vec{k'})\,(\,1 + f(\vec{k'})\,) \,\right) \,, 
\\ && \hspace*{1cm} 
r_{\chi }(|\omega| \,, \vec{q}) = \frac{\lambda ^{2}}{16\pi ^{2}}\,
\left( \,\theta (|\omega| - q) + \frac{2}{\beta q}\,
\ln \frac{1 - e^{-\beta (|\omega| + q)/2}}
{1 - e^{-\beta |\,|\omega| - q\,|/2}}\,\right) 
\,.
\end{eqnarray}
One can separate the temperature independent ($f^{(0)}$)
and dependent ($f^{(T)}$) pieces 
for the third contribution by the presence of the $\beta $ factor;
prior to the subtraction of the counter term,
\begin{eqnarray}
&& \hspace*{0.5cm} 
\delta f_{3}^{(i)} = 
\frac{\lambda ^{2}}{128\pi ^{4}\,k\omega _{k}}\,
\int\,dk'\frac{k'}{\omega _{k'}}\,\int_{-\infty }^{\infty }\,d\omega \,
\frac{1}{(\omega - \omega _{k})^{2} + \Gamma _{k}^{2}/4}\,
\nonumber 
\\ && \hspace*{2cm} 
\cdot 
\int_{|k - k'|}^{k + k'}\,dq\,q\,
\left( \,
G_{i}(\omega \,, \omega _{k'} \,, q) 
- F_{i}(\omega_{k'} - \omega _{k}, q)\,f(k') 
\,\right)
\,, \label{unsubtracted off-shell 3} 
\\ && \hspace*{-1.5cm}
G_{i}(\omega \,, \omega _{k'} \,, q) = 
F_{i}(-\,\omega - \omega _{k'}, q)\,(\,1 + f(k')\,)
+ F_{i}(-\,\omega + \omega _{k'} , q)\,f(k') 
\,, 
\\ && 
F_{0}(x\,,  q) = \theta (x - q) 
\,, \hspace{0.5cm} 
F_{T}(x \,, q) =  \frac{2T}{q}\,\ln \frac{1 - e^{-\beta (|x| + q)/2}}
{1 - e^{-\beta |\,|x| - q\,|/2}} \,.
\end{eqnarray}

There are divergent terms of the form,
\begin{equation}
\frac{A + B\,\vec{k}^{2}}{4\omega _{k}^{2}} \,, 
\end{equation}
which are cancelled by the mass and the wave function counter terms
in the proper self-energy $\Pi (\omega \,, \vec{k})$.
To see this, let us focus on the term both independent of
the temperature and of the distribution function $f$,
which can be worked out by the $\omega $ integration explicitly;
\begin{eqnarray}
&& 
f_{3}^{(0\,, 0)}(k) \approx \frac{\lambda ^{2}}{128\pi ^{4}}\,
\frac{1}{k\omega _{k}}\,\int_{0}^{\Lambda }\,dk'
\nonumber \\ && 
\hspace*{0.5cm} \cdot \frac{k'}{\omega _{k'}}\,
\left( \, k + k' - |k - k'|
- (\omega _{k'} + \omega _{k})\,
\ln \frac{\omega _{k'} + \omega _{k} + k + k'}
{\omega _{k'} + \omega _{k} + |k - k'|}\,\right)
\,,
\end{eqnarray}
where $\Lambda $ is the momentum cutoff.
By expanding around the momentum $\vec{k} = 0$, 
we may identify 
\begin{eqnarray}
&&
A = \frac{\lambda ^{2}}{16\pi ^{4}}\,M\,\int_{0}^{\Lambda }\,dk'\,
\frac{k'\,^{2}}{\omega _{k'}\,(\omega _{k'} + k' + M)}
\,, 
\\ && 
B = \frac{A}{2\,M^{2}} - \frac{\lambda ^{2}}{32\pi ^{4}}\,
\int_{0}^{\Lambda }\,dk'\,
\frac{k'\,^{2}}{\omega _{k'}\,(\omega _{k'} + k' + M)^{2}}
\,. \label{wave function renorm 1} 
\end{eqnarray}

The logarithmic infinity, 
the second term of eq.(\ref{wave function renorm 1}),
is related to the two-loop infinity of the self-energy diagram of Fig.15.
By a straightforward calculation this diagram gives a wave
function renormalization factor,
\begin{equation}
\delta Z = -\,\frac{\lambda ^{2}}{256\pi ^{4}}\,\ln \Lambda ^{2} \,.
\end{equation}
One sees that this corresponds to the logarithmic infinity of 
eq.(\ref{wave function renorm 1}), confirming the relativistic
covariance of the wave function renormalization.

Removal of infinities thus works with renormalization.
We are not much interested in the remaining, temperature independent
finite part, since this is absorbed by a new definition of the vacuum
in interacting field theory.

We thus next turn to the finite, temperature dependent part
$\delta f_{3}^{(T)}$.
For some part of this integral 
it is not difficult to count the mass $M$ dependence
in the low temperature limit.
Thus, the integrated number density from this part
is either of the form,
\( \:
\lambda ^{2}\,(T/M)^{4}\,n_{\varphi }
\: \)
or 
\( \:
\lambda ^{2}\,(T/M)^{4}\,T^{3} \,.
\: \)
These are smaller by a positive power of $\frac{T}{M}$ than
$\delta n_{1}$ and $\delta n_{2}$ we already considered.
The remaining piece has the spectrum 
\( \:
\theta (\omega _{k'} - \omega )\,r_{\chi }^{(T)}(|\omega - \omega _{k'}|
\,, \vec{k} - \vec{k'})
\: \)
with $r_{\chi }^{(T)}$ the temperature dependent two-body spectrum.
This contribution can be combined with $f_{1}(\vec{k})$ of 
eq.(\ref{dominant off-shell d-function 1}), to give the following
$f-$dependent distribution,
\begin{eqnarray}
&& \hspace*{-0.5cm}
f_{f}(\vec{k}) = 
\frac{\lambda ^{2}}{64\pi ^{4}}\,\frac{T}{k\omega _{k}}\,
\int_{0}^{\infty }\,dk'\,
\frac{k'}{\omega _{k'}}\,f(k')\,\int_{|k - k'|}^{k + k'}\,dq\,
\int_{-\infty }^{\infty }\,d\omega \,
\nonumber \\ && \hspace*{-1cm} \cdot 
\frac{1}{(\omega - \omega _{k})^{2} + \Gamma _{k}^{2}/4}\,
\left( \,\frac{1}{e^{\beta (\omega - \omega _{k'})} - 1}\,
\ln \frac{1 - e^{-\beta (|\omega - \omega _{k'}| + q)/2}}
{1 - e^{-\beta |\,|\omega - \omega _{k'}| - q\,|/2}}
- (\omega \rightarrow \omega _{k}) \,\right) \,.
\label{f-indepent off-shell f} 
\end{eqnarray}
We have numerically computed this $k'$ integrand, which turns out
well in proportion to $k'\,^{2}$ in the non-relativistic limit.
Thus, the integrated number density is of order,
\( \:
\lambda ^{2}(\frac{T}{M})^{2}\,n_{\varphi } \,,
\: \)
of the same order as $n_{1}^{f}$ of eq.(\ref{inverse scat to n}).

In summary, the dominant term of the off-shell contribution is
\begin{equation}
\delta f_{{\rm eq}}(\vec{k}) \approx f_{f}(\vec{k}) + f_{2}^{0}(\vec{k}) \,, 
\end{equation}
where $f_{f}$ is given by eq.(\ref{f-indepent off-shell f}) or
its numerically better alternative, and
$f_{2}^{0}$  by eq.(\ref{dominant inverse f}).
The use of the more precise numerical result for the $f-$dependent
term $f_{f}$ such as (\ref{f-dependent n numerical}) is
not necessary. since it is shown in Section 4 that the $f-$dependent
term $f_{2}^{0}$ dominates in the evolution equation for the
number density, thus 
\( \:
\delta f_{{\rm eq}} \approx f_{2}^{0}
\: \).

\newpage
\begin{Large}
\begin{center}
{\bf Figure caption}
\end{center}
\end{Large}

\vspace{0.5cm} 
\hspace*{-0.5cm}
{\bf Fig.1}

Tadpole diagrams.

\vspace{0.5cm} 
\hspace*{-0.5cm}
{\bf Fig.2}

Two-body spectral function $r_{\chi }(\omega \,, k)$ for 
two massless $\chi \chi $ state.
Two choices of the momentum $k$ relative to the temperature are shown.

\vspace{0.5cm} 
\hspace*{-0.5cm}
{\bf Fig.3}

Four different processes contributing to the Boltzmann equation.
The solid lines are for the $\varphi $ particle, while the dotted
are for the $\chi $ particle.

\vspace{0.5cm} 
\hspace*{-0.5cm}
{\bf Fig.4}

Singularity structure of the self-energy function $F(z)$.
The dotted cross gives poles in the second Riemann sheet,
while the wavy line is the branch-cut singularity.

\vspace{0.5cm} 
\hspace*{-0.5cm}
{\bf Fig.5}

Functions of the integrand in the energy integral,
eq.(\ref{integral formula f-independent f}) in the text.
The dotted line is the Planck distribution of $T = 0.01\,M$, 
while the dashed
is the rest of the Breit-Wigner form, with the solid line
giving product of these two. Parameter values taken are
\( \:
k = 0.01\,M \,, k' = 0.005\,M \,, \lambda = 0.01 \,.
\: \)

\vspace{0.5cm} 
\hspace*{-0.5cm}
{\bf Fig.6}

Time evolution of the yield $Y$ for $\varphi $ mass of $10\,GeV$,
and indicated couplings.
For a comparison the on-shell Boltzmann result is also shown.

\vspace{0.5cm} 
\hspace*{-0.5cm}
{\bf Fig.7}
Time evolution of the yield $Y$ for the $\lambda = 10^{-3}$
and a few $\varphi $ mass values.

\vspace{0.5cm} 
\hspace*{-0.5cm}
{\bf Fig.8}

Comparison of the S-wave and P-wave annihilation in the time evolution.

\vspace{0.5cm} 
\hspace*{-0.5cm}
{\bf Fig.9}

The parameter region for the off-shell and the on-shell dominance
in the S-wave boson pair annihilation model.
The contour of the closure density, $\rho _{c} = 10^{4}\,eV\,cm^{-3}$,
is also shown.
The parameter relations are
\( \:
\delta = 5.4\times 10^{-6}\,\lambda ^{2} \,, 
\delta /\eta = 1.2\times 10^{-22}\,M/GeV \,.
\: \)

\vspace{0.5cm} 
\hspace*{-0.5cm}
{\bf Fig.10}

Contour lines for the present mass density in the unit of
$\rho _{c} = 10^{4}\,eV\,cm^{-3}$ (S-wave model).

\vspace{0.5cm} 
\hspace*{-0.5cm}
{\bf Fig.11}

Similar contour lines to Fig.10 for the P-wave
annihilaation. 

\vspace{0.5cm} 
\hspace*{-0.5cm}
{\bf Fig.12}

Diagram for two-body spectral function $r_{\chi }$.

\vspace{0.5cm} 
\hspace*{-0.5cm}
{\bf Fig.13}

Equilibrium distribution function for the unstable particle decay
at $T = 0.05\,M$.
The exact computation (solid) is compared with the approximate one
(dotted), and the Planck formula (dashed).
In the inset result is given in linear scale.

\vspace{0.5cm} 
\hspace*{-0.5cm}
{\bf Fig.14}

Equilibrium number density;
exact result (solid) is compared with the on-shell result (dotted).

\vspace{0.5cm} 
\hspace*{-0.5cm}
{\bf Fig.15}

Self-energy diagram.

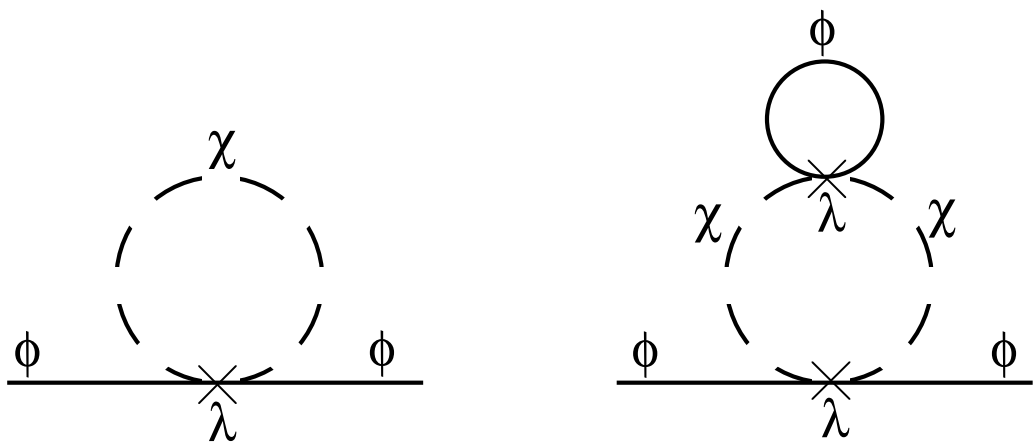
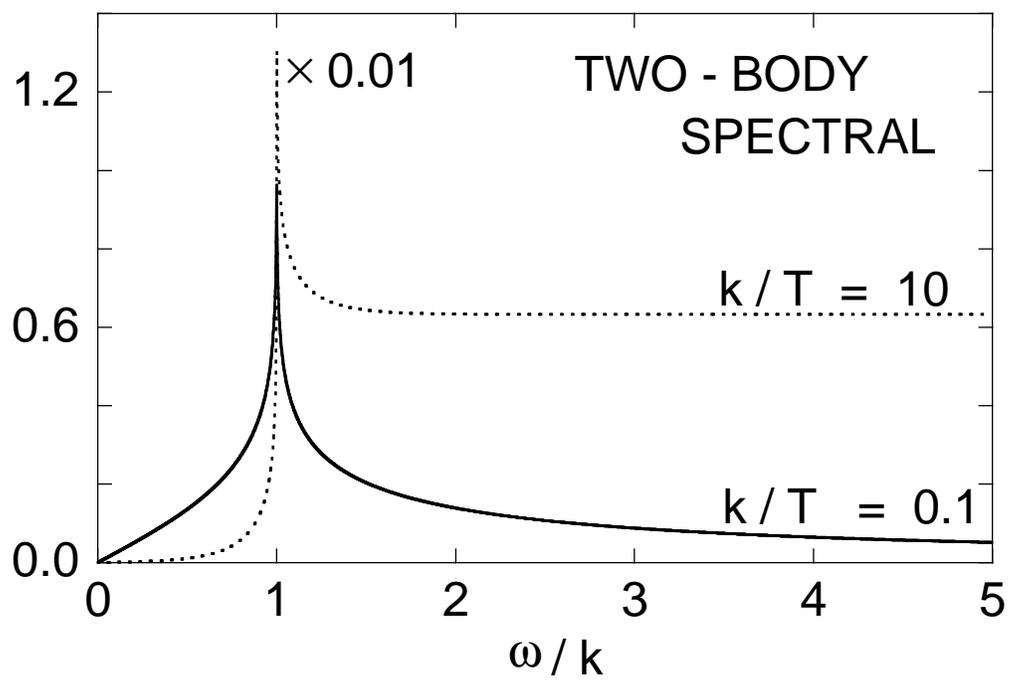
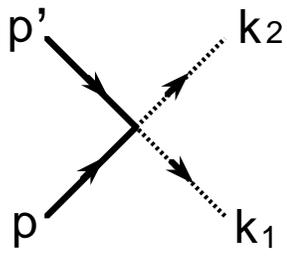
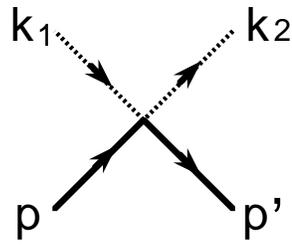
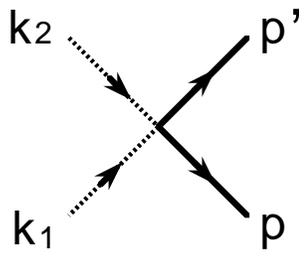
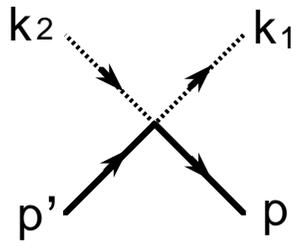
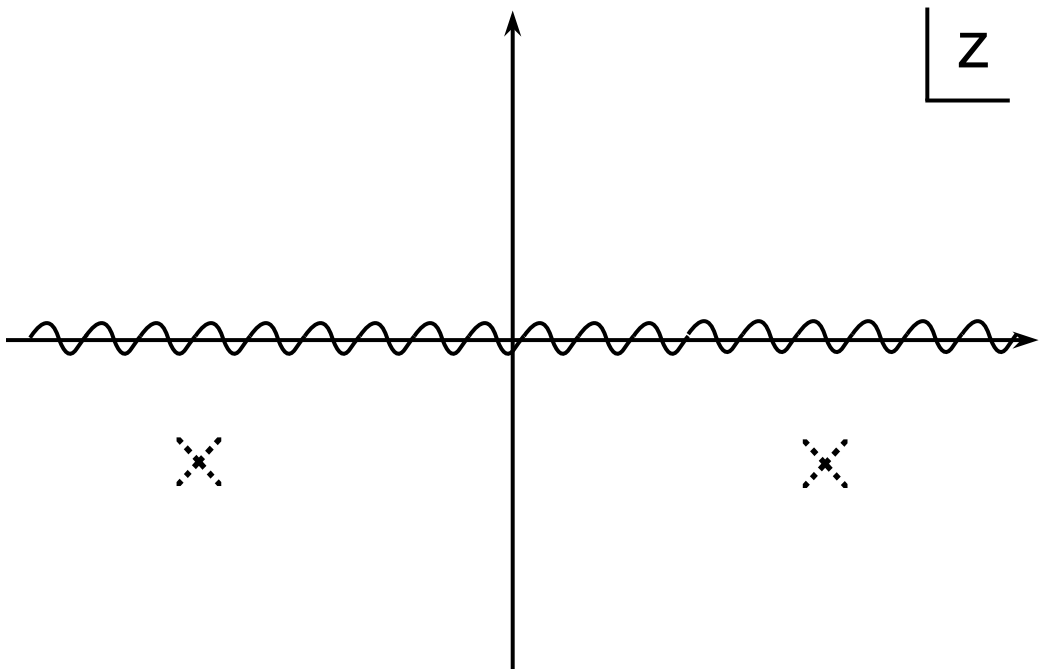
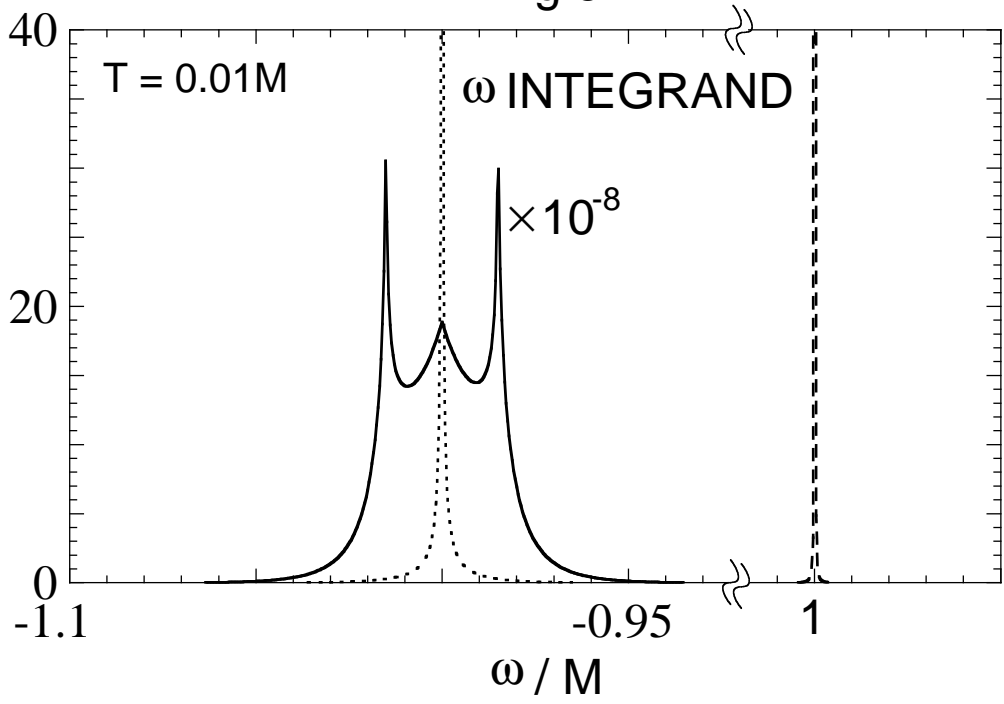
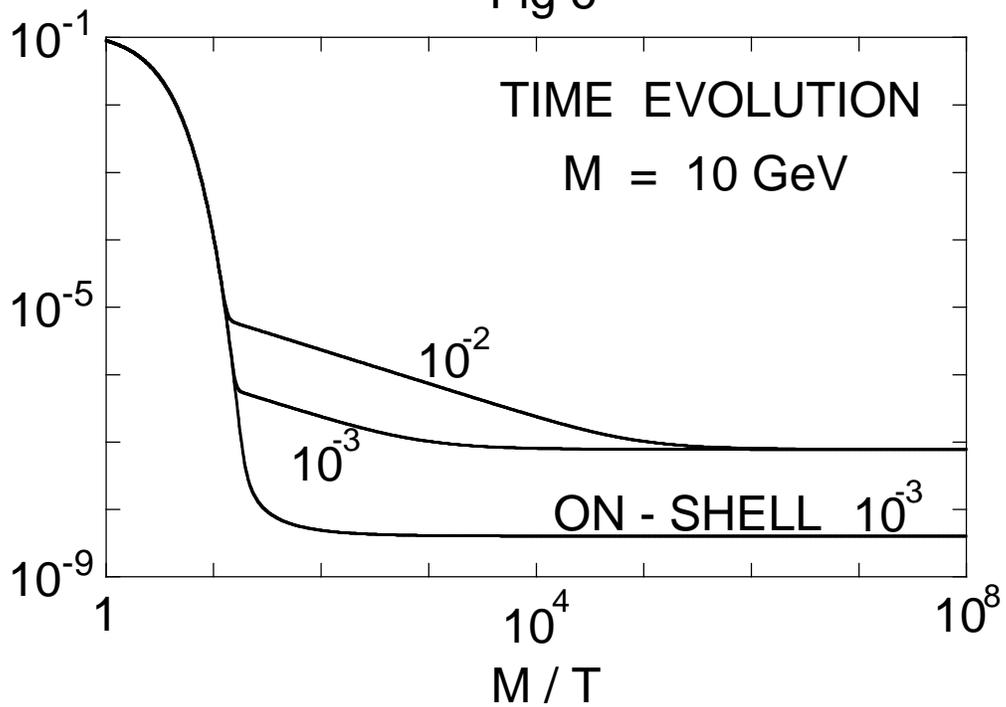
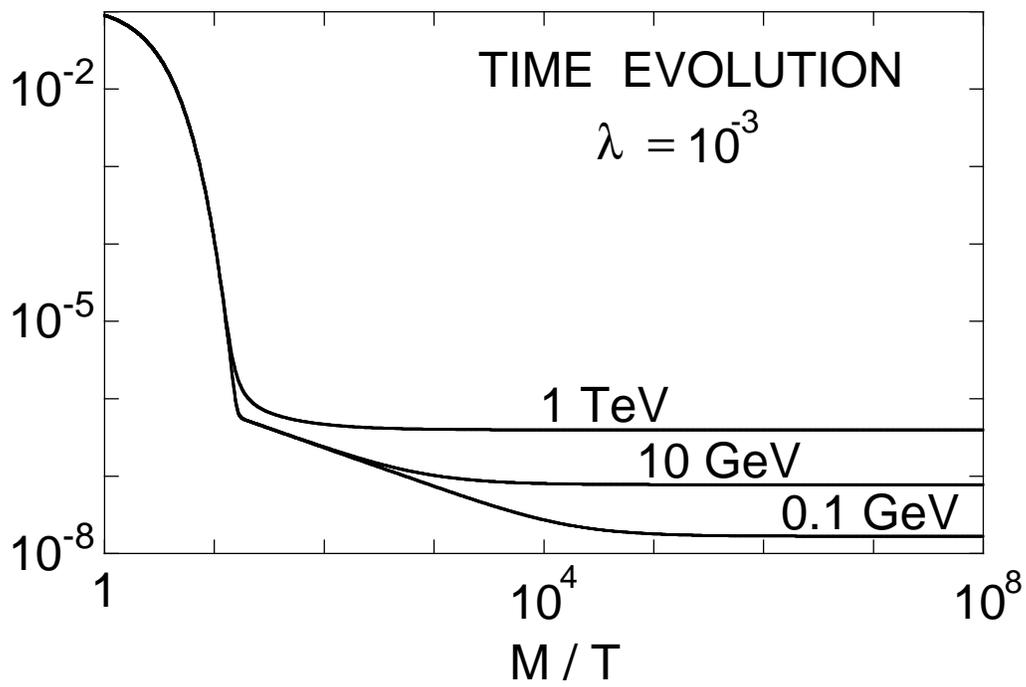
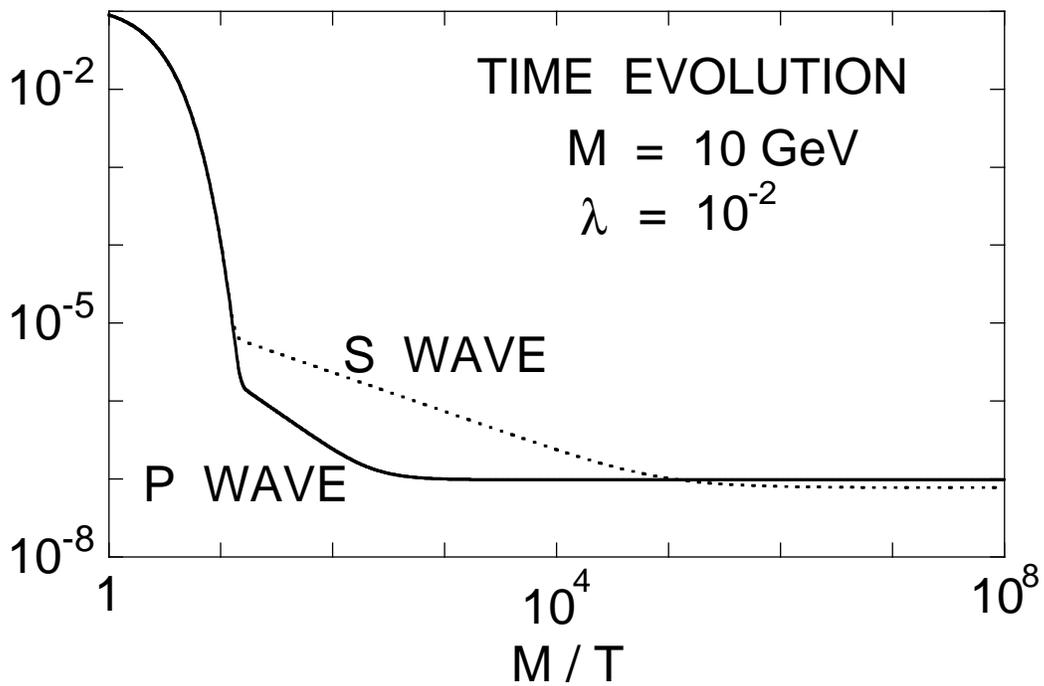
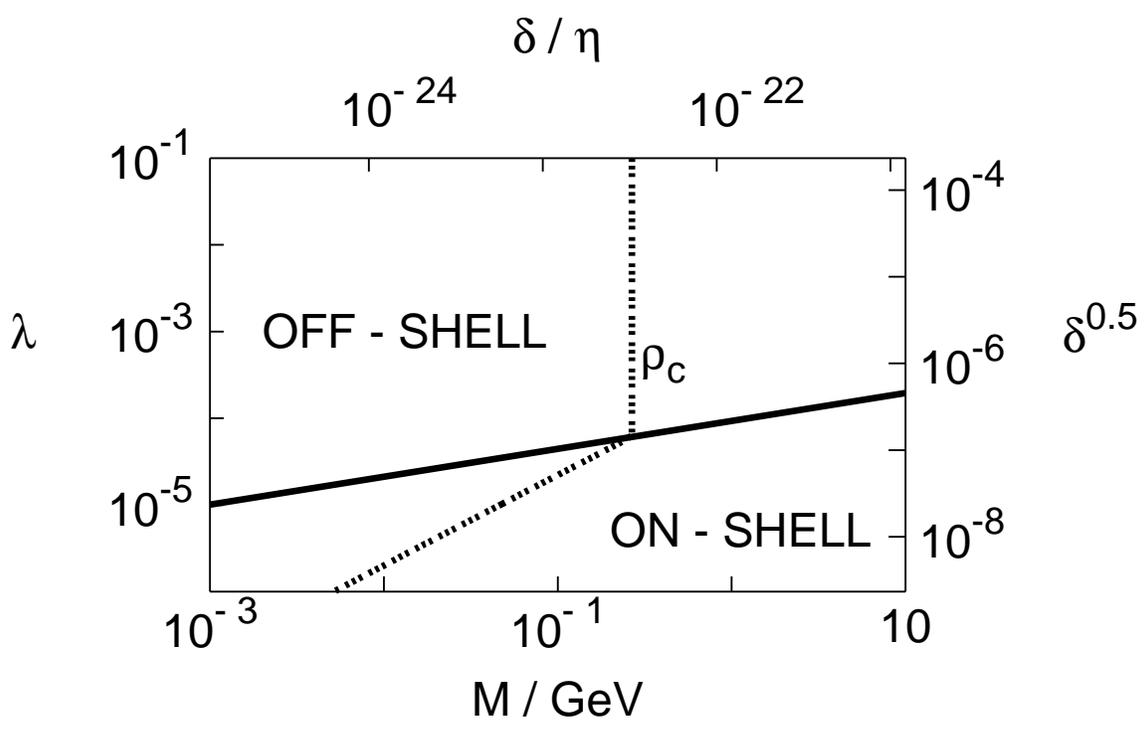
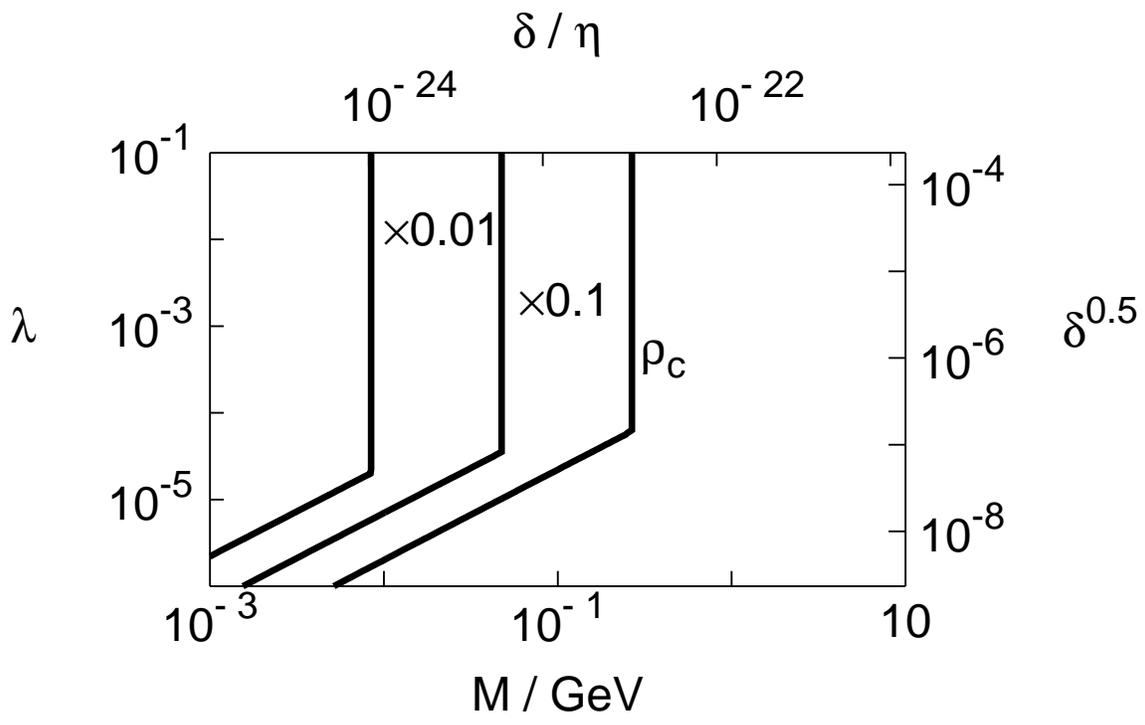
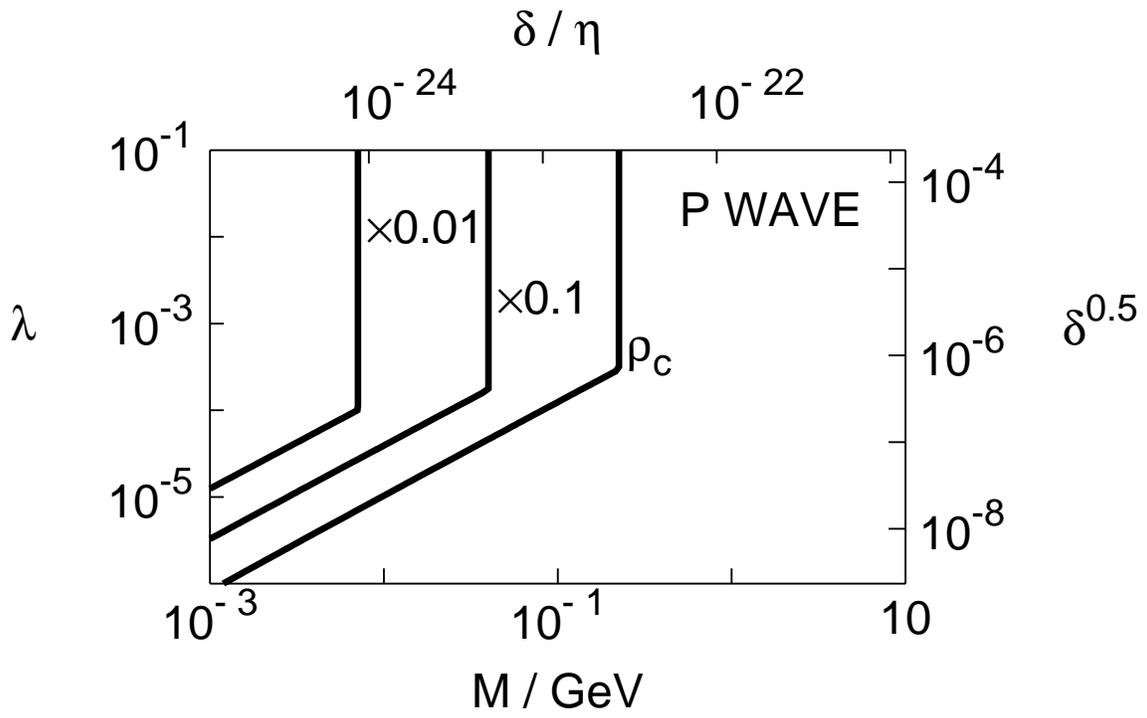
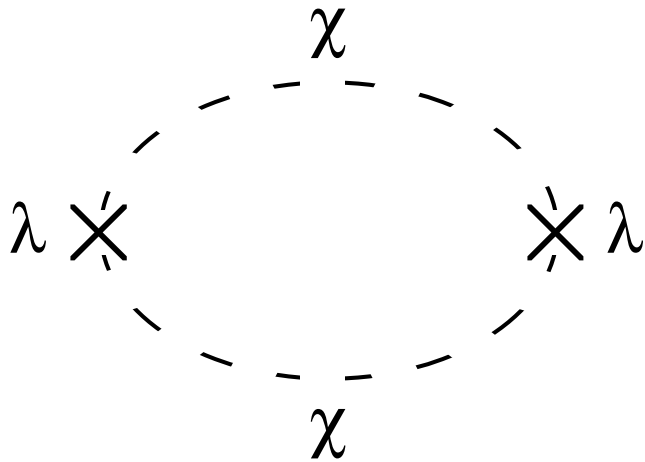
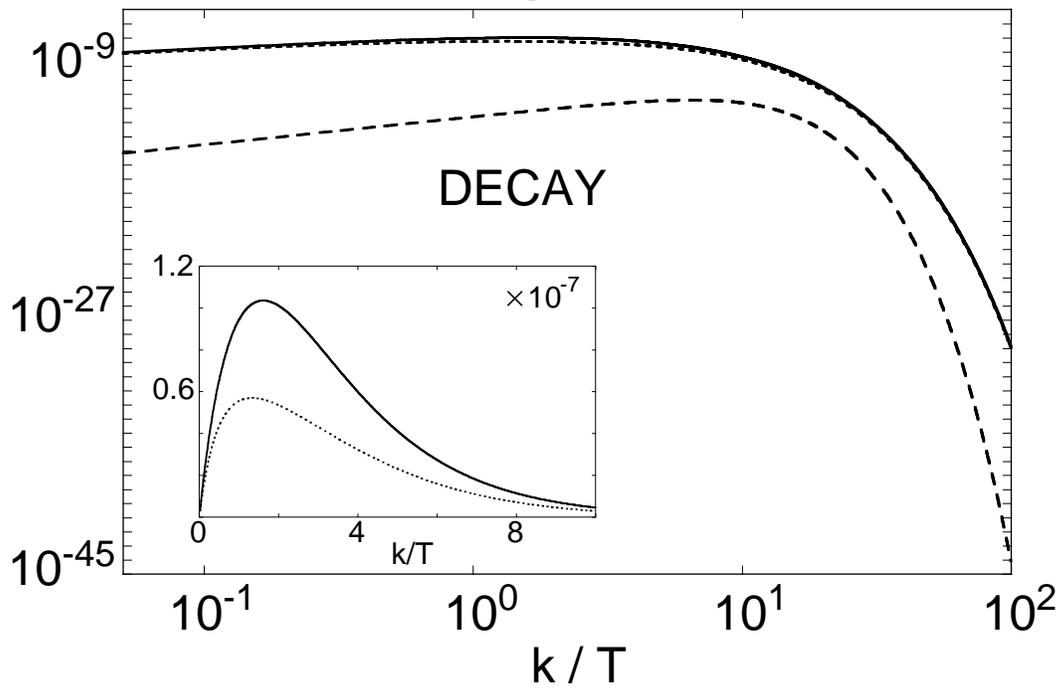
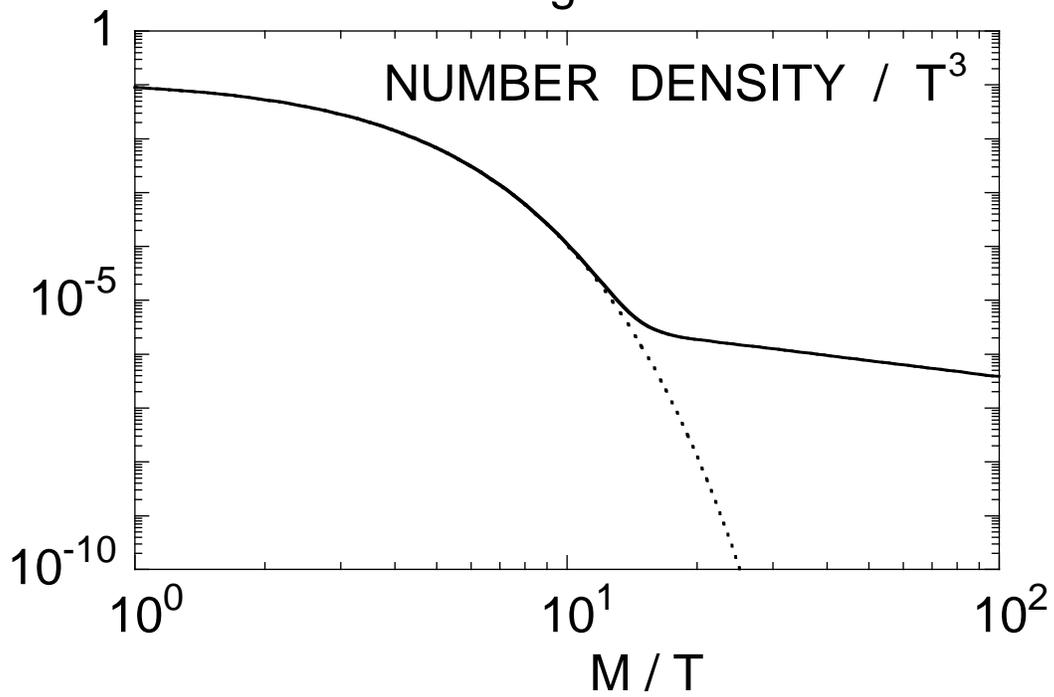
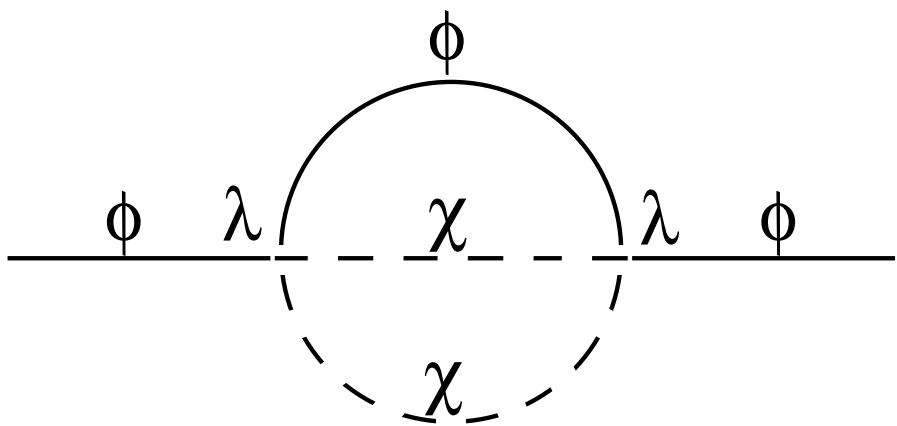
\end{document}